\documentclass[12pt,preprint]{aastex}
\shorttitle{}
\shortauthors{Nesvorn\'y et al.}
\begin{document}
\title{Origin and Evolution of Short-Period Comets}
\author{David Nesvorn\'y$^1$, David Vokrouhlick\'y$^2$, Luke Dones$^1$, Harold F. Levison$^1$, \\
Nathan Kaib$^3$, Alessandro Morbidelli$^4$}
\affil{(1) Department of Space Studies, Southwest Research Institute, 1050 Walnut St., \\Suite 300, 
Boulder, CO 80302, USA} 
\affil{(2) Institute of Astronomy, Charles University, V Hole\v{s}ovi\v{c}k\'ach 2, CZ--18000 Prague 8,
Czech Republic} 
\affil{(3) HL Dodge Department of Physics and Astronomy, University of Oklahoma, Norman, OK 73019, USA}
\affil{(4) D\'epartement Cassiop\'ee, University of Nice, CNRS, Observatoire de la C\^ote d'Azur, Nice, 
06304, France}
\begin{abstract}
Comets are icy objects that orbitally evolve from the trans-Neptunian region 
into the inner Solar System, where they are heated by solar radiation and become 
active due to sublimation of water ice. Here we perform simulations in which cometary 
reservoirs are formed in the early Solar System and evolved over 4.5 Gyr. The gravitational 
effects of Planet~9 (P9) are included in some simulations.
Different models are considered for comets to be active, including a simple assumption that comets 
remain active for $N_{\rm p}(q)$ perihelion passages with perihelion distance $q<2.5$ au. The orbital 
distribution and number of active comets produced in our model is compared to observations. The 
orbital distribution of ecliptic comets (ECs) is well reproduced in models with $N_{\rm p}(2.5)\simeq500$ 
and without P9. With P9, the inclination distribution of model ECs is wider than the observed one. 
We find that the known Halley-type comets (HTCs) have a nearly isotropic inclination distribution. 
The HTCs appear to be an extension of the population of returning Oort-cloud comets (OCCs) to shorter 
orbital periods. The inclination distribution of model HTCs becomes broader with increasing $N_{\rm p}$, 
but the existing data are not good enough to constrain $N_{\rm p}$ from orbital fits. 
$N_{\rm p}(2.5)>1000$ is required to obtain a steady-state population of large active HTCs that is 
consistent with observations. To fit the ratio of the returning-to-new OCCs, by contrast, our model 
implies that $N_{\rm p}(2.5)\lesssim 10$, possibly because the detected long-period comets are smaller 
and much easier to disrupt than observed HTCs.  
\end{abstract}
\keywords{comets:general}
\section{Introduction}
Comets are icy objects that reach the inner Solar System after leaving distant reservoirs beyond 
Neptune and dynamically evolving onto elongated orbits with very low perihelion distances (see Dones 
et al. (2015) for a review). Their activity, manifesting itself by the presence of a dust/gas coma 
and characteristic tail, is driven by solar heating and sublimation of water ice. Comets are 
short-lived, implying that they must be resupplied from external reservoirs (Fern\'andez 1980, 
Duncan et al. 1988). The goal of this work is to model the formation of cometary reservoirs early 
in Solar System history, follow their evolution to the present time, and see how observations of comets 
can be used to constrain the orbital structure of trans-Neptunian populations. Our main focus is the 
short-period comets (SPCs), because the population of comets with short orbital periods ($P<200$ yr) is 
relatively well characterized from observations and allows us to meaningfully constrain the model. We aim at 
better understanding of the origin and dynamical/physical evolution of SPCs.     

Levison \& Duncan (1997, hereafter LD97) considered the origin and evolution of ecliptic comets 
(ECs; see Section 2 for a definition and their relationship to the Jupiter-family comets, JFCs). 
The Kuiper belt at 30-50 au was assumed in LD97 to be the main source of ECs. They
showed that small Kuiper belt objects (KBOs) reaching a Neptune-crossing orbit can be slingshot, 
by encounters with different planets, to very low perihelion distances ($q<2.5$~au), at which 
point they are expected to become active and visible. The new ECs, reaching $q<2.5$~au for the 
first time, have a narrow inclination distribution in the LD97 model, because their orbits
were assumed to start with low inclinations ($i<5^\circ$) in the Kuiper belt, and the inclinations 
stayed low during the orbital transfer. LD97 pointed out that the inclination distribution of ECs 
becomes wider over time due to scattering encounters with Jupiter. 
The best fit to the observed inclination distribution of ECs was obtained in LD97 when it was assumed 
that ECs remain active for $\simeq$12,000 years after first reaching $q<2.5$ au.

The escape of bodies from the classical Kuiper belt at 30-50 au (hereafter classical KB) is driven by 
slow chaotic processes in various orbital resonances with Neptune. Because these processes affect only 
part of the belt, with most orbits in the belt being stable, questions arise about the 
overall efficiency of comet delivery from the classical KB. Duncan \& Levison (1997), concurrently 
with the discovery of the first Scattered Disk Object (SDO; (15874) 1996 TL66,  Luu et al. 1997),
suggested that the scattered disk should be a more prolific source of ECs than the classical KB
(see Gladman et al. (2008) for a formal definition of these dynamical classes). This is because 
SDOs can approach Neptune during their perihelion passages and be scattered by 
Neptune to orbits with shorter orbital periods. The scattered disk should thus produce more ECs 
than the unstable part of the classical KB.

The inclination distribution of SDOs is wider than the one used as the source of ECs in LD97. 
This is because the scattered disk is in contact with Neptune, and the orbital inclinations of
SDOs increase over time by scattering encounters with Neptune. In fact, the inclination distribution 
of SDOs (median $i\simeq25^\circ$; Nesvorn\'y et al. 2016) is much broader than that of ECs (median
$i\simeq13^\circ$; Section 2). This raises a question whether the scattered disk can produce 
the narrow inclination distribution of ECs (Rickman et al. 2017). Di Sisto et al. (2009) tested 
this issue but assumed that SDOs have lower inclinations (Di Sisto \& Brunini 2007; median 
$i\simeq 15^\circ$; their Fig. 3) than they have in reality (Kaib \& Sheppard 2016, Nesvorn\'y et 
al. 2016). 

The Halley-type comets (HTCs) have longer orbital periods and larger inclinations than do most ECs. 
The HTCs population is not well characterized from the existing observations, because the observational 
biases for long orbital periods are more severe than for ECs (see Section 2). Levison et al. (2001) 
studied the origin and evolution of HTCs. They suggested that HTCs evolve into the inner Solar System 
from an inner, presumably flattened part of the Oort cloud. This theory was motivated by the 
inclination distribution of HTCs, which at the time of the Levison et al. work was thought to be flattened 
with a median of $\simeq$45$^\circ$. Later on,  Levison et al. (2006) considered the scattered disk as 
the main source of HTCs and showed that some SDOs can evolve into the Oort cloud and back, thus 
providing an anisotropic source of HTCs. Back in 2006, the median orbital inclination of HTCs was thought 
to be $\simeq$55$^\circ$, somewhat larger than in 2001, but still clearly anisotropic. This turns 
out to be part of a historical trend (Wang \& Brasser 2014 and Section 2).

Our understanding of the origin and evolution of comets is incomplete in part because the presumed source 
populations of trans-Neptunian objects with cometary sizes ($\sim$1-10 km) are not well characterized 
from observations. It is therefore difficult to establish whether there are enough small objects  
in any trans-Neptunian reservoir to provide the source of comets (e.g., Volk \& Malhotra 2008). 
Turning this argument around, previous studies typically assumed that a specific type of comets come from a specific reservoir
(e.g., ECs from the scattered disk), reconstructed that reservoir from observations and theoretical 
considerations, and modeled comet delivery from the reservoir to infer how many comet-size objects 
currently need to be in the reservoir to explain the observed population of comets. 

For example, Brasser \& Morbidelli (2013; hereafter BM13) found that the scattered disk needs to contain $\sim2 \times 10^{9}$ 
bodies with diameter $D>2.3$ km to provide an adequate source of ECs, and the Oort cloud needs to have 
$\sim4\times10^{10}$ to $\sim10^{11}$ bodies with $D>2.3$ km to explain the flux of new Oort cloud comets (OCCs). 
Rickman et al. (2017) suggested there are $\sim10^9$ SDOs with $D>2$ km from modeling of ECs. Levison et al. 
(2006), on the other hand, required that there are $\sim 3 \times 10^{9}$ SDOs with $D>10$ km to produce the 
observed population of HTCs, assuming no physical evolution (more SDOs would be needed if they accounted 
for physical disruption/fading of HTCs). 

While some of these estimates may appear to be high, it is not obvious whether they are 
implausibly high, because we just do not know from observations how many small objects there are in 
the distant regions. Another approach to this problem, which we pursue here, is to perform 
end-to-end simulations in which cometary reservoirs are produced in the early Solar System
and evolved over 4.5~Gyr (see also BM13). The number of comets produced in 
the model at $t=4.5$~Gyr can be inferred from the number of comets in the original trans-planetary disk, 
which in turn can be calibrated from the number of Jupiter Trojans (Nesvorn\'y \& Vokrouhlick\'y 2016, hereafter NV16). 
This is because the Trojan implantation efficiency from the original disk is well-determined 
(Nesvorn\'y et al. 2013; see also Morbidelli et al. 2005) and because the size distribution 
of Trojans is well characterized from observations (e.g., Wong \& Brown 2015, Yoshida et al. 2017). 
If the calibration works, the model should match the observed number of comets, and the number of comet-sized 
objects in the distant reservoirs can be inferred from it (Section~4.7). 

This approach, to be reliable, requires that we have a good model for the early evolution of the 
Solar System. Here we use the model developed in Nesvorn\'y \& Morbidelli (2012; hereafter NM12),
which was inspired by many previous studies (e.g, Tsiganis et al. 2005, Morbidelli et al. 2007). 
NM12 performed a large number of self-consistent simulations of early
planetary migration/instability in an attempt to identify the initial configuration and dynamical 
evolution of planets that would lead to the Solar System as we know it now. This includes the number 
and orbits of the outer planets and survival of the terrestrial planets. The identified solutions 
were scrutinized against various constraints from small body populations such as the asteroid and Kuiper belts, 
Jupiter Trojans, regular and irregular moons of the outer planets (see NV16 and references 
therein), showing the general applicability of the NM12 model to various problems. Here we use 
the NM12 model to study cometary populations. 

In section 2, we discuss the dynamical classification of SPCs, their orbital distribution and
physical properties. Our model is explained in Section 3. The results are reported in Section 4
and discussed in Section 5. We confirm the conclusion of previous studies that the scattered disk 
is the main source of ECs, and find that the Oort cloud is the main source of HTCs. Planet 9 (hereafter P9), 
hypothesized to exist on a wide orbit around the Sun (Trujillo \& Sheppard 2014, Batygin \& Brown 2016a),
is included in some of our simulations (see Section 3) to test its influence on the structure 
of the trans-Neptunian region and comet delivery. We find that P9 would enhance the flux 
of HTCs by $\sim$30\%. The inclination distribution of ECs can be matched in a straightforward 
manner in a model without P9, but when P9 is included, it acts to increase the inclination 
dispersion of SDOs. This propagates into the inclination distribution of ECs, which then appears 
to be too broad to match observations. 
\section{Properties of known SPCs}
SPCs are defined as bodies showing cometary activity and having short orbital periods 
($P<200$ yr).\footnote{The main belt comets or ``active'' asteroids (Jewitt et al. 2015) are not considered here.}
The period range is arbitrary, because there is nothing special about the boundary at the 200-yr period,
and the orbital period distribution of known comets appears to continue smoothly across this boundary. 
With $P<200$ yr, SPCs are guaranteed to have at least one perihelion passage in modern history, 
with many being observed multiple times. This contrasts with the situation for the long-period comets 
(LPCs; $P>200$ yr), which can be detected only if their perihelion passage coincides with the present epoch.

\begin{figure}
\epsscale{0.8}
\plotone{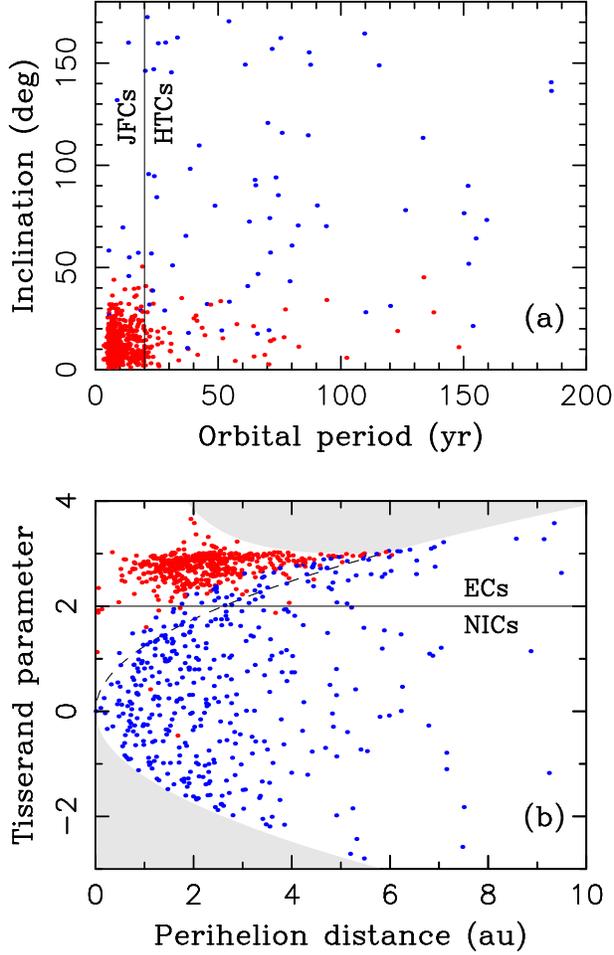}
\caption{The orbital distribution of known SPCs. The thin lines show the division between 
JFCs and HTCs (panel a; $P=20$ yr), and between ECs and NICs (panel b; $T_{\rm J}=2$). The color 
indicates the relationship between different categories. In panel a, the red dots denote
ECs with $T_{\rm J}>2$, and the blue dots denote NICs with $T_{\rm J}<2$. In panel b, the red 
dots denote JFCs with $P<20$ yr, and the blue dots denote comets with $P>20$ yr and $a<10,$000~au. 
The gray areas in panel b cannot be reached by orbits. The dashed line in panel b is 
$T_{\rm J}=2\sqrt{2q}$, which is an approximate boundary of prograde orbits evolving from 
$a \gg a_{\rm J}$ and $e \sim 1$.}
\label{spc}
\end{figure}

Figure \ref{spc} shows the orbital distribution of known SPCs. We obtained these data from the 
JPL Small-Body Database Search Engine\footnote{\tt http://ssd.jpl.nasa.gov/sbdb\_query.cgi} in January 2017.
To reduce the influence of observational biases we only show comets with {\it known} total (nucleus and coma) 
absolute magnitude, $H_{\rm T}$. Comets that do not have $H_{\rm T}$ reported in the JPL database are discarded,
because their detection circumstances are unclear. Paired bodies, such as fragments of disrupted 
comet 73P/Schwassmann-Wachmann, were removed, leaving only one data point for each parent comet 
in Fig. \ref{spc}. 

The orbital distribution of SPCs shows clear evidence for two populations, which are 
historically known as JFCs and HTCs. The JFC 
population is a tightly concentrated group in orbital space with short orbital periods and low 
inclinations. The HTC population is more dispersed in period-inclination space. These characteristics 
hint at different origins of JFCs and HTCs.

The distinction between JFCs and HTCs is blurred, at least to some degree, because the two populations
partially overlap in orbital space. A traditional approach to the classification problem is to define 
JFCs as comets with $P<20$ yr. This definition is motivated by the fact that the dense clump of comets 
in Fig. \ref{spc}a shows periods $5\lesssim P \lesssim20$ yr, which is similar to the orbital period 
of Jupiter ($P_{\rm J}=11.9$ yr). Another approach to this problem would be to use a criterion based 
on inclinations and define JFCs as comets with orbital inclinations lower than some threshold.  

LD97 opted, instead, to use the Tisserand invariant of the circular restricted three-body problem (Tisserand 1889), 
which conveniently combines the comet's orbital period (or, equivalently, the semimajor axis $a$) and inclination 
into a single expression. 
The Tisserand parameter with respect to Jupiter, $T_{\rm J}$, is defined as 
\begin{equation}
T_{\rm J}={a_{\rm J} \over a} + 2 \sqrt{(1-e^2) {a \over a_{\rm J}}} \cos i \ ,
\end{equation}      
where $a_{\rm J}$ is Jupiter's semimajor axis, and $e$ and $i$ are the comet's orbital eccentricity and 
inclination. Since Jupiter's orbit is slightly eccentric, $T_{\rm J}$ is not strictly conserved
and changes over time. Still, the evolution of $T_{\rm J}$ is much slower than the evolution of $P$,
and $T_{\rm J}$ therefore represents a better classification parameter than $P$. It can be shown 
that $U/v_{\rm J}=\sqrt{3-T_{\rm J}}$, where $v_{\rm J}$ is the orbital speed of Jupiter, and $U$ is the 
encounter speed between Jupiter (assumed to move on a strictly circular orbit) and a comet.
Comets with $T_{\rm J}$ just below 3 can therefore have low-velocity encounters with Jupiter.

LD97 classified comets into two categories: (1) ECs with $2<T_{\rm J}<3$, and (2) nearly isotropic comets (NICs) 
with $T_{\rm J}<2$. This classification is highlighted in Fig. \ref{spc}b. Most JFCs ($P<20$ yr) are ECs 
($2<T_{\rm J}<3$), and vice versa, and most HTCs ($20 < P < 200$ yr) are NICs ($T_{\rm J}<2$). The NIC category 
is broad, however, and includes LPCs as well. Orbits with $T_{\rm J}>3$ are generally 
not Jupiter crossing, and are therefore typically not classified as cometary (but note that those with $T_{\rm J}$ 
only slightly exceeding 3 can still cross the eccentric orbit of Jupiter). The Encke-type comets with 
$T_{\rm J}>3$ and aphelion distances $Q = a(1+e) \lesssim 4.2$ au are not considered here.

There are two problematic regions of orbital space where the definitions based on the orbital period or 
Tisserand parameter have contradictory implications. First, several known comets with $P<20$ yr have 
large orbital inclinations, or even retrograde orbits (Figure \ref{spc}a). They could be 
classified as JFCs and grouped with other low-inclination JFCs, which would be confusing, because they 
do not seem to be part of the JFC population. Instead, they appear to be a low-period extension of NICs. 
Second, the SPC population of low-inclination orbits extends from $P<20$ yr to $P>20$ yr, while the formal 
definition of JFCs based on the orbital period does not allow for that.\footnote{As a side note, we point
out that there are only a very few orbits with $2\sqrt{2q}<T_{\rm J}<2$ and $q<2$ au in Fig. \ref{spc}b.
This is because NICs evolve from $a \gg a_{\rm J}$ and $e \sim 1$, and thus have $T_{\rm J}<2\sqrt{2q}$.}   

Ideally, a good classification scheme should reflect the distinct origin and evolution of different kinds 
of comets. Since comets evolve into the inner Solar System from distant reservoirs beyond Neptune, the 
flattened inclination distribution of JFCs/ECs requires a flattened source such as the Kuiper belt and/or 
scattered disk (Fern\'andez 1980, Duncan et al. 1988, LD97, Di Sisto et al. 2009, BM13). Because any association 
with a specific source depends on a comparison of the orbital distributions obtained in a model with observations,
we first investigated how sensitive the orbital distribution of {\it known} JFCs/ECs is to different 
assumptions. To limit potential pollution from HTCs/NICs, we only used the cometary orbits with $P<20$~yr 
and $2<T_{\rm J}<3$ in our tests.

We found that the orbital distribution of JFCs/ECs is reasonably well defined. To demonstrate that, we 
tested various ranges of $H_{\rm T}$, and plotted orbital distributions for several subsets of JFCs/ECs. 
The main idea behind these tests is that brighter comets are more easily detected and their orbital 
distribution should therefore be less affected by observational biases. We found that the distributions 
of semimajor axis, inclination, and Tisserand parameter are nearly independent of any $H_{\rm T}$ cut. 
The biggest dependence on $H_{\rm T}$ is seen in the distribution of perihelion distances, $q=a(1-e)$, 
where limiting the sample to the brightest comets results in a distribution with slightly larger values 
of $q$. This is expected because brighter comets can be detected at larger perihelion distances.

\begin{figure}
\epsscale{0.8}
\plotone{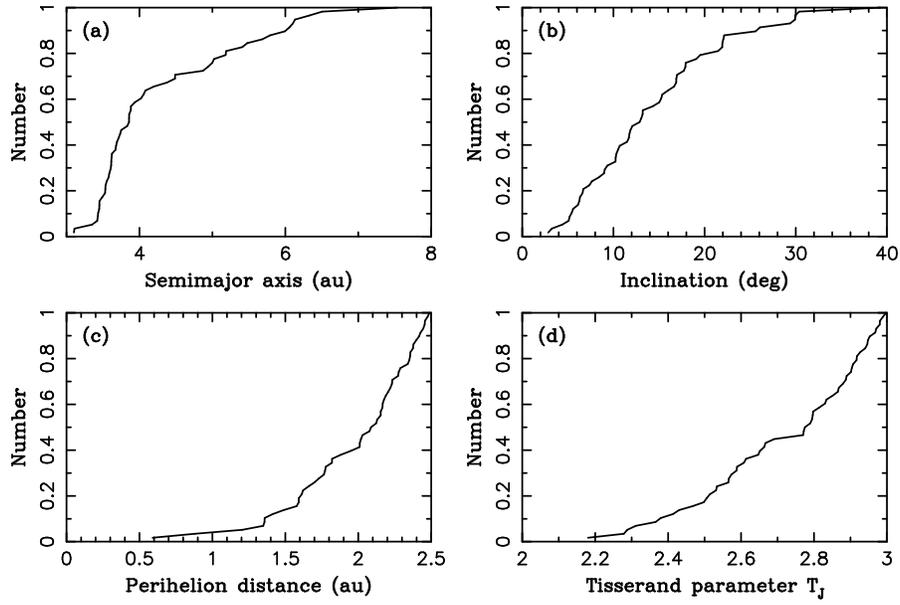}
\caption{The cumulative orbital distributions of known JFCs/ECs with $P<20$ yr, $2<T_{\rm J}<3$, 
$q<2.5$ au, and $H_{\rm T}<10$. All distributions shown here were normalized to 1.}
\label{real1}
\end{figure}

Figure \ref{real1} shows the cumulative orbital distributions of JFCs/ECs for $P<20$ yr, $2<T_{\rm J}<3$,
$q<2.5$ au, and $H_{\rm T}<10$. There are 58 known comets that satisfy these criteria. 
Increasing the $H_{\rm T}$ cutoff results in better statistics (for example, there are  
115 known JFCs/ECs with $H_{\rm T}<12$) but the perihelion distribution starts to shift toward smaller 
values. We therefore opted for using the 58 known JFCs/ECs with $H_{\rm T}<10$ as a base for our 
model comparisons in Section 4.\footnote{Alternatively, to limit the effects of the perihelion distance 
bias, we could have adopted a cut of $q<1.5$ au (e.g., Di Sisto et al. 2009), 
where the distributions are (nearly) independent of $H_{\rm T}$. We find this unnecessary with the new 
data and prefer to use a broader range of $q$.}    

A similar analysis was performed for HTCs/NICs, where the observational biases are expected to be 
more severe, mainly because HTCs/NICs have longer orbital periods and larger inclinations,
both of which act to make their detection more difficult. To minimize potential pollution of the sample 
from JFCs/ECs, we selected cometary orbits with $20<P<200$ yr and $T_{\rm J}<2$. There were several 
surprises. First, the number of known HTCs/NICs increased substantially from the previous analyses 
of data in Levison et al. (2001, 2006). This is contributed by new detections from several ongoing 
near-Earth object (NEO) surveys. Second, the HTC/NIC population with $20<P<200$ yr and $T_{\rm J}<2$ 
shows a very nearly isotropic distribution of inclinations with a median inclination near 80$^\circ$. 

\begin{figure}
\epsscale{0.8}
\plotone{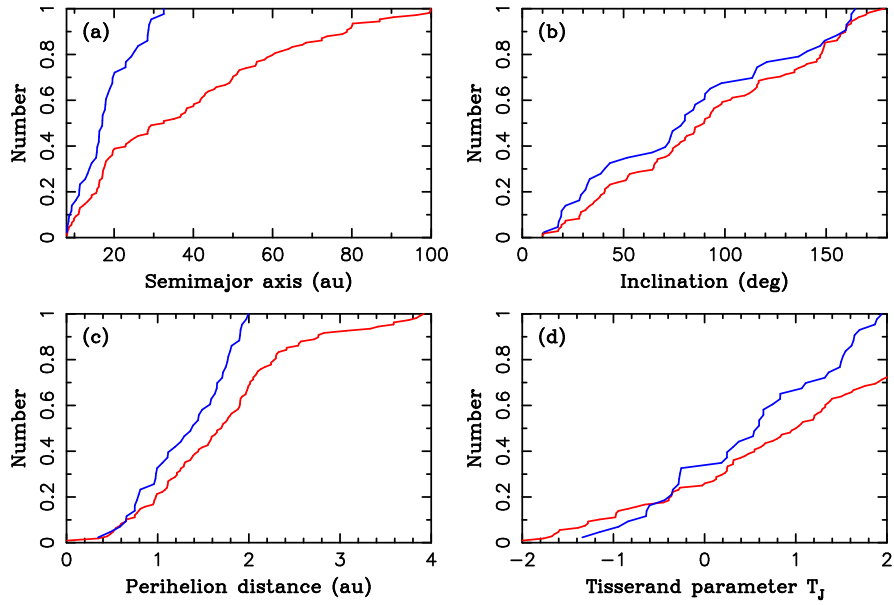}
\caption{The cumulative orbital distributions of known comets with $P>20$ yr, $a<100$ au, $q<4$~au, and 
$T_{\rm J}<2$ (red), and HTCs/NICs with $20<P<200$ yr, $q<2$ au, and $T_{\rm J}<2$ (blue).}
\label{real2}
\end{figure}

The orbital distribution of HTCs/NICs with $P>20$ yr and $T_{\rm J}<2$ is shown in Figure \ref{real2}.
Two sets are shown: a broader one with $a<100$ au and $q<4$ au (Set 1; 108 known comets), and a narrower 
one with $P<200$ yr and $q<2$ au (Set 2; 48 known comets). This is done to to highlight several things. 
First of all, in panel (b) of Figure \ref{real2}, the inclination distribution of both sets 
is indeed nearly isotropic (Wang \& Brasser 2014) with the median inclination $\simeq$90$^\circ$ 
for Set 1 and $\simeq$80$^\circ$ for Set 2. In contrast, using the cometary catalog available back 
in 2001 and criteria similar to our Set 2, Levison et al. (2001) found a  median inclination of 
only $\simeq$45$^\circ$. This led them to consider a flattened source of HTCs such as the inner,
presumably flattened part of the Oort cloud, and later, in Levison et al. (2006), the scattered disk.  

The perihelion distance distribution of Set 2 (Fig. \ref{real2}c) shows a sharp transition 
from $q<2$ au to $q>2$ au, indicating that the population of known comets is strongly incomplete for 
$q>2$ au (ECs do not show a similar sharp transition at $q\simeq2$ au). In fact, 
roughly 80\% of known comets with $P>20$ yr and $a<100$ au have $q<2$ au, and only $\simeq$20\% have 
$2<q<4$ au. This is probably related to the dependence of cometary activity on $q$, with comets reaching 
$q<2$ au becoming active and readily detectable. To limit the influence of unknown observational 
biases when comparing our model with observations in Section 4, we will only consider HTCs/NICs with 
$q<2$ au.

Another important observational bias that must strongly affect the semimajor axis distribution 
in Fig. \ref{real2}a is the period dependence of comet detectability. This bias arises because comets
with very long orbital periods spend most of the time at large heliocentric distances, where they 
show little or no activity and are not detected. This may explain the bulging profile in Fig. \ref{real2}a, 
where roughly 40\% of known comets have $a<20$ au, while this percentage should presumably be much smaller 
in the underlying distribution, if the incompleteness of the known sample for long orbital periods 
were accounted for. To limit the effects of the orbital period bias, we only consider HTCs/NICs with 
$a<20$ au ($P<89.4$ yr) in Section 4.

In summary, when comparing our model with observations in Section 4, we adopt the following
ranges: $q<2.5$ au, $2<T_{\rm J}<3$, $P<20$ yr and $H_{\rm T}<10$ for JFCs/ECs, and $q<2$ au, 
$T_{\rm J}<2$ and $20<P<89.4$ yr for HTCs. For the reasons explained above we believe 
that this sample is best suited for comparisons with our model, because it does not 
appear to contain any obvious signs of observational biases (that does not mean it is 
bias free). We use different ranges of perihelion distances for JFCs/ECs ($q<2.5$ au) and 
HTCs/NICs ($q<2$ au) because these are the widest ranges of $q$ that we can use without 
running into immediate problems with observational biases. In section 4.5, where we discuss
the joint model for JFCs/ECs and HTCs, we use $q<2$ au for both cometary populations.
\section{Numerical model of SPCs}
In our previous work (NM12), we developed a numerical model of the early evolution of the Solar System. 
The NM12 model follows Neptune's migration into a massive planetesimal disk ($M_{\rm disk}\simeq15$-$20$ 
Earth masses, $M_{\oplus}$) between $\sim$22 and 30 au. As the disk is dispersed during planetary 
migration, planetesimals are ejected from the Solar System, impact planets, or end up in long-lived 
reservoirs such as the asteroid belt (Levison et al. 2009, Vokrouhlick\'y et al. 2016), Jupiter 
Trojans (Morbidelli et al. 2005, Nesvorn\'y et al. 2013), irregular satellites (Nesvorn\'y et al. 2007, 2014), 
Kuiper belt (Malhotra 1993, Gomes 2003, Hahn \& Malhotra 2005; Levison et al. 2008; Dawson \& Murray-Clay 
2012; Nesvorn\'y 2015a,b; NV16), scattered disk (BM13, Kaib \& Sheppard 2016, Nesvorn\'y 
et al. 2016), and Oort cloud (Brasser et al. 2006, 2007, 2008; Levison et al. 2010; Kaib et al. 2011; 
BM13). The NM12 model explains various properties of small body reservoirs in the 
Solar System. Here we use the NM12 model, in its parametrization described in NV16, to study the origin of 
comets.   
\subsection{Integration method} 
Our numerical integrations consist of tracking the orbits of the four giant planets 
(Jupiter to Neptune) and a large number of small bodies representing the outer planetesimal disk. 
The terrestrial planets are not included. To set up 
an integration, Jupiter and Saturn are placed on their current orbits. Uranus and Neptune are placed 
inside of their current orbits and are migrated outward. The initial semimajor axis $a_{\rm N,0}$, 
eccentricity $e_{\rm N,0}$, and inclination $i_{\rm N,0}$ define Neptune's orbit before the main stage 
of migration/instability. Here we use $a_{\rm N,0}=22$ au, $e_{\rm N,0}=0$ and $i_{\rm N,0}=0$. 

The {\tt swift\_rmvs4} code, part of the {\it Swift} $N$-body integration package (Levison \& Duncan 1994), 
is used to follow the orbital evolution of all bodies. 
The code was modified to include artificial forces that mimic the radial migration and damping of 
planetary orbits. These forces are parametrized by the exponential e-folding timescales, $\tau_a$, 
$\tau_e$ and $\tau_i$, where $\tau_a$ controls the radial migration rate, and $\tau_e$ and $\tau_i$ 
control the damping rates of $e$ and $i$ (NV16). We set $\tau_a=\tau_e=\tau_i$ because such roughly 
comparable timescales were found in NM12. 

The numerical integration is divided into two stages with 
migration/damping timescales $\tau_1$ and $\tau_2$ (NV16).  The first migration stage is stopped 
when Neptune reaches $a_{\rm N,1}\simeq27.7$~au. Then, to approximate the effect of planetary encounters 
during the dynamical instability in NM12, we apply a discontinuous change of Neptune's semimajor axis and 
eccentricity, $\Delta a_{\rm N}$ and $\Delta e_{\rm N}$. Motivated by the results of NM12 and Nesvorn\'y (2015b), 
we set $\Delta a_{\rm N}=0.5$~au and $\Delta e_{\rm N}=0.1$. The second migration stage starts with Neptune 
having the semimajor axis $a_{\rm N,2}=a_{\rm N,1}+\Delta a_{\rm N}$. We use the {\tt swift\_rmvs4} code, 
and migrate the semimajor axis (and damp the eccentricity) on an e-folding timescale $\tau_2$. The 
migration amplitude was adjusted such that the planetary orbits obtained at the end of the simulations 
were nearly identical to the real orbits.\footnote{The migration and damping timescales of Uranus were
assumed to be the same as those of Neptune. In NM12 simulations, Uranus's orbit was not much affected 
by the instability. We therefore used $\Delta a_{\rm U}=0$ and $\Delta e_{\rm U}=0.05$.}

We found from NM12 that the orbital behavior of Neptune during the first and second migration 
stages can be approximated by $\tau_1\simeq10$ Myr and $\tau_2\simeq30$~Myr for a disk mass $M_{\rm disk}=20$ 
$M_{\rm \oplus}$, and $\tau_1\simeq20$~Myr and $\tau_2\simeq50$ Myr for $M_{\rm disk}=15$~$M_{\rm \oplus}$. The 
real migration slows down, relative to a simple exponential, at late stages. Here we therefore set 
$\tau_1=10$-30 Myr and $\tau_2=30$-100 Myr. All migration simulations were run to 0.5~Gyr. They were 
extended to 4.5~Gyr with the standard {\tt swift\_rmvs4} code (i.e., without migration/damping after 
0.5~Gyr). 
\subsection{Migration graininess}
We developed an approximate method to represent the jitter that Neptune's orbit experiences due to close
encounters with massive planetesimals. The method has the flexibility to use any smooth migration history 
of Neptune as an input, include any number of massive planetesimals in the original disk, and generate 
a new migration history where the random element of encounters with the massive planetesimals is taken into account. 
This approach is useful, because we can easily control how grainy the migration is while preserving the 
global orbital evolution of planets from the smooth simulations. See NV16 for a detailed description of 
the method. Here we set the mass of massive planetesimals to be equal to that of Pluto. We motivate this 
choice by the fact that two Pluto-class objects are known in the Kuiper belt today (Pluto and Eris). 

The migration graininess is included in the present integrations, because NV16 showed that grainy migration
of Neptune is required to get the right proportion of resonant populations in the classical belt. The migration
graininess may not be important for cometary reservoirs, but we include it in the present work for 
completeness.  
\subsection{Planetesimal disk} 
The planetesimal disk was divided into two parts. The part from just outside Neptune's initial orbit 
($\simeq$22 au) to $r_{\rm edge}$ represents the massive inner part of the disk (NM12). We used $r_{\rm edge}=28$-30 au, 
because our previous simulations in NM12 showed that the massive disk's edge must be at 28-30 au for Neptune to stop 
at $\simeq$30~au (Gomes et al. 2004). The estimated mass of the planetesimal disk below 30~au is 
$M_{\rm disk}\simeq15$-20~$M_{\rm \oplus}$ (NM12). The massive disk has a crucial importance here, because it was
the main source from which cometary reservoirs formed (Dones et al. 2015). The planetesimal disk had an outer, 
low-mass extension reaching from $r_{\rm edge}$ to at least $\simeq$45 au. The disk extension is needed to explain 
why the Cold Classicals (hereafter CCs) have several unique physical and orbital properties, but it should not substantially 
contribute elsewhere, because of the small original mass of the extension. 

Specifically, Fraser et al. (2014) estimated that the current CC population at 42-47 au represents only 
$\sim$0.0003 Earth masses. Assuming that CCs did not lose much of its original population during planetary migration 
(e.g., Nesvorn\'y 2015 found that the primordial population of CCs was reduced by a factor of $\sim$2 during 
early stages), this shows that the surface density of solids must have dropped substantially from 30 au to 42 au. The profile
of the surface density at 30-42 au is not well constrained, but given that Neptune stopped at 30 au, it is reasonable
to assume that the density decreased immediately beyond 30 au. Here we therefore choose to ignore the outer 
extension of the disk at $>$30 au. Including it would probably not change our results substantially. Presumably,
the dynamics of objects starting at 30-35 au during Neptune's migration would be similar to those starting
at $<$30 au. They would become scattered by Neptune (Levison et al. 2008) with a small fraction
ending in the scattered disk and Oort cloud, but this contribution should presumably be minor compared to that 
of the inner, more massive disk. A detailed investigation of this subject is beyond the scope of the present work.  
  
Each of our simulations included one million disk bodies distributed from outside Neptune's initial orbit to 
$r_{\rm edge}$. The radial profile was set such that the disk surface density $\Sigma \propto 1/r$, where $r$ is 
the heliocentric distance. The initial eccentricities and initial inclinations of disk bodies in 
our simulations were distributed according to the Rayleigh distribution with $\sigma_e=0.05$ and $\sigma_i=2^\circ$, 
where $\sigma$ is the usual scale parameter of the Rayleigh distribution (the mean of the Rayleigh distribution 
is equal to $\sqrt{\pi/2}\sigma$). The disk bodies were assumed to be massless, such that their gravity did not 
interfere with the migration/damping routines.
\subsection{Effects of other planets}
The gravitational effects of the hypothetical fifth giant planet (Nesvorn\'y 2011, Batygin et al. 2012, NM12) 
on the disk planetesimals were ignored. The fifth giant planet orbit probably crossed into the trans-Neptunian 
reservoirs only briefly, during some $\sim$10$^5$ yr, before it was ejected by Jupiter into interstellar space. 
It likely did not cause major perturbations of orbits in the trans-Neptunian region, although this may depend 
on how exactly planets evolved during the instability (Batygin et al. 2012). In any case, it is 
difficult to account for the fifth giant planet with the numerical scheme used here, because its orbit 
evolves chaotically during the instability and cannot be easily parametrized. To include the fifth planet in 
a simulation, the orbital histories of planets would need to be taken directly from the self-consistent 
simulations of planetary instability/migration (e.g., Nesvorn\'y et al. 2013). 

P9 is hypothesized, via its gravitational shepherding effects, to produce the non-uniform distribution of orbital
angles of known extreme KBOs ($a>150$ au, $q>35$ au; Trujillo \& Sheppard 2014, Batygin \& Brown 2016a). Its existence
could also explain the tilt of the plane of the Solar System with respect to the solar equator (Bailey et al. 2016,
Lai 2016, Gomes et al. 2017), and the high inclinations of large semimajor axis Centaurs (Gomes et al. 2015, Batygin 
et al. 2016b). The predicted parameters of P9 are: $M_9 \gtrsim 10$ $M_\oplus$, $400 \lesssim a_9 \lesssim 900$~au, 
$0.4 \lesssim e_9 \lesssim 0.8$ and $i_9 \lesssim 30^\circ$. In this work, we performed several simulations with P9 
to investigate the influence of P9 on the dynamical structure of distant cometary reservoirs and on comet
delivery. The goal was to diagnose properties of P9 from orbital characteristics of the cometary populations. 

To construct an adequate model with P9 we first need to know how and when P9 reached its current orbit.  
We considered several possibilities (e.g., Kenyon \& Bromley 2016, Li \& Adams 2016). It seems to us, for 
example, that P9 cannot be related to the hypothesized fifth giant planet (Nesvorn\'y 2011, Batygin et al. 2012, 
NM12). This is especially clear if the planetary instability happened late, because in that case it is hard to 
imagine any plausible mechanism that could have stabilized the fifth planet on a wide orbit. Instead, we find it 
more plausible that P9 reached its wide orbit well before the epoch of planetary instability (i.e.,  before 
Neptune dispersed the massive planetesimal disk below 30 au; e.g., Izidoro et al. 2015). This would mean that the 
dynamical origin of the trans-Neptunian populations, including distant cometary reservoirs, postdates the chain of 
events that ended with the P9 on its wide orbit. Working under this assumption, here we performed several 
simulations where P9 was included as a massive perturber since $t=0$.
\subsection{Galactic tide and stellar encounters}
We assumed that the Galaxy is axisymmetric and the Sun follows a circular orbit of radius $R_0$ in the
Galactic midplane. The Sun's angular speed about the Galactic center is $\Omega_0$. Then, from Levison et 
al. (2001) (see also Heisler \& Tremaine 1986, Wiegert \& Tremaine 1999), we have that
the Galactic tidal acceleration is given by:
\begin{equation}
\mathbf{F}_{\rm tide}=\Omega_0^2\left[ (1-2\delta) x \mathbf{e}_x - y \mathbf{e}_y -\left( {4 \pi G \rho_0 
\over \Omega_0^2}-2\delta\right) z \mathbf{e}_z \right]\ , 
\label{tide} 
\end{equation}
where $\delta=-(A+B)/(A-B)$, $A$ and $B$ are the Oort constants, $G$ is the gravitational constant, and
$\rho_0$ is the mass density in the solar neighborhood. Here, the coordinate system $(\mathbf{e}_x,
\mathbf{e}_y,\mathbf{e}_z)$ was chosen such that $\mathbf{e}_x$ points away from the Galactic center,
$\mathbf{e}_y$ points in the direction of the Galactic rotation, and $\mathbf{e}_z$ points toward the 
south Galactic pole. Rotations were applied to move between $(\mathbf{e}_x, \mathbf{e}_y,\mathbf{e}_z)$
and the reference system of our integrations, which has the $Z$-axis pointing along the initial 
angular momentum vector of the planets. 

Numerically, we used $\rho_0=0.1$ $M_{\odot}$ pc$^{-3}$ ($M_{\odot}$ is the solar mass) in most of our 
simulations, and tested $\rho_0=0.2$ $M_{\odot}$ pc$^{-3}$ in one case as well. The Oort constants were 
set to be $A=14.82$ km s$^{-1}$  kpc$^{-1}$ and $B=-12.37$ km s$^{-1}$  kpc$^{-1}$, giving $\delta=-0.09$. 
Also, $\Omega_0=A-B=2.78\times10^{-8}$ yr$^{-1}$.

The effect of stellar encounters was modeled in the $N$-body code by adding a star at the beginning 
of its encounter with the Sun and removing it after the encounter was over. The stars were released 
and removed at the heliocentric distance of 1 pc (206,000~au). We used the model of Heisler et al. 
(1987) to generate stellar encounters. The model accounts for 20 species of main sequence stars and 
white dwarfs. The stellar mass and number density of different stellar species were computed following the
procedure outlined in Heisler et al. (1987). For each species, the velocity distribution was
approximated by an isotropic Maxwellian with one-dimensional variance. The number of stellar 
encounters below perihelion distance $q$ therefore followed $N(<\!\!q) \propto q^2$. The dynamical effects of passing 
molecular clouds were ignored.

The early stages, when the Sun presumably interacted with other stars in an embedded globular cluster 
(Adams 2010), are not considered here. On one hand, the effect of stellar encounters during these stages 
may be needed to explain the detached orbits of some extreme KBOs (e.g., Sedna and VP113; Levison et al. 2004).
On the other hand, cometary-size disk planetesimals have been strongly affected by aerodynamic gas drag
during the early stages (before the nebular gas was removed by photoevaporation). Instead of being ejected 
to large heliocentric distances, the orbits of small bodies were probably circularized by gas drag on inside 
and outside of planetary orbits (e.g., Brasser et al. 2007). If so, these early stages would not substantially 
contribute to the formation of cometary reservoirs.
\subsection{Summary of our simulations}
We performed 14 simulations in total (Table 1). Two reference simulations were performed without P9, the Galactic tide
or stellar encounters. They differed in the timescale of Neptune's migration: $\tau_1=30$ Myr and 
$\tau_2=100$ Myr (CASE 1 or C1 for short) and $\tau_1=10$~Myr and $\tau_2=30$ Myr (CASE 2 or C2). Other
simulations used the same migration parameters as C1 or C2, but also included some combination of
P9, Galactic tide and/or stellar encounters. Three simulations were performed in C1 with no P9. 
In one job, we used $\rho_0=0.1$ $M_{\odot}$ pc$^{-3}$ (C1G1) and no stellar encounters. The remaining 
two jobs were done with $\rho_0=0.1$ $M_{\odot}$ pc$^{-3}$ and $\rho_0=0.2$ $M_{\odot}$ pc$^{-3}$, and
stellar encounters (C1G1S and C1G2S, respectively). 

In addition, we performed nine simulations with different masses and orbits of P9. We used $M_9=10$, 15, 
20 and 30 $M_{\oplus}$, $a_9=500$, $700$ or 900 au, and $i_9=0$, 15$^\circ$ or 30$^\circ$. The eccentricity 
of P9 was set in each case from the solar obliquity constraint (Bailey et al. 2016, Lai 2016, Gomes
et al. 2017), except for one case with $i_9=0$ where we used $M_9=20$ $M_{\oplus}$, $a_9=700$ au, $e_9=0.6$, 
and C1 migration parameters. None of these simulations, except one, included effects of the Galactic tide 
or stellar encounters. Our most complete job with C1 migration parameters included P9 with $M_9=15$ 
$M_{\oplus}$, $a_9=700$ au, $e_9=0.6$, $i_9=30^\circ$, Galactic tide with $\rho_0=0.1$ $M_{\odot}$ pc$^{-3}$
and stellar encounters.  

The simulations were performed on NASA's Pleiades supercomputer (10 jobs in C1, 1 in C2) and Prague 
cluster Tiger (3 jobs in C2). One C1 simulation over 4.5 Gyr required about 600 hours on 25 Ivy Bridge nodes (20 cores 
each) of Pleiades, totaling over 34 CPU-years per simulation.  
\subsection{Comet production runs} 
The last integration segment, between $t=3.5$ Gyr and $t=4.5$ Gyr ($\Delta T=1$ Gyr; time $t$ is defined 
such that $t=0$ at the start of our integrations about 4.5 Gyr ago and $t$ runs forward in time to $t=4.5$ 
Gyr at the current epoch), was performed with a code specialized for the analysis of cometary orbits. 
First, we used cloning to improve the statistics of orbits reaching below Saturn's orbit. This was done by 
monitoring the heliocentric distance, $r$, of each body at each timestep (0.5 yr). If $r<9$ au for the first time, 
the body was cloned 100 times producing 100 new (cloned) orbits. The cloned orbits were generated by small 
random perturbation of the velocity vector of the original orbit. A similar method was used in BM13.

Second, in addition to the normal output of {\it Swift},
we modified the code to output the orbital elements of comets in 100-yr intervals. The information
was written in separate output files if $a<35$~au, corresponding to $P\lesssim200$ yr, and $q<5.2$ au.
SPCs with perihelion distances beyond the orbit of Jupiter were not recorded in the file, but some 
statistics for them can be obtained from the standard {\it Swift} output. The orbital elements written
in the output file were rotated to the reference plane defined by the instantaneous angular momentum
vector of the four outer planets (Jupiter to Neptune). This is required because P9, included in some
simulations, acts to tilt the angular momentum vector of the Jovian planets by several degrees over 4.5 Gyr.

Third, a detailed output was implemented for LPCs. This was done by monitoring the heliocentric distance
of each body in the simulation, including clones. If a body reached $r<5.2$ au, we recorded the 
body's and planetary state vectors into a special `LPC' file. A separate output was written in the LPC 
file for each perihelion passage with $q<5.2$ au. After the whole simulation was over, in a subsequent set 
of simulations, we used the LPC file to set up backward integrations such that we can determine the 
orbital elements of each comet {\it before} it entered into the planetary region. The orbital elements were
calculated near orbital aphelion, if the aphelion distance $Q<200$ au, or near $r=200$ au, if $Q>200$ au
(and for hyperbolic orbits). 
     
Nongravitational forces on comets were ignored. We used a relatively long timestep, 0.5~yr, in all main 
integrations, which is roughly 1/20 of Jupiter's orbital period, and verified that using a shorter timestep
(we tried 0.25, 0.1 and 0.05 yr) does not significantly affect the results. The simulation results were 
used to build a steady state model of comets. Initially, we used the full length of the last integration 
segment ($\Delta T=1$ Gyr) to obtain the best statistics. Subsequently, we also tested how the results 
depend on the length of the time segment used for the analysis. Using a short time segment $\Delta T$ near 
the current epoch should more closely reflect the present population of comets. On the other hand, the 
statistics become inadequate if $\Delta T$ is too short, especially for HTCs and if a short physical 
lifetime is assumed (see the next section). We did not find any significant differences in the results 
and used $\Delta T=1$ Gyr, which has the best statistics, in the rest of this work. Note that the transfer
time of comets from $q<9$ au to $q<2.5$ au is short ($\sim$6 Myr) compared to $\Delta T=1$ Gyr. Thus, 
not cloning bodies that reached $q<9$ au just before $t=3.5$ Gyr is an adequate approximation. 
\subsection{Physical lifetime of comets}
The steady state model of SPCs is compared to observations in Section 4. In addition to the 
distribution of orbital elements $a$, $q$, and $i$, we also consider the distribution of $T_{\rm J}$. 
To do this comparison correctly, as pointed out in LD97, we must account for the physical lifetime of 
active comets (i.e., how long comets remain active). We considered three different parametrizations 
of the physical lifetime:
\begin{description}
\item{\bf Number of perihelion passages with $q<2.5$ au.} In our simplest parametrization of the physical 
lifetime, we count the number of perihelion passages with $q<2.5$ au, $N_{\rm p}(2.5)$, and assume that
a comet becomes inactive if $N_{\rm p}(2.5)$ exceeds some threshold. The threshold is determined by orbital 
fits to observations. 
\item{\bf Time spent with $r<2.5$ au.} We determine the time spent by each body with $r<2.5$~au,
$T(2.5)$ and assume that a comet becomes inactive if $T(2.5)$ exceeds some threshold. Relative to the
$N_{\rm p}(2.5)$ criterion, $T(2.5)$ penalizes orbits with low $q$ and/or low $a$ values, because 
bodies with these orbits spend more time below 2.5 au.
\item{\bf Heliocentric distance weighted effective erosion time.} Comets reaching low heliocentric distances are heated 
by solar irradiation and are expected to erode faster than more distant comets. The nature of the 
relationship between the effective\footnote{Including normal cometary activity, splitting events, etc.} 
erosion rate and heliocentric distance is uncertain. Here we assume, 
motivated by the heliocentric distance dependence of the water ice sublimation rate (Marsden 1973), that 
the erosion rate is proportional to $r^{-2}$ if $r<2.5$ au. The time spent at each $r$ is then weighted 
by $r^{-2}$ and accumulated in $T_{\rm e}(2.5)$. Relative to $T(2.5)$, $T_{\rm e}(2.5)$ penalizes bodies 
reaching low heliocentric distances.
\end{description} 
The parametrizations described above are a compromise between complexity and realism. More complex 
models, such as the splitting model of Di Sisto et al. (2009), are not considered. These models may 
be more realistic but have more parameters and are therefore difficult to constrain. We do not 
use LD97's parametrization of physical lifetime (see also BM13), because their 
parametrization was developed for ECs, and is not applicable to HTCs or LPCs, which have much longer orbital periods. 
Two possibilities exist for a comet to become inactive: it either becomes dormant or it disrupts and 
disappears. We do not distinguish between these different possibilities in this work and attempt 
to constrain our model from observations of {\it active} comets.
\section{Results}
\subsection{Orbital structure of the trans-Neptunian region}
\begin{figure}
\epsscale{0.32}
\plotone{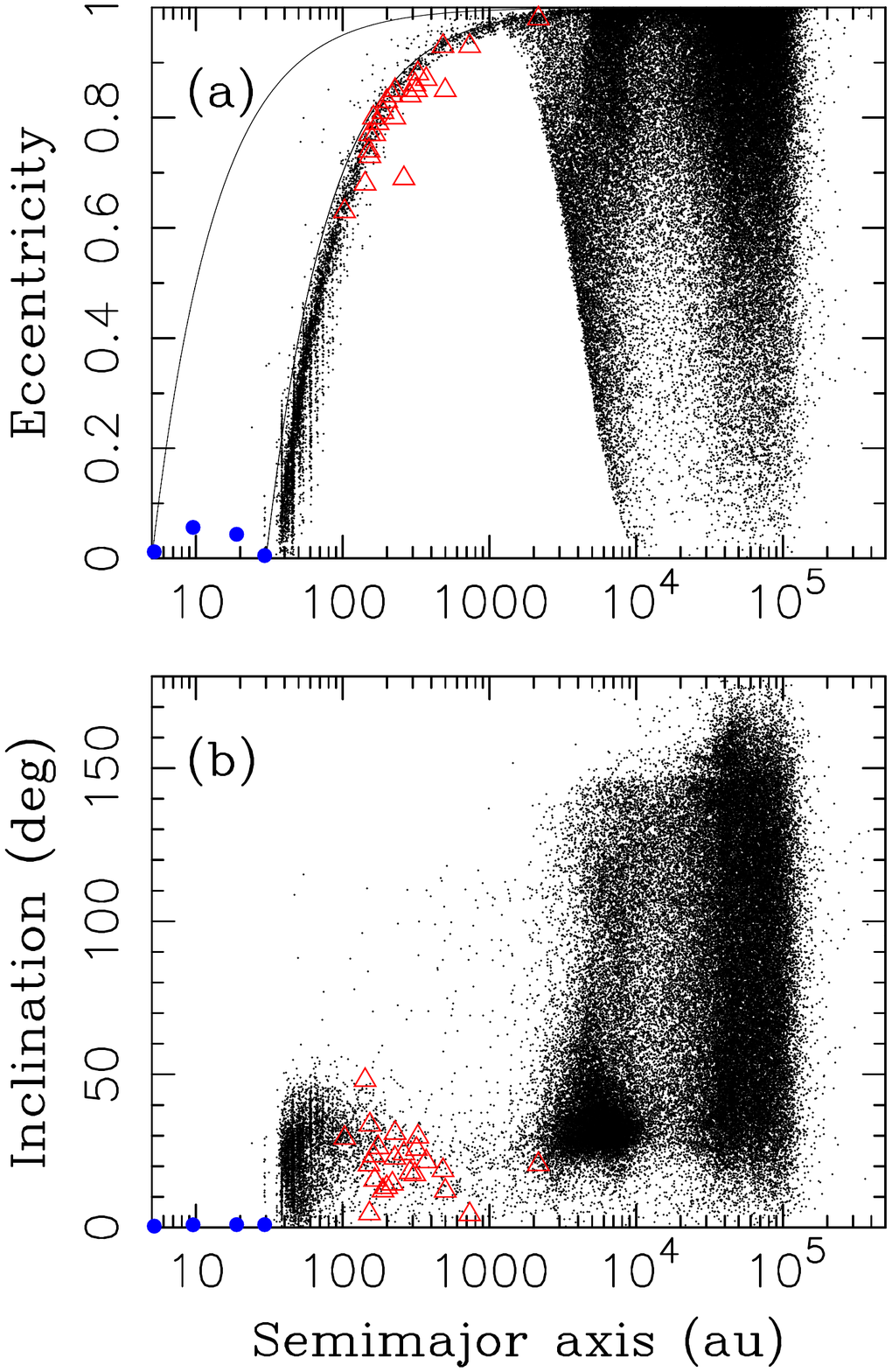}
\epsscale{0.32}
\plotone{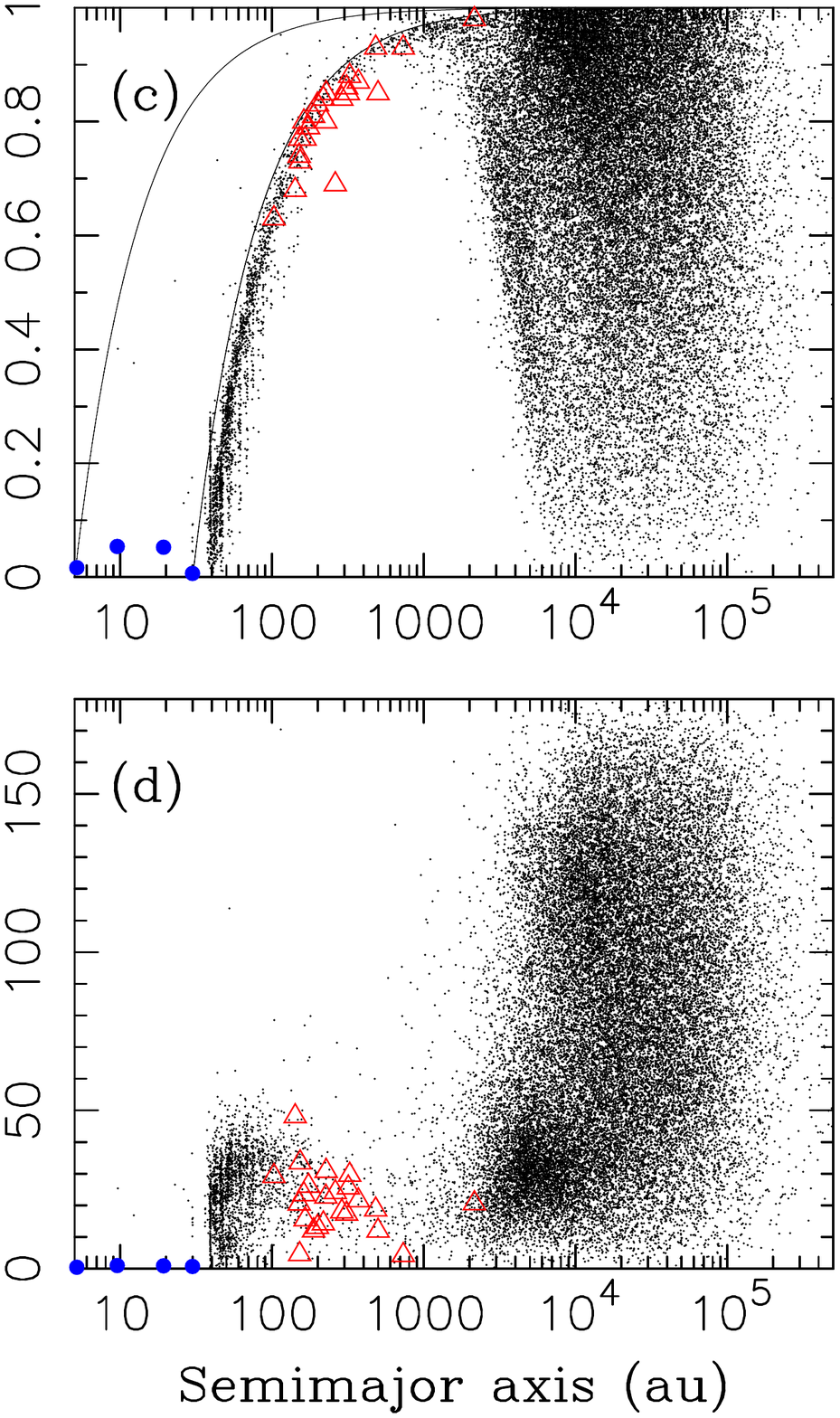}
\hspace{0.0mm}
\plotone{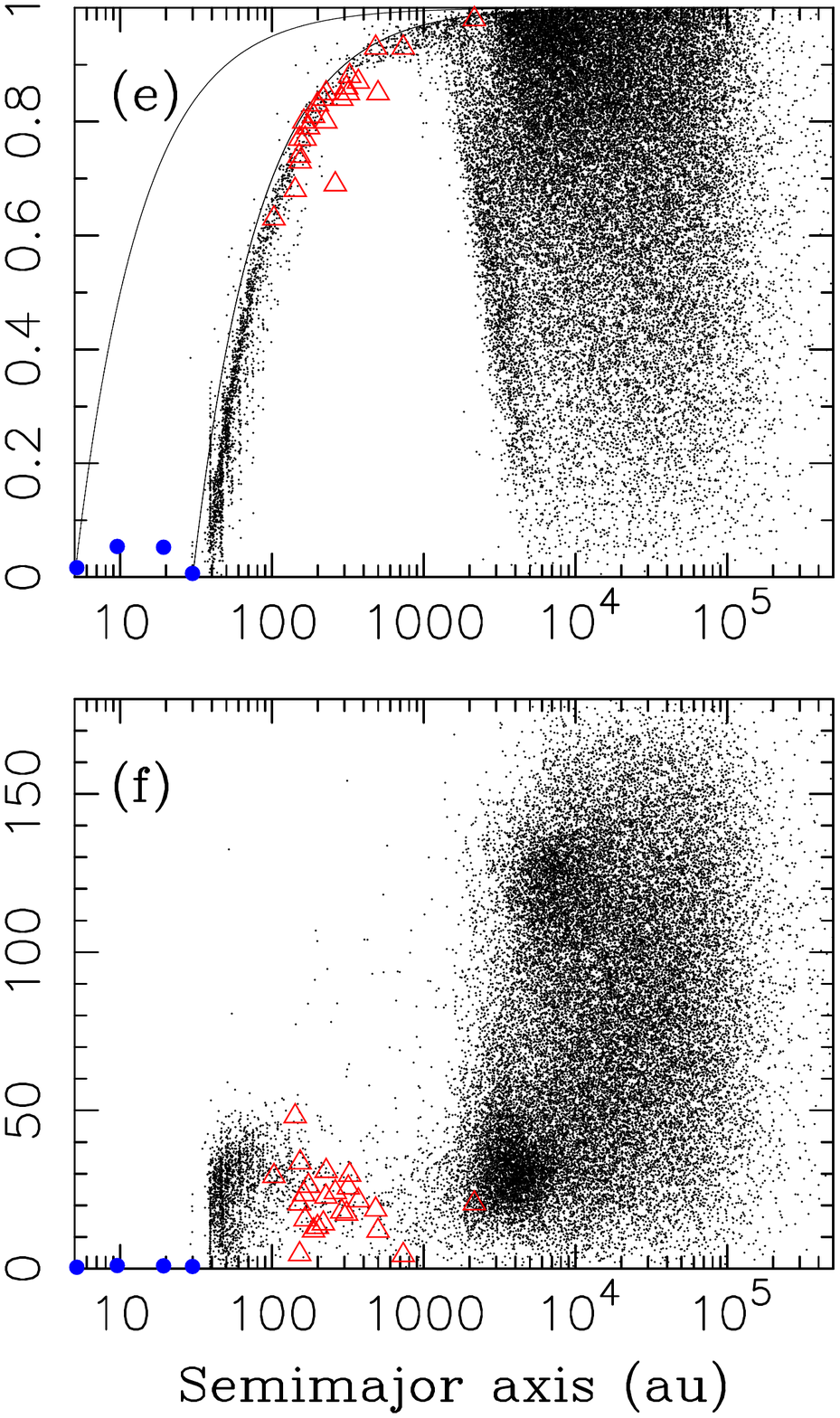}
\caption{The orbital distribution of bodies produced in our Case 1 simulations ($\tau_1=30$~Myr, $\tau_2=100$ Myr, 4000 Plutos)
at $t=4.5$ Gyr. From left to right, the panels show results obtained in different models of external perturbations:
(1) Galactic tide with $\rho_0=0.1$ $M_\odot$ pc$^{-3}$ and no stellar encounters (C1G1; panels a and b), 
(2) stellar encounters and Galactic tide with $\rho_0=0.1$ $M_\odot$ pc$^{-3}$ (C1G1S; panels c and d), and
(3) stellar encounters and Galactic tide with $\rho_0=0.2$ $M_\odot$ pc$^{-3}$ (C1G2S; panels e and f). In all cases,
we sub-sampled the model distributions such that the orbital structures beyond 1,000 au are shown with more clarity.
The thin lines in the upper panels show orbits with $q=5$ au and $q=30$ au. Planetary orbits are denoted by blue dots.
The red triangles show the orbits of known extreme KBOs. The inclination in the bottom panels is given with respect
to the plane of the Jovian planets.}
\label{distr2}
\end{figure}

\begin{figure}
\epsscale{0.32}
\plotone{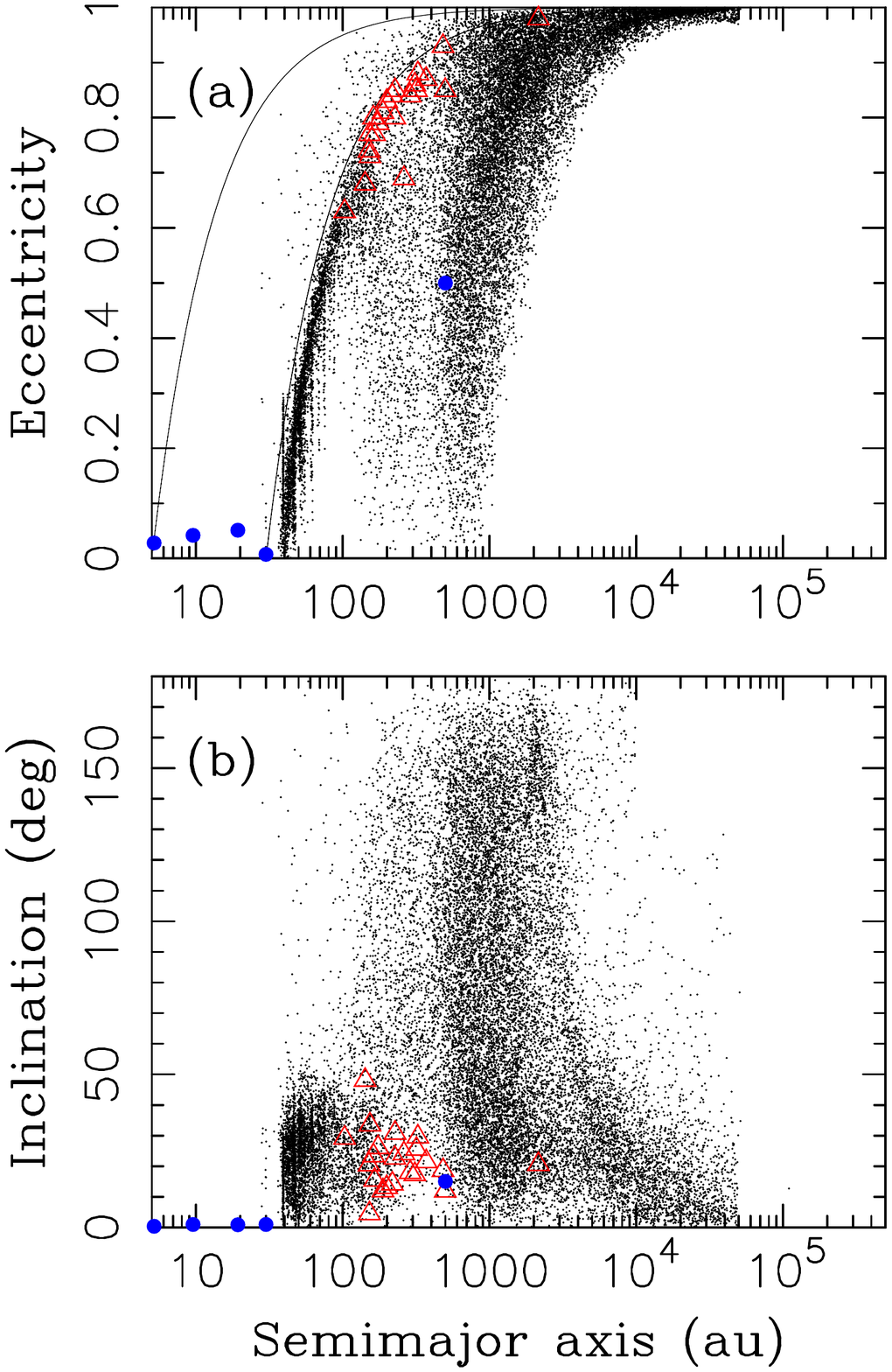}
\epsscale{0.32}
\plotone{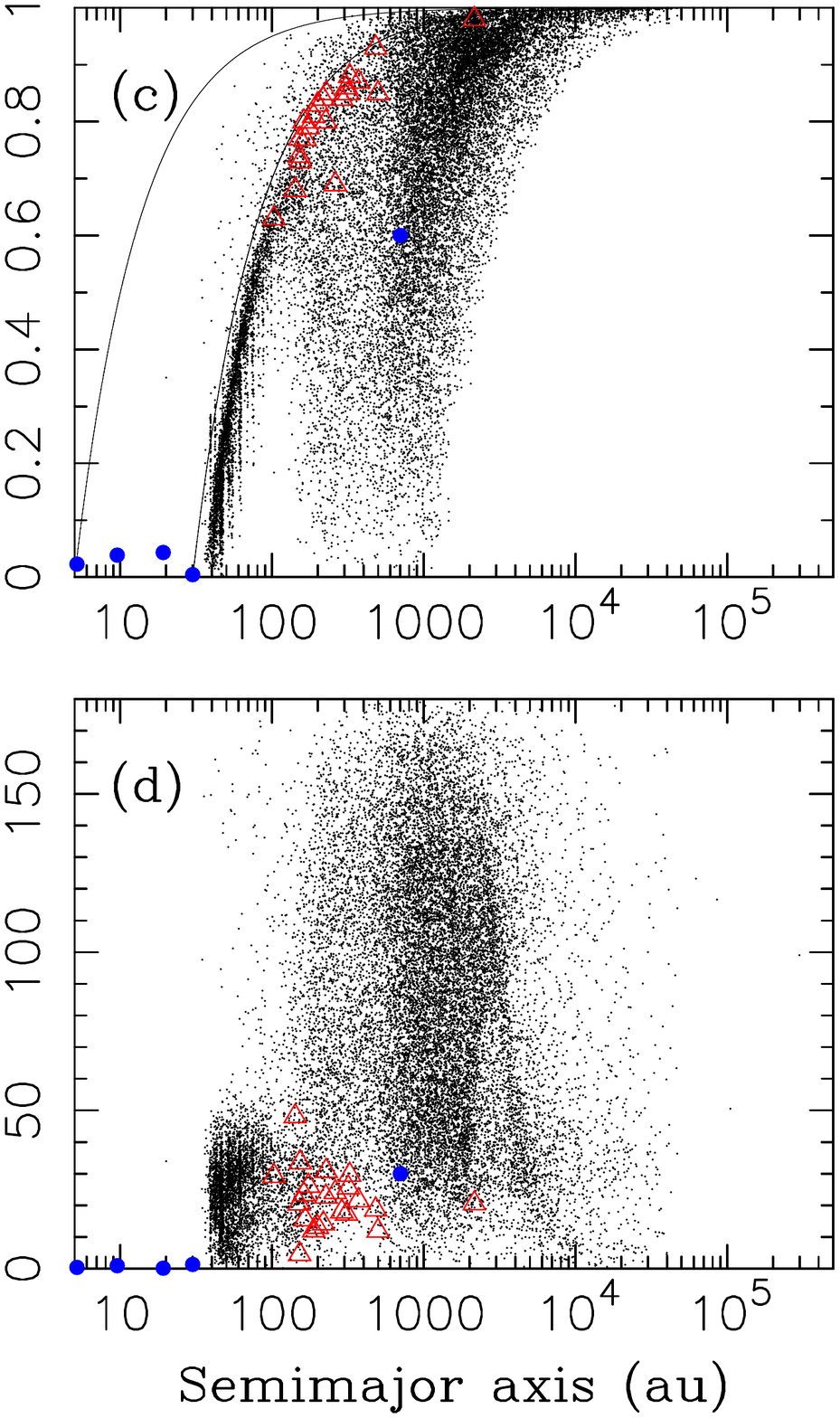}
\hspace{0.0mm}
\plotone{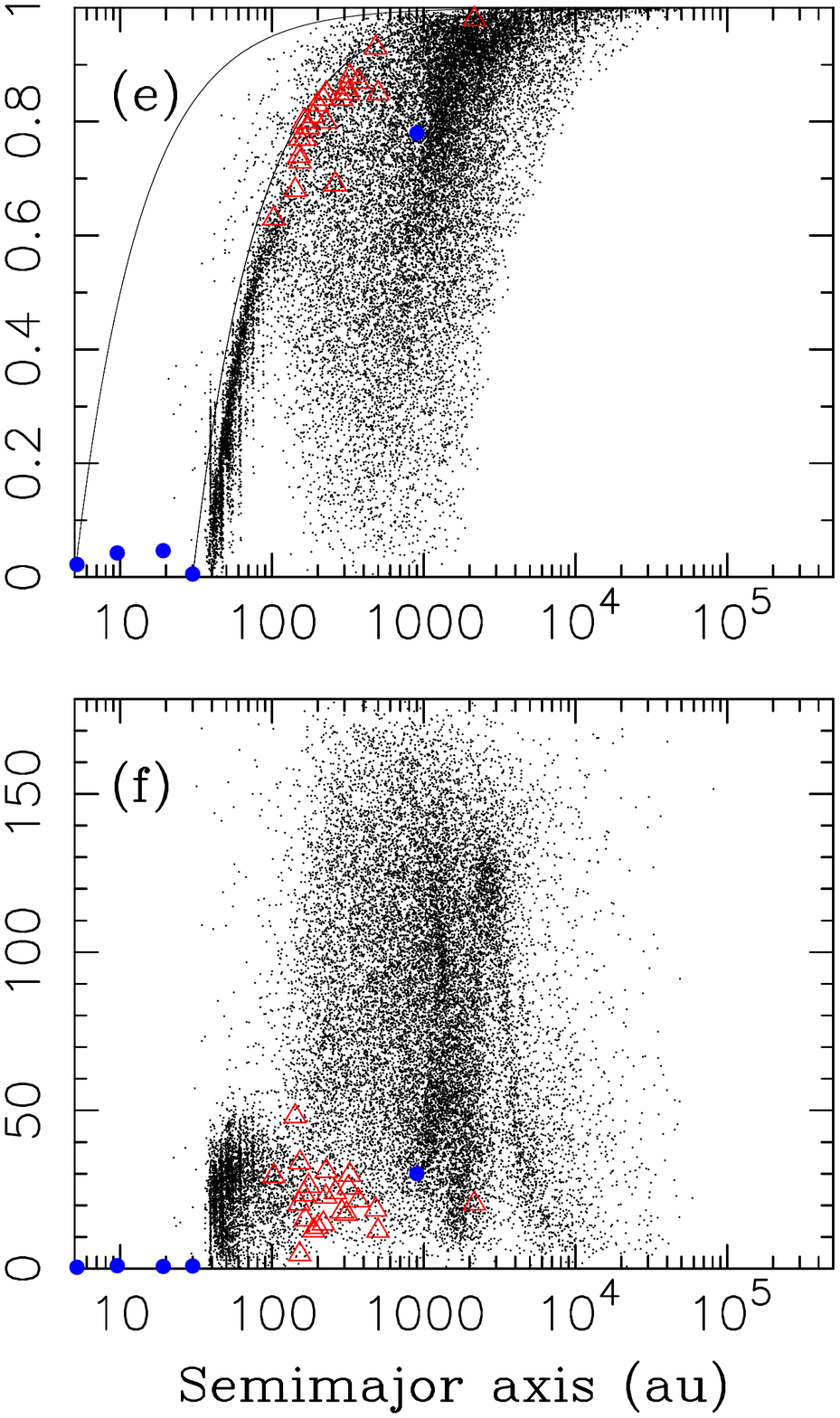}
\caption{The orbital distribution of bodies produced in our Case 1 simulations ($\tau_1=30$~Myr, $\tau_2=100$ Myr, 4000 Plutos)
at $t=4.5$ Gyr. The Galactic tide and stellar encounters were not included here. 
P9 was included with the following parameters: (1) $M_9=20$~M$_\oplus$, $a_9=500$ au, $q_9=250$ au, and $i_9=15^\circ$ 
(C1M20a; panels a and b), (2) $M_9=20$~M$_\oplus$, $a_9=700$ au, $q_9=280$ au, and $i_9=30^\circ$ 
(C1M20b; panels c and d), and (3) $M_9=20$~M$_\oplus$, $a_9=900$ au, $q_9=200$ au, and $i_9=30^\circ$ 
(C1M20c; panels e and f). See Fig. 4 for the meaning of different symbols.}
\label{distr3}
\end{figure}

\begin{figure}
\epsscale{0.32}
\plotone{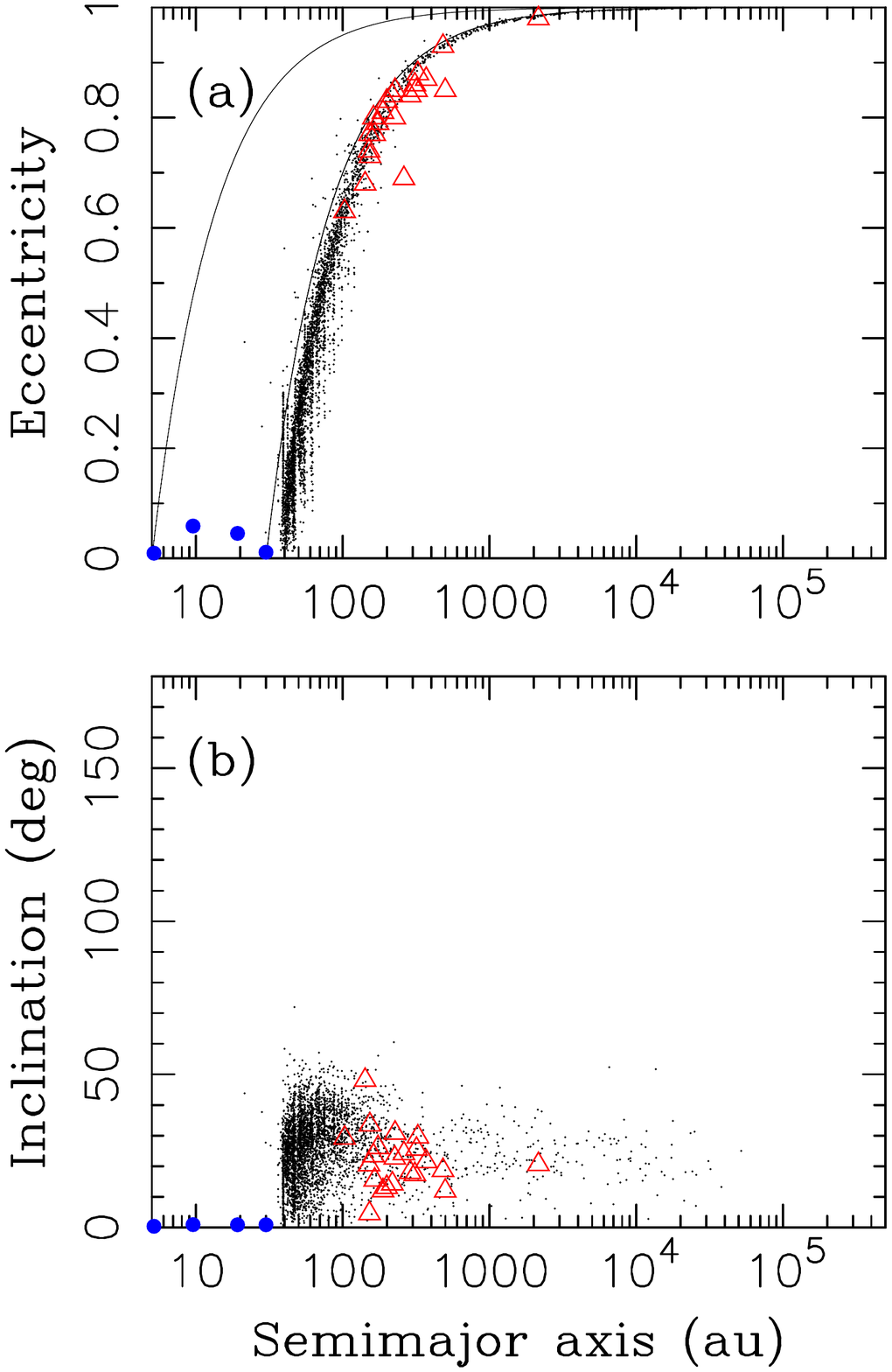}
\epsscale{0.32}
\plotone{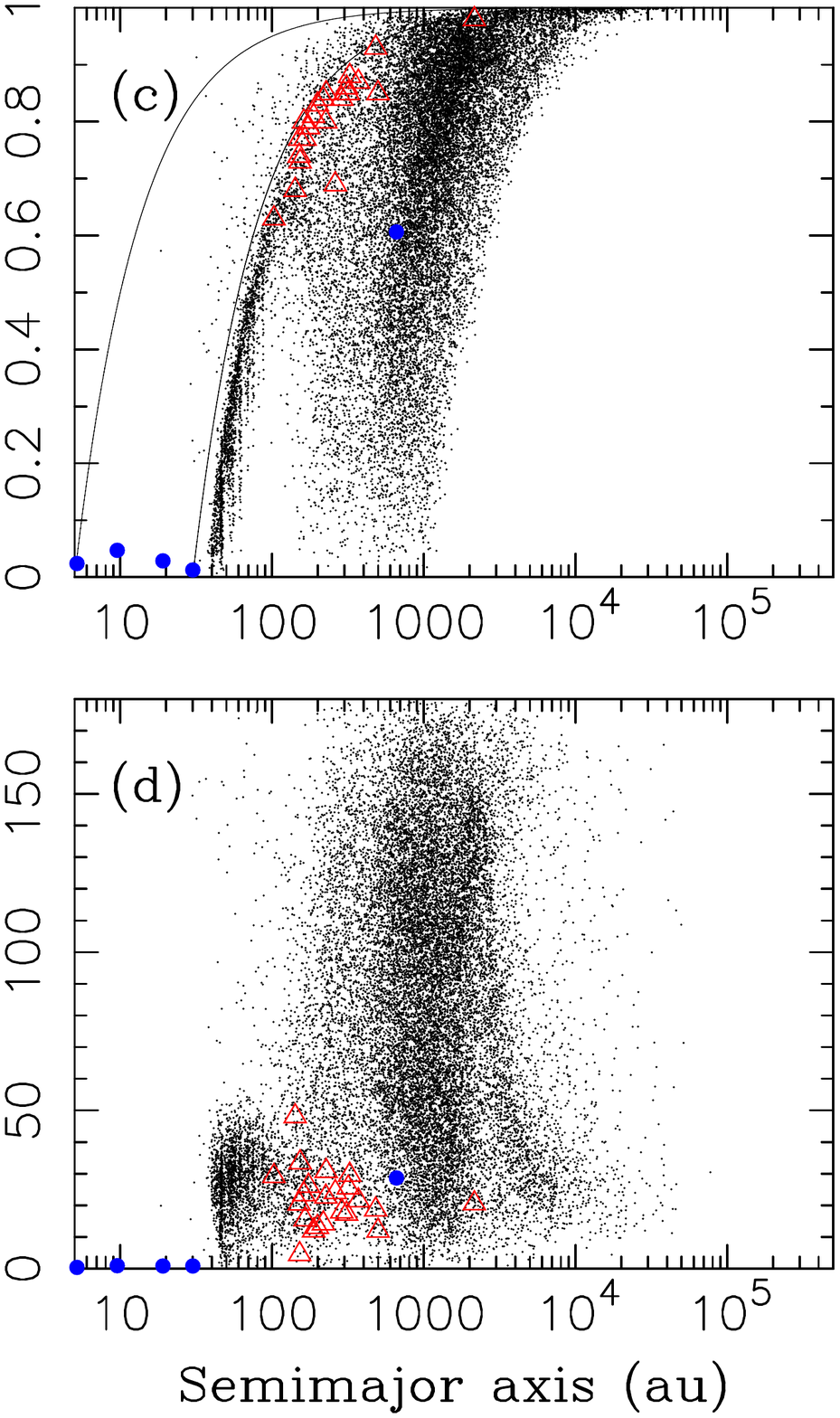}
\hspace{0.0mm}
\plotone{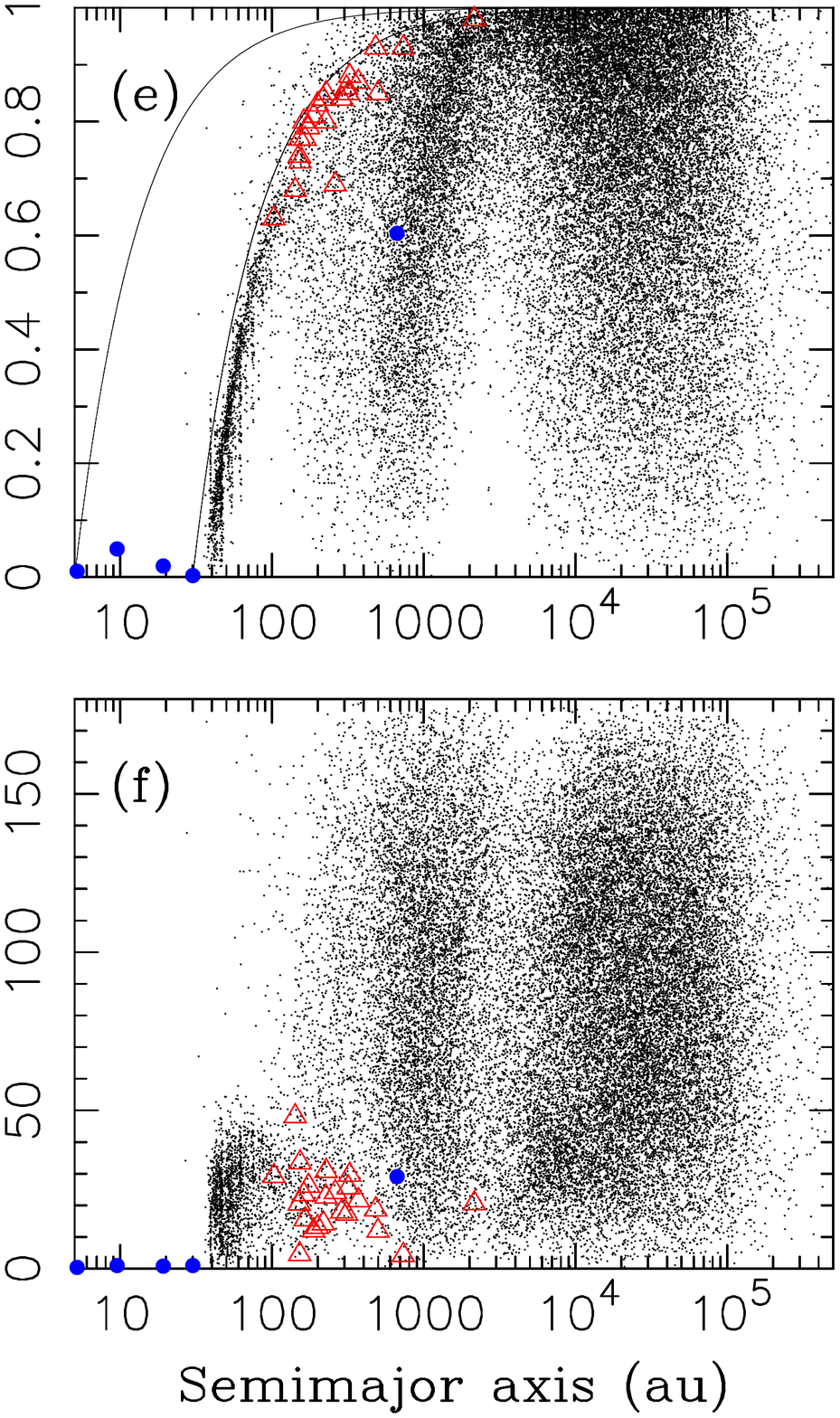}
\caption{The orbital distribution of bodies produced in our Case 1 simulations ($\tau_1=30$~Myr, $\tau_2=100$ Myr, 4000 Plutos)
at $t=4.5$ Gyr. From left to right, the panels show results obtained in different models: 
(1) no P9, Galactic tide or stellar encounters (C1; panels a and b), 
(2) no Galactic tide or stellar encounters, P9 with $M_9=15$ M$_\oplus$, $a_9=700$ au, $q_9=280$~au, and $i_9=30^\circ$ 
(C1M15; panels c and d), and
(3) Galactic tide with $\rho_0=0.1$ $M_\odot$ pc$^{-3}$, stellar encounters and P9 with $M_9=15$ M$_\oplus$, 
$a_9=700$ au, $q_9=280$ au, and $i_9=30^\circ$ (C1ALL; panels e and f). See Fig. 4 for the meaning of different symbols.}
\label{distr4}
\end{figure}

Figures \ref{distr2}, \ref{distr3} and \ref{distr4} show the orbital structure of the trans-Neptunian 
region at the end of our simulations ($t=4.5$ Gyr). The results of the simulations without P9 and with 
different assumptions about external perturbations from the Galaxy and passing stars are compared 
in Fig. \ref{distr2}. There are two notable structures in Fig. \ref{distr2}. The first one is the scattered 
disk that extends along the Neptune-crossing line to $\sim$1000 au. The scattered disk is created 
when bodies encounter Neptune and are scattered along the $q\sim30$ au line to very large heliocentric 
distances. The detailed orbital structure of the inner part of the scattered disk, and how it depends on 
Neptune's migration, was recently discussed in Kaib \& Sheppard (2016) and Nesvorn\'y et al. (2016). 
 
We find that the scattered disk, here defined as orbits with $50<a<1000$ au, contains 
$\sim$3000 bodies at $t=4.5$ Gyr, of which $\simeq$80\% have $50<a<200$ au (hereafter the inner 
SDOs). This means that the population of inner SDOs is much larger, roughly 4 times larger, than the 
population of outer SDOs ($200<a<1000$ au). The total number of SDOs with $50<a<1000$ au represents 
a fraction $\simeq 3\times 10^{-3}$ of the original $10^6$ disk bodies at $t=0$, or, in terms of mass, 
$\simeq$0.06 $M_\oplus$ for $M_{\rm disk}=20$ $M_\oplus$. This estimate is consistent with the results 
reported in Nesvorn\'y et al. (2016). BM13 found from their simulations 
that SDOs should represent a fraction $\simeq6\times10^{-3}$ of the original disk, which is a $\simeq$2 
times larger value than we found here. The difference can be attributed to different orbital evolution 
of planets (BM13 adopted orbital histories of planets from the original Nice model and Levison et al. (2008)).

In our simulations, $\simeq$1500 bodies ended on stable orbits in the classical Kuiper belt with $a<50$ au
(this includes hot classicals and resonant populations), corresponding to the fraction $\simeq1.5\times10^{-3}$ 
of the original disk, or $\simeq$0.03 $M_\oplus$. According to these results, the scattered disk should 
presently be $\sim$2 times more populous/massive than the classical Kuiper belt. For comparison, Trujillo
et al. (2001) estimated from observations that the mass of the scattered disk is $\sim$0.03 $M_\oplus$,
which is a 2 times lower value than the one found here, but their 1 sigma uncertainty admits masses as 
high as $\sim$0.06 $M_\oplus$, which would be in a good agreement with our work. Trujillo et al. (2001) also 
suggested that the mass of the scattered disk is similar to that of the classical Kuiper belt with $a<50$ au,
while Fraser et al. (2014) found instead that the mass of the classical Kuiper belt should only be 
$\sim$0.01 $M_\oplus$, which is 3 times lower than Trujillo's estimate for the scattered disk. Thus, while 
there is general agreement to within a factor of few among different works, a better characterization 
of the trans-Neptunian population from observations will be needed to test our model in detail.   
 
The orbital structure of the inner scattered disk is very different from that of the outer scattered disk. 
Most inner SDOs ($\simeq$80\%) are fossilized, meaning that their (barycentric) semimajor axis did not change 
by more than 1.5 au over the last 1 Gyr. This includes objects that interacted with Neptune's orbital resonances 
in the past and subsequently decoupled from Neptune by various dynamical processes (Kaib \& Sheppard 2016, 
Nesvorn\'y et al. 2016). The remaining $\simeq$20\% of inner SDOs are being actively scattered by Neptune 
(hereafter the scattering SDOs; Gladman et al. 2008). Nearly all outer SDOs, on the other hand, are 
on scattering orbits (i.e., their semimajor axis changes by more than 1.5 au in the last 1 Gyr). 
Thus, even though the inner scattered disk is more massive than the outer one, the number of scattering 
objects in each population is roughly the same.  

The second notable feature in Fig. \ref{distr2} is the Oort cloud. In Fig. \ref{distr2}a,b, where the stellar
encounters were ignored, the Oort cloud has a well defined structure with inner ($a<20,000$ au) and
outer parts ($a>20,000$ au). The outer part of the Oort cloud forms first and is present in our simulations 
already in the first 10 Myr. By checking on the orbital histories of outer Oort-cloud bodies we found that 
most of them reached $a>20,000$ au after having encounters with Saturn (and typically without having 
encounters with Jupiter; Dones et al. 2004). In addition, a significant fraction of outer Oort-cloud bodies 
reached their distant orbits by being scattered by Uranus or Neptune (and without having encounters with 
Jupiter or Saturn). The inner Oort cloud formed as a `wave front' of orbits in our simulations that was 
moving from outside in as time advanced. Most bodies that ended up in the inner Oort cloud were scattered to 
$a>1000$ au by Neptune (some after having encounters with Uranus, but rarely with Jupiter/Saturn). 

The gap between the inner and outer parts of the Oort cloud at $a\simeq20,000$ au also formed gradually in 
our simulations as orbits were slowly removed from this region. The removal process is controlled by the period 
of Kozai cycles produced by the Galactic tide. For $a<10,000$ au, the Kozai period is longer than the age of 
the Solar System (Higuchi et al. 2007), and orbits that become decoupled by the Galactic tide from the Jovian  
planets do not have time to complete one Kozai cycle. These orbits persist to the end of the simulations. The 
orbits with $a>10,000$ au, on the other hand, have shorter Kozai periods and can complete one or more Kozai 
cycles. Once the semimajor axis is above $a \sim 20,000$ au, however, the time for a comet to cycle from $q>30$ 
au to $q<30$ au to back to $q>30$ au is substantially shorter than its orbital period. Thus, even if $q$ drops 
below 30 au, the comet may never make a passage near the planets, and the planets are thus less efficient at 
influencing orbits with $a>20,000$ au.

The inner Oort cloud has an anisotropic distribution of inclinations. Two features can be noted in 
Fig. \ref{distr2}b: (1) the retrograde orbits with $a<20,000$ au generally do not have $i>150^\circ$, 
and (2) there is a concentration of prograde orbits with $i\sim30^\circ$. Issue (1) is related to the fact
that the Galactic tide below 20,000 au can be closely approximated by the quadrupole term. In the 
quadrupole approximation, the orbits that start with inclinations $i'<90^\circ$ with respect to the Galactic 
plane cannot swap to $i'>90^\circ$ (e.g., Naoz et al. 2013). The highest inclination that the inner Oort cloud 
orbits can reach with respect to the Solar System plane is thus $\simeq90^\circ+60^\circ$, where $\simeq60^\circ$ is the angle
between the Galactic and Solar System planes, or $\simeq150^\circ$ in total. Issue (2) is also related 
to Kozai dynamics. The concentration for $i\sim30^\circ$ appears because the inner Oort cloud orbits 
spend the most time with $i'\simeq 90^\circ$ (e.g., Higuchi et al. 2007), meaning that they are perpendicular 
to the Galactic plane. 

The orbital features discussed above are smeared when it is accounted for stellar encounters, but they are
still visible in Fig. \ref{distr2}c-f. With $\rho_0=0.2$ $M_\odot$ pc$^{-3}$ (simulation C1G2S; panels e and 
f), the inner Oort cloud extends to slightly lower semimajor axes, but its overall structure remains the 
same.  We find that $\simeq$6.5\% of the original disk bodies starting at 22-30~au at $t=0$ end up in the Oort 
cloud ($a>1000$ au) at $t=4.5$ Gyr. This is similar to the results of BM13 and 
somewhat higher than estimates obtained in previous dynamical models (3-5\% e.g., Dones et al. 2004, 
Kaib \& Quinn 2008, Brasser et al. 2010, Kaib et al. 2011). With $M_{\rm disk}=20$ $M_\oplus$, we therefore 
find that the total mass of today's Oort cloud should be $\sim$1.3 $M_\oplus$. Of this, roughly 60\% 
should be in the inner Oort cloud ($1000<a<20,000$ au) and $\simeq$40\% in the outer Oort cloud ($a>20,000$ au).
The Oort-cloud-to-scattered-disk ratio obtained in our simulations is found to be $\simeq$20, while BM13 
reported $\simeq$12 from their simulations.

The orbital structure of the trans-Neptunian region dramatically changes when P9 is included in the model. 
To start with, we first discuss our P9 models without the Galactic tide or stellar encounters (Fig. \ref{distr3}).
The dynamical effects of P9, mainly the Kozai resonance, act to decouple SDOs from Neptune and produce a nearly 
isotropic cloud of bodies roughly centered at P9's semimajor axis location. In the following, we call this 
hypothetical structure the P9 cloud. 

Figure~\ref{distr3} shows how the orbital structure of P9 cloud depends 
on the orbital parameters of P9. We find that $\simeq1.7\times10^4$ bodies end up in the P9 cloud 
($200<a<1000$ au) at $t=4.5$ Gyr, corresponding to 1.7\% of the original $10^6$ disk bodies at $t=0$, 
or $\simeq0.34$~$M_\oplus$ for $M_{\rm disk}=20$ $M_\oplus$. If real, the P9 cloud would represent a 
$\sim$5 times larger population than the classical KB and scattered disk below 200 au combined. The number 
of inner SDOs with $50<a<200$ au, and the number and orbital structure of the classical KB are not affected by P9.

Figure \ref{distr4} summarizes the structure of the trans-Neptunian region in different models. Without 
any external perturbations, only the scattered disk is present (panels a and b), and the orbits of 
extreme SDOs, such as Sedna and 2012 VP113, are not obtained in the model. With P9 (panels c-f),
the P9 cloud forms with an estimated mass of $\simeq0.3$-0.4 $M_\oplus$. This would provide an explanation 
for the high-$q$ orbits of Sedna and 2012 VP11. Shankman et al. (2017), however, claimed that the detection of 
known extreme KBOs would imply a very massive P9 cloud (tens of $M_\oplus$). This exceeds, by roughly 
two orders of magnitude, the P9-cloud masses inferred from our dynamical modeling. From the simulations 
with P9, we find that orbits similar to Sedna and 2012 VP11 have perihelion longitudes $\varpi$ concentrated 
near $\varpi-\varpi_9=180^\circ$ (Batygin \& Brown 2016a). This concentration, however, is not strong enough,
at least for the P9 parameters investigated here, to explain the current observations. We fail to identify 
any anisotropy in the distribution of nodal longitudes, $\Omega$, and perihelion arguments, $\omega$.  

The orbital distribution obtained in the C1ALL model with P9 ($M_9=15$ $M_\oplus$), Galactic tide ($\rho_0=0.1$ 
$M_{\odot}$ pc$^{-3}$) and stellar encounters is shown in Figure \ref{distr4}e,f. Both the P9 and Oort clouds 
form in this model with an approximate division between them at $a\simeq3000$ au. We find that the populations
of the classical Kuiper belt ($a<50$ au), inner scattered disk ($50<a<200$ au) and P9 cloud 
($200<a<3000$ au) represent fractions $1.3 \times 10^{-3}$, $3.2 \times 10^{-3}$ and 0.017, respectively,
which is very similar to the fractions reported for other models above. The Oort cloud population 
($a>3000$ au) in C1ALL is somewhat smaller than in the models without P9, representing a fraction $0.044$ of 
the original disk (while we found a larger fraction of 0.060 for $a>3000$ au in the C1G1S model). This probably 
means that the presence of P9 makes it somewhat more difficult for bodies to reach the Oort cloud.   
\subsection{Orbits of Ecliptic Comets}
\begin{figure}
\epsscale{0.8}
\plotone{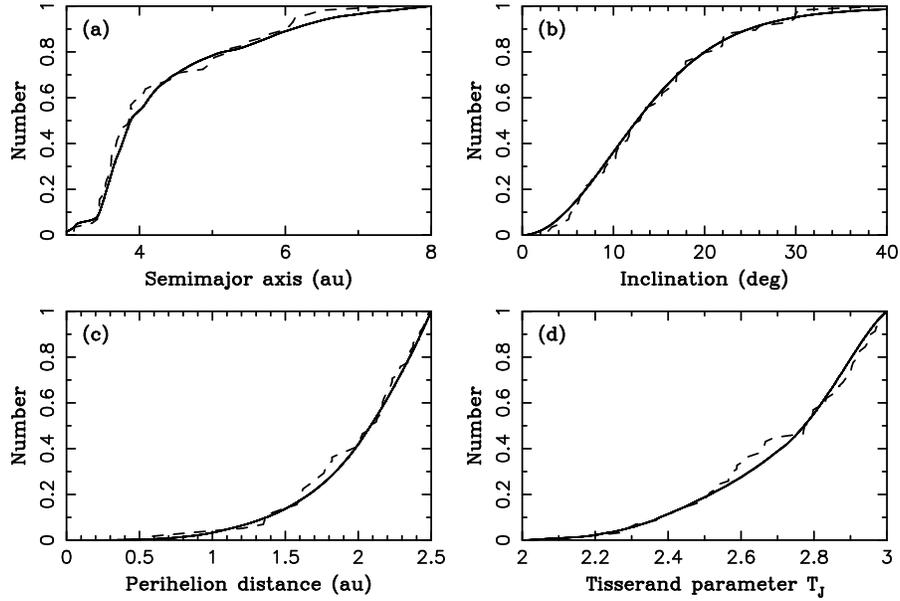}
\caption{The cumulative orbital distributions of ECs with $P<20$ yr, $2<T_{\rm J}<3$ and $q<2.5$ au. 
The model results from C1G1S (solid lines) are compared to the distribution of known JFCs (dashed lines;
$H_{\rm T}<10$). In the model, we assumed that ECs remain active and visible for $N_{\rm p}(2.5)=500$ perihelion 
passages with $q<2.5$ au.} 
\label{jfcm1}
\end{figure}

Using the methods described in Sections 3.7 and 3.8, we determined the orbital distribution of ECs in  
our models. Here we first discuss the results obtained without P9. Figure \ref{jfcm1} shows the best result 
from the C1G1S model (Table 1). To limit the effect of observational biases discussed in Section 2, here 
we considered comets with $P<20$ yr, $2<T_{\rm J}<3$, $q<2.5$~au, and $H_{\rm T}<10$. The best fit
shown in Figure \ref{jfcm1} was obtained with $N_{\rm p}(2.5)=500$ (Section 3.8). It turns out that similarly 
good fits can be obtained with other parametrizations of the physical lifetime described in Section 3.8
(e.g., $T(2.5)=400$ yr or $T_{\rm e}(2.5)=100$~yr). A realistic range of $N_{\rm p}(2.5)$, as determined 
by the Kolmogorov-Smirnov (K-S) test, is 300-800. The models with $N_{\rm p}(2.5)<300$ or $N_{\rm p}(2.5)>800$ do 
not fit the observed inclination distribution of ECs. The model distributions are narrower for 
$N_{\rm p}(2.5)<300$ and broader for $N_{\rm p}(2.5)>800$ than the observed distribution.
  
The physical lifetime of comets is consistent with the results of Di Sisto et al. (2009) who found 
$N_{\rm p}(2.5)=$300-450 (and $N_{\rm p}(1.5)=$170-200).
Our results are also consistent with LD97, where the physical lifetime of ECs was parame\-tri\-zed by the time, 
$T_{\rm act}$, during which a comet remained active after first becoming visible (i.e., after first 
reaching $q<2.5$ au). Specifically, LD97 found that $T_{\rm act}\simeq12$,000 yr was required to fit the 
inclination distribution of ECs (see also BM13). 
Since, according to Fig. \ref{jfcm1}a, the median orbital period 
of ECs is $\simeq$8 yr, $N_{\rm p}(2.5)=500$ implies $T_{\rm act} \simeq 4$,000 yr. This is a factor 
of $\simeq$3 shorter than LD97's best estimate of $T_{\rm act}$, mainly because the source of ECs in our
model is the scattered disk with a wide inclination distribution (while LD97 considered the classical KB 
with $i<5^\circ$). The new ECs, reaching orbits with $q<2.5$ au for the first time, thus have a slightly
wider inclination distribution in our model than in LD97. This implies shorter 
$T_{\rm act}$.\footnote{Rickman et al. (2017) found $T_{\rm act} \sim 0$ from their study of JFCs, which used 
the simulations of Bro\v{z} et al. (2013). These simulations followed the early stages of the outer 
planetesimal disk dispersal. They may not be adequate for the JFC population observed at the present epoch.} 

The best fit result shown in Fig. \ref{jfcm1} is very good.  We do not need to invoke observational biases to
obtain a good fit. This is satisfactory, because it leaves $N_{\rm p}(2.5)$ (or, equivalently, $T(2.5)$ or $T_{\rm e}(2.5)$)
as the only significant free parameter that needs to be adjusted in the model. The orbital distribution of ECs
is independent of the timescale of Neptune's migration (models C1 and C2 produce similar results) and
of whether the Galactic tide or stellar encounters are included in the model (models C1, C1G1 and C1G1S
produce the same result). Our model is thus identified as the simplest physical/dynamical model that is 
capable of matching the orbital distribution of active ECs. Other, more elaborate physical models have
been developed in the past (e.g., Di Sisto et al. 2009, Rickman et al. 2017), but these models have more 
parameters and are more difficult to constrain. 

Note that our physical model must be, to some degree, unrealistic, because many known active ECs have 
$q>2.5$ au. It is therefore not true that ECs can be active only when $q<2.5$~au. When we consider ECs 
with $q>2.5$ au, however, we immediately run into a problem with observational biases. First, many ECs 
with $q>2.5$ au are probably undetected, even if they become active, because they appear too faint 
for a terrestrial observer. Second, only a fraction of ECs probably become active when reaching, say, 
$2.5<q<5$ au. Accounting for the observational incompleteness is tricky and we do not feel confident that 
expanding the model in this direction would produce meaningful results. Still, for the sake of argument, 
we attempted to match the observed distribution of active ECs with $q<5$ au. We found that acceptable 
fits can be found, for example, with $N_{\rm p}(2.5)=500$, and assuming that all comets with $q<2.5$ au 
are detected, while only a fraction $(q/2.5)^{-\gamma}$ of those with $2.5<q<5$ au are detected, where 
$\gamma \sim 5$. This would indicate that only $\sim$3\% of ECs are detected when they reach 
$q\sim5$ au, and this fraction becomes $\sim$100\% for $q\lesssim2.5$ au. 

\begin{figure}
\epsscale{0.8}
\plotone{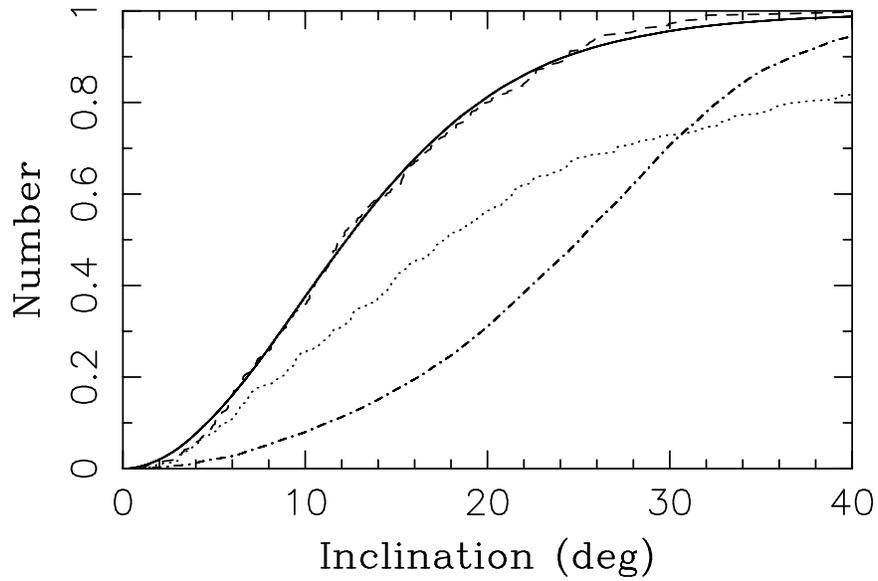}
\caption{The inclination distribution of ECs with $P<20$ yr, $2<T_{\rm J}<3$ and $q<2.5$ au.
The model result (solid line; C1G1S, $N_{\rm p}(2.5)=500$) is compared to the inclination distribution of 
known ECs (dashed lines). For reference, the plot shows the inclination distribution of 
model SDOs at $t=4.5$ Gyr (dot-dashed line) and known Centaurs (dotted line).}
\label{incl}
\end{figure}

Figure \ref{incl} compares the inclination distribution of ECs to those of SDOs and Centaurs. For 
SDOs, we used the the C1G1S simulation results at $t=4.5$ Gyr, and selected orbits with $50<a<200$ au 
and $q<35$ au. This approximates the source region of ECs in our model (see Section 4.7).
For Centaurs, we plotted the inclination distribution of known Centaurs from the JPL Small-Body Database 
Search Engine. Figure \ref{incl} shows that both the SDO and Centaur inclination distributions are
significantly wider than the inclination distribution of ECs. This may seem surprising because the  
dynamical processes that mediate the delivery of ECs from the scattered disk, mainly the scattering 
encounters with planets, should act to increase the orbital inclinations and not to decrease them.  
By testing this we found that the handover of bodies from the Neptune-crossing orbits toward Jupiter
favors orbits with the Tisserand parameter with respect to Neptune, $T_{\rm N}$, somewhat smaller, but 
not much smaller, than 3. This naturally selects the low inclination orbits (see discussion in LD97). 
In addition, $2<T_{\rm J}<3$, used here to define ECs, also favors the low-inclination orbits. Both 
these effects therefore contribute to create an unfamiliar situation, where the inclination 
distribution of the target population (ECs) is narrower than that of the source (SDOs).  
\subsection{EC orbits with P9}
\begin{figure}
\epsscale{0.8}
\plotone{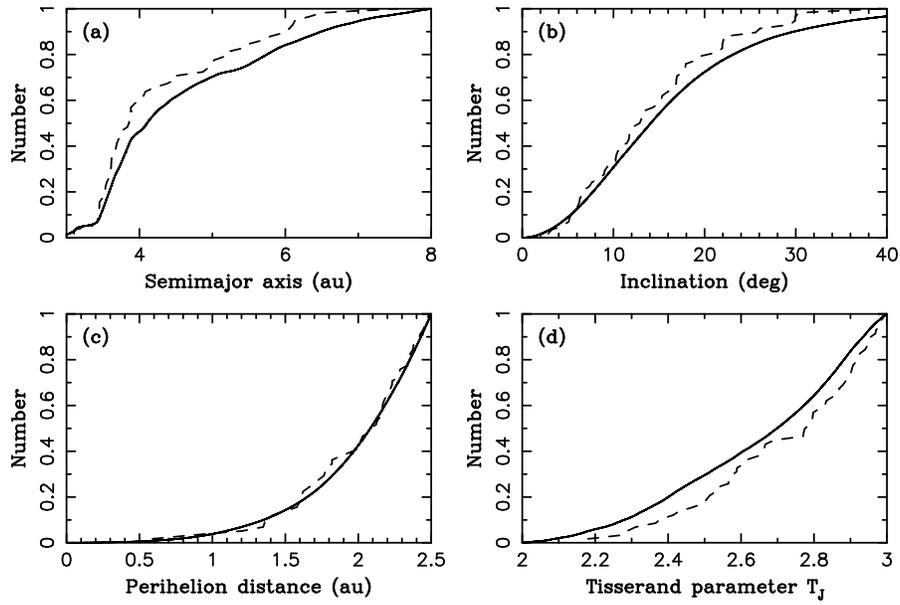}
\caption{The same as Fig. \ref{jfcm1} but with P9 included in the simulation (C1M15; $M_9=15$~M$_\oplus$, 
$a_9=700$ au, $q_9=280$ au, and $i_9=30^\circ$).}
\label{jfcm3}
\end{figure}

Figure \ref{jfcm3} shows the orbital distribution of ECs obtained in the model with P9. This is the 
best result that we were able to obtain with P9 in the C1M15 simulation. Other simulations with P9 
produced similar results. The fit is not as good as the one in Fig. \ref{jfcm1}, because in this 
case the inclination distribution of model ECs is somewhat broader than the inclination distribution 
of real ECs.\footnote{The orbital inclinations of ECs shown in Fig. \ref{jfcm1} and discussed in 
the following text are given with respect to the plane of the Jovian planets. A rotation to this
reference plane was applied at every integration output.} We applied the K-S test to understand how 
significant the difference is. Because, as we explained in Section 2, the inclination distribution 
of known ECs is not sensitive to the $H_{\rm T}$ cutoff, here we did not use any $H_{\rm T}$ 
cutoff to maximize the statistics. We found that the K-S probability with P9 is $p_{\rm K-S}=0.008$, 
which is to be compared to $p_{\rm K-S}=0.91$ obtained in the model without P9.

\begin{figure}
\epsscale{0.8}
\plotone{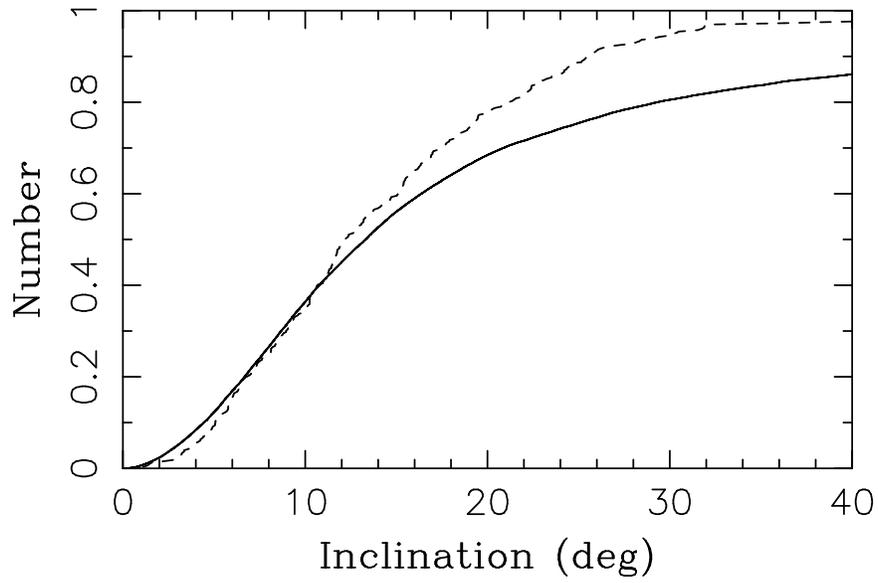}
\caption{The same as Fig. \ref{jfcm3}b but with $0<T_{\rm J}<3$ and $N_{\rm p}(2.5)=100$.}
\label{incl2}
\end{figure}

The difference therefore seems to be significant, indicating that ECs could be used to provide
a useful constraint on P9. Related to that, we note that the difference of the inclination distributions
in Fig. \ref{jfcm3} is somewhat diminished by selecting orbits with $2<T_{\rm J}<3$. If, instead, 
we compare cometary populations with $0<T_{\rm J}<3$ (and $P<20$ yr and $q<2.5$~au, as usual), both the 
real and model inclination distribution became broader, but the discrepancy becomes more significant, 
because the model distribution with P9 has many high-$i$ orbits with $P<20$ yr. This issue cannot be 
resolved by considering $N_{\rm p}(2.5)<300$. This is because, with $N_{\rm p}(2.5)<300$, the model 
distribution with P9 has a different profile than the observed distribution (Figure \ref{incl2}). 
We found that $p_{\rm K-S}=6\times10^{-7}$ for the model with P9, $0<T_{\rm J}<3$ and $N_{\rm p}(2.5)=100$, 
while $p_{\rm K-S}=0.89$ for the model without P9 and $0<T_{\rm J}<3$ and $N_{\rm p}(2.5)=500$.
Adding to that, with $N_{\rm p}(2.5)\sim100$, there are not enough active/visible ECs in the model to 
explain the observed population (Section 4.6).

The problems with P9 discussed above are related to the fact that the scattering disk
at $50<a<200$ au, which is the main source reservoir of ECs (Section 4.7), has significantly larger inclinations 
than in the model without P9. This happens because many SDOs interact with P9, with their 
orbital inclination being excited, and then return into the scattering disk with $a<200$ au. 
Specifically, we find that all our simulations without P9 show similar inclinations distributions 
of inner SDOs ($50<a<200$ au) with 20-25\% of scattering orbits having $i>30^\circ$, and 
only 5-7\% of scattering orbits having $i>40^\circ$. Thus, the scattering disk without P9 is 
relatively flat. With P9, instead, roughly 60\% of scattering orbits with $50<a<200$ au have 
$i>30^\circ$ and roughly 50\% of scattering objects with $50<a<200$ au have $i>40^\circ$. 
The scattering disk with P9 is thus apparently puffed up by dynamical effects of P9 (see Fig. 
\ref{distr3}). This is reflected by the broad inclination distribution of ECs obtained with P9.

There are several possible solutions to this problem, some of which we were able to rule out.  
For example, we tested P9 with zero orbital inclination with respect to the invariant plane of 
the Solar System, and found that the inclination distribution of ECs obtained in this model
(C1I0; Table 1) is practically the same as in models with $i_9 > 0$. We also verified that the same results
were obtained when we used a shorter integration timestep. Another possibility would be to consider
P9 on an orbit with $q_9 > 300$ au, such that the effect of P9 on inner SDOs is diminished
($200<q_9<300$ au in all models investigated here) and/or P9 with a lower mass. It is not clear, 
however, whether these cases could match other constraints such as the orbital alignment of 
extreme KBOs, solar obliquity, etc. A detailed investigation of this is beyond the scope of this work. 
Here we found that cases with $M_9<15$ $M_\oplus$ (and $200<q_9<300$ au) did not produce a sufficiently 
strong orbital alignment of extreme KBOs, and cases with $M_9 \gtrsim 15$ $M_\oplus$ produced a 
plausible alignment in $\varpi$, but not in $\Omega$ or $\omega$. 
\subsection{Orbits of Halley-type comets}
\begin{figure}
\epsscale{0.8}
\plotone{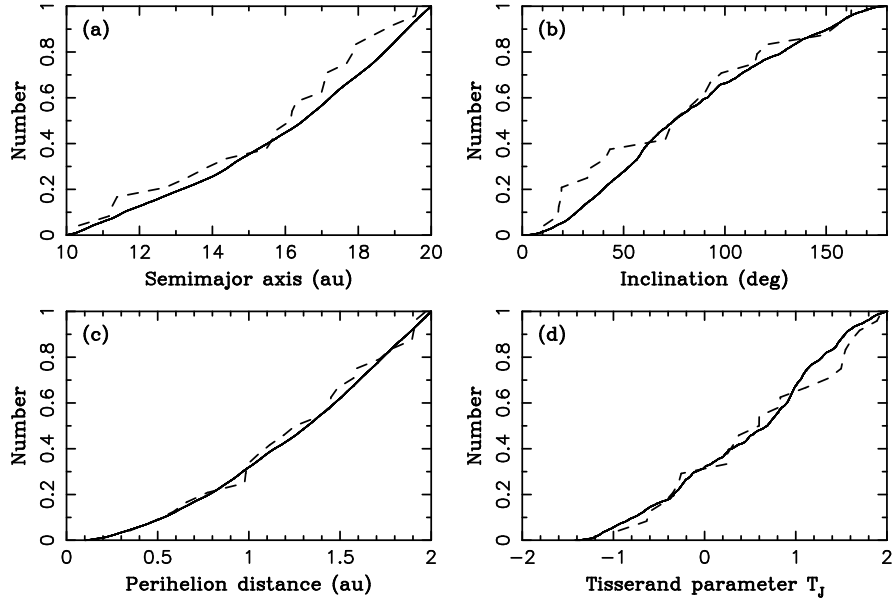}
\caption{The cumulative orbital distributions of HTCs with $10<a<20$ au, $T_{\rm J}<2$ and $q<2$~au.
The model results (C1G1S; solid lines) are compared to the distribution of known HTCs (dashed 
lines). Here we assumed that $N_{\rm p}(2.5)=3000$.}
\label{htcm1}
\end{figure}

Figure \ref{htcm1} shows the orbital distribution of HTCs obtained in the C1G1S model (Galactic tide
and stellar encounters included, no P9). HTCs are produced from the Oort cloud in this model. They 
are a low orbital period extension of the returning Oort cloud comets (e.g., Nurmi et al. 2002).
The range of orbital parameters in Fig. \ref{htcm1} is restricted to 
$q<2$ au and $a<20$ au, such that we avoid issues with the observational incompleteness of known 
HTCs with $q>2$ au and/or $a>20$~au (Section 2). In addition, here we only consider
the HTC orbits with $a>10$ au to limit potential pollution of the sample from ECs (both HTCs and ECs 
are considered in the next section). While the restricted range of orbital elements is probably the 
best characterized part of the HTC population, the orbital distributions within this range may still 
be affected by observational biases. Thus, as a word of caution, we note that the comparison of 
model results with observations in this section is subject to some uncertainty.  
  
With these precautions in mind, Figure \ref{htcm1} appears to show a relatively good agreement. HTCs 
produced from the Oort cloud have a nearly-isotropic inclination distribution with a slight 
preference for prograde orbits. The model distributions of $a$, $q$ and 
$T_{\rm J}$ look good as well. To obtain these results we assumed $N_{\rm p}(2.5)=3000$, which is a 
factor of $\simeq$6 higher value than what was needed to fit the inclination distribution of ECs. 
For HTCs, the orbital distributions are not very sensitive to $N_{\rm p}(2.5)$ and 
$N_{\rm p}(2.5)=500$ gives qualitatively similar results to those shown in 
Fig. \ref{htcm1}. The value of 
$N_{\rm p}(2.5)$ is thus not very well constrained by the fit to the observed orbital distribution 
of HTCs. Instead, $N_{\rm p}(2.5)>1000$ is driven by the requirement to produce a number of active 
HTCs that is consistent with observations (to be discussed in Section 4.6).   

We tested different parametrizations of the physical lifetime of comets described in Section 3.8 
and found that they do not help to solve this problem. There are several other possibilities:
(1) For some reason, $N_{\rm p}(2.5)\simeq500$ derived from the fit to the inclination distribution 
of ECs is too low. For example, our scattered disk (in the simulations without P9) may be too excited 
in inclinations, which could then drive $N_{\rm p}(2.5)$ to low values (because the new ECs 
would already have 
a broad inclination distribution). It is hard to imagine that this might be the case, because the 
scattered disk is gradually excited by encounters of SDOs with Neptune. The excitation therefore does 
not depend on some simulation detail. (2) The number of HTCs obtained in our model is too low, by a 
factor of several. This possibility is discussed in Section 5 together with our preferred resolution 
of this problem, where the physical lifetime of a comet is a function of the comet's size.    

\begin{figure}
\epsscale{0.8}
\plotone{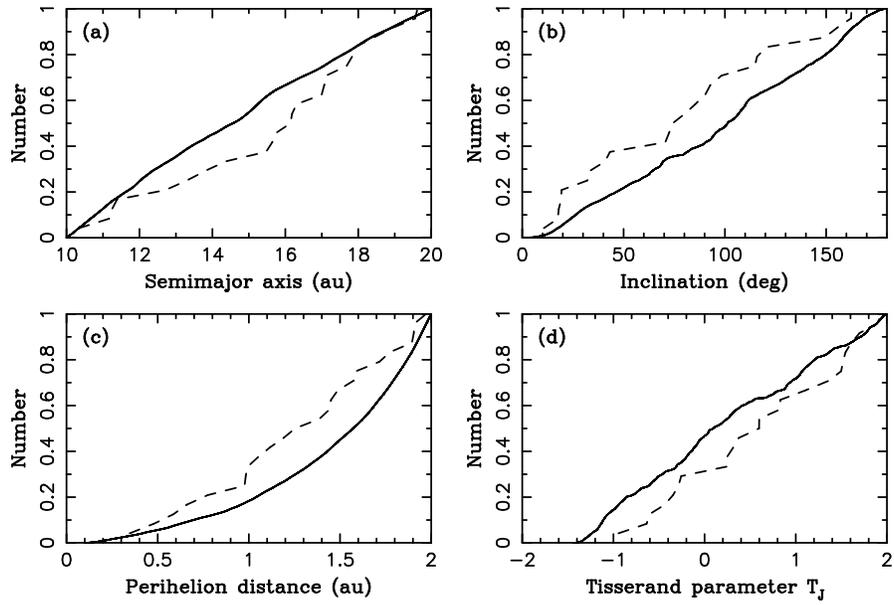}
\caption{The same as Figure \ref{htcm1} but for a model with P9 (C1M15). }
\label{htcm2}
\end{figure}

Figure \ref{htcm2} shows the orbital distributions of HTCs obtained in the C1M15 model (P9 with
$M_9=15$ $M_\oplus$, and no Galactic tide or stellar encounters; Table 1). The results obtained with
other parameters of P9 were similar. In this model, HTCs are produced from the P9 cloud that is 
roughly centered at the semimajor axis of P9 (Figure \ref{distr3}). The delivery of HTCs from 
the P9 cloud is a two step process. First, the secular effects of P9 act to decrease the perihelion 
distance of an object in the P9 cloud. Subsequently, when $q<30$ au, the orbital period of SDOs 
can be shortened by the gravitational effects of the Jovian planets. Since the period of 
secular cycles in the P9 cloud is $>$100 Myr, the first step is a very slow process. This is an important 
difference with respect to the delivery of HTCs from the Oort cloud, where bodies can be placed 
on orbits with very low perihelia in one orbital period.

The HTC population obtained from the P9 cloud model shows similarities to the observed population 
(Figure \ref{htcm2}), but it does not fit the orbital distribution as well as the Oort cloud model (Fig. \ref{htcm1}).
The inclination distribution of model HTCs in Figure \ref{htcm2}b is very nearly isotropic 
with a small preference for the retrograde orbits (median inclination $\simeq$100$^\circ$).
While this preference does not seem to be reflected in the existing observational data, it cannot be 
ruled out either. The perihelion distance distribution of model HTCs does not fit the data 
too well, showing a convex profile with the number of orbits below $q$ proportional 
to $q^2$. The observed distribution is flatter. This may be a consequence of larger observational 
incompleteness for orbits with higher perihelion distances. 

So far we discussed the population of HTCs with $10<a<20$ au and $q<2$ au. This is because this part
of orbital space should be best characterized from observations (Section~2). In a recent paper, 
Fern\'andez et al. (2016) opted to use the full range $7.4<a<34.2$~au (corresponding to $20<P<200$ yr) 
and $q<1.3$ au (16 known comets), and compared the orbital distribution of HTCs to those obtained from LPCs 
and Centaurs. They argued that HTCs should have had at least one perihelion passage 
in a modern history and should thus have a good chance of being detected. They found that the distribution of 
orbital energy (or equivalently, of semimajor axis) obtained from LPCs does not fit the distribution of 
HTCs. Instead, they argued that the immediate source of HTCs are Centaurs. Here we confirm that HTC
orbits evolving from the Oort cloud have a cumulative semimajor axis distribution $N(<\!\!a) 
\propto a^2$, while HTCs from the P9 cloud would have a flatter distribution (that better fits 
the known HTCs with $20<P<200$ yr and $q<1.3$ au). A careful characterization of the HTC population 
with $a>20$ au will be needed before these arguments can be placed on a firmer ground. 

In summary, we find that both the Oort and P9 clouds are potential sources of HTCs, but the Oort 
cloud model fits the existing orbital data of HTCs better than the P9 cloud model.  
As we will discuss in Section 4.6, the Oort cloud is a more prolific source of HTCs than the P9 cloud
(by a factor of $\sim$2-3). This shows that P9 is not required from the considerations based on HTCs. 
On the other hand, inclusion of P9 in a model does not harm the orbital distribution of HTCs.
\subsection{Joint model for ECs and HTCs}
\begin{figure}
\epsscale{0.8}
\plotone{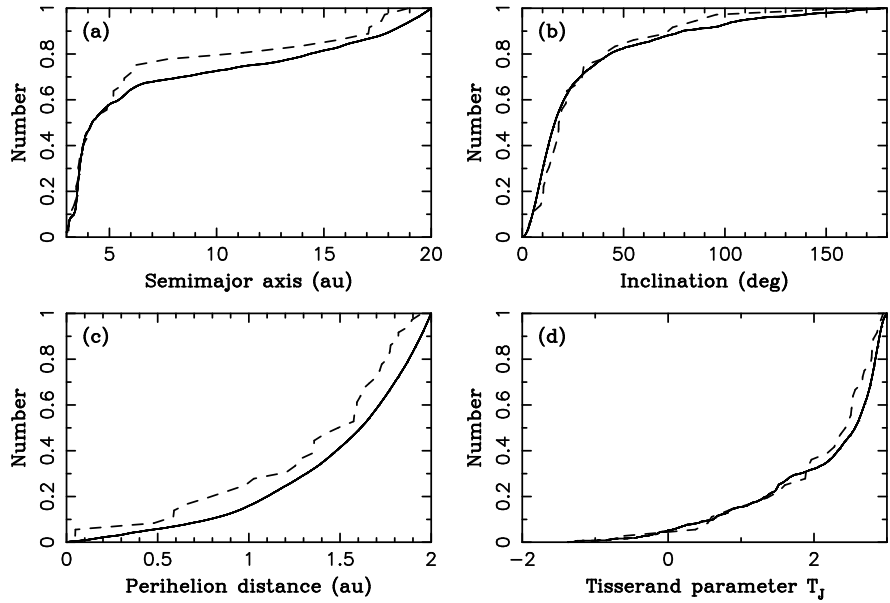}
\caption{The cumulative orbital distributions of SPCs with $a<20$ au, $T_{\rm J}<3$ and $q<2$~au. 
The model results (solid lines; C1G1S) are compared to the distribution of known SPCs (dashed lines). 
In the model, we assumed that $N_{\rm p}(2.5)=500$.}
\label{spcm1}
\end{figure}

Here we remove the distinction between ECs and HTCs and attempt to fit the orbits of all 
SPCs together. Figure \ref{spcm1} shows the result for the C1G1S simulation (no P9) and 
$N_{\rm p}(2.5)=500$, which we established in Section 4.1 to be the best-fit value for ECs. 
The model distributions in Fig. \ref{spcm1} have profiles similar to the observed distributions
but do not fit them very well. The distribution in panel (a) can be interpreted as 
an evidence that we have too many HTCs in our model, relative to the EC population. This would 
be surprising, however, because with $N_{\rm p}(2.5)=500$ the number of HTCs is severely reduced 
(Section 4.6). If $N_{\rm p}(2.5)=3000$ instead, which is the preferred value to obtain the right 
number of HTCs, the problem in Fig. \ref{spcm1}a would appear to be much worse.    

We believe that this problem is related to the observational incompleteness of comets with long 
orbital periods. We find that the semimajor axis distributions in Fig. \ref{spcm1} would match
when it is assumed that $\sim$70\% of HTCs with $8<a<20$ au have been discovered so far (while the 
population of ECs is assumed to be nearly complete). Note that, however, because of the issues with 
$N_{\rm p}$ discussed above, the intrinsic population of HTCs may in reality be larger than shown 
in Fig.~\ref{spcm1}. If so, a larger incompleteness of HTCs would need to be invoked to bring the 
model into agreement with observations (but see discussion in Section 5, where we argue that
$N_{\rm p}$ is a function of comet size and $N_{\rm p}(2.5) \sim 500$ should apply to km-sized 
comets in general). 
\subsection{Model expectation for the number of SPCs}  
The simulations presented here allow us to link the number of ECs and HTCs to the number of 
planetesimals in the original trans-Neptunian disk below 30 au. This has not been done before, 
except for BM13, at least not in a self-consistent dynamical model that 
was also shown to reproduce many properties of other small-body populations in the Solar System. 
In previous works, ECs and HTCs were considered separately and different schemes were developed 
to deal with different comet categories. In some cases, the number of ECs was linked, through 
a chain of multiplicative factors, to the number of objects in the present scattered disk. 
In other cases, the number of HTCs was calibrated by the number of observed new LPCs (e.g., 
Rickman et al. 2017). While all these works have their own merits, here we prefer to emphasize 
the link to the original planetesimal disk, which presumably is the common source of ECs and 
HTCs. This is done as follows. 

NV16 calibrated the number of bodies in the original disk. For that, they assumed that the size 
distribution of disk planetesimals followed the size distribution of today's Jupiter Trojans, which is 
well characterized down to at least $D\simeq5$ km (Wong \& Brown 2015, Yoshida et al. 2017). 
This assumption is based on previous modeling efforts, which showed that Jupiter Trojans were implanted from the
original planetesimal disk (Morbidelli et al. 2005, Nesvorn\'y et al. 2013) and that the 
collisional evolution of Jupiter Trojans after their implantation was not strong enough to
substantially modify their size distribution (e.g., Wong \& Brown 2015). To absolutely 
calibrate the number of original disk planetesimals, NV16 used the estimate of the implantation 
probability determined in Nesvorn\'y et al. (2013), who showed that a fraction of 
$\simeq7\times10^{-7}$ of the disk planetesimals becomes implanted on stable Trojan orbits.

The uncertainty of this estimate is not well established. In the three simulations presented 
in Nesvorn\'y et al. (2013), the implantation probability was found to vary only by $\simeq$15\%. 
On the other hand, our new simulations with very slow planetary migration rates reveal how the 
survival rate of Jupiter Trojans depends on the migration timescale. Some of the results with the 
longest migration timescales show probabilities as low as $\simeq3\times10^{-7}$. Here we therefore 
choose to adopt the implantation probability $5\times10^{-7}$, which is in the middle of the 
values discussed above, and use this value to calibrate the number of planetesimals in the 
original disk.   

Additional constraints on the size distribution come from the mass of the original disk 
needed to generate plausible dynamical evolution of the planetary system ($M_{\rm disk}\simeq20$ 
M$_\oplus$; NM12, Deienno et al. 2017), the expected number of Pluto-class objects in the original disk 
($N_{\rm Pluto}=1000$-$4000$; NV16), and various Kuiper belt constraints (see NV16). 
Figure~\ref{sfd} shows the reconstructed cumulative size distribution of disk planetesimals. 
This figure indicates that there were approximately $6 \times 10^9$ disk planetesimals with 
$D>10$ km. This estimate is uncertain by a factor of $\sim$2, mainly due to the uncertainty 
in the implantation probability of Jupiter Trojans, and its dependence on planetary migration.

The number of comets expected in a steady state, $N_{\rm com}(>\!\!D)$, can be computed from
\begin{equation}
N_{\rm com}(>\!\!D)=N_{\rm rec} {N_{\rm disk}(>\!\!D) \over N_{\rm sim}} {\Delta t \over \Delta T}\ , 
\label{number}
\end{equation}
where $N_{\rm rec}$ is the number of cometary orbits recorded in $\Delta T$, $N_{\rm disk}(>\!\!D)$ is 
the number of original disk planetesimals larger than $D$, $N_{\rm sim}=10^8$ stands for the number of 
bodies used in our simulations ($10^6$ original bodies times 100 for cloning), 
$\Delta t=100$ yr is the sampling interval, and $\Delta T$ is the time interval
used to accumulate good statistics. Here we use $\Delta T=1$~Gyr ($\Delta T<1$ Gyr leads to 
similar results but the statistics are worse). Note that $N_{\rm rec}$ depends on the assumed physical 
lifetime of comets.

In the following text, we compare $N_{\rm com}$ with the number of known comets with $D>10$~km. 
There are several reasons behind this choice, perhaps the most important being that comets 
with small nuclei are probably more difficult to detect than those with $D>10$~km. The 
known population of small comets is therefore incomplete and biased in uncertain ways. 
In previous work, the total absolute magnitude $H_{\rm T}$ was often taken as a proxy for the nuclear 
size of a comet, but we do not find any correlation when the nuclear size of comets determined in 
Fern\'andez et al. (2013) is compared to their $H_{\rm T}$ (or their nuclear magnitude) reported at 
the JPL site. We therefore believe that attempts to relate $H_{\rm T}$ to the nuclear size may be 
misguided. 

We searched various catalogs to determine the number of known ECs and HTCs with $D>10$ km. There are
two large ECs listed in the JPL database: 10P/Tempel 2 ($D=10.6$~km) and 28P/Neujmin 1 ($D = 21.4$ km)
(see Lamy et al. (2004) for discussion). 
In addition, Fern\'andez et al. (2013) reported four additional ECs with $D>10$ km, two
of which have $q<2.5$ au at the present time. These are 162P/Siding Spring ($D=14.1$ km) and 
315P/LONEOS ($D=10.8$ km).\footnote{172P/Yeung is not counted here because it currently has 
$q=3.34$ au. 315P/LONEOS, also known as P/2004 VR9 or P/2013 V6, has $q=2.43$ au and is just barely 
below the 2.5-au limit.} So, together, there appear to be 4 known ECs with $D>10$ km and 
$q<2.5$ au. 

It is not known how solid this estimate is. On one hand, some size estimates were 
obtained from an assumed visual albedo (often taken as low as $\sim2$\%). These determinations are less 
reliable than those derived from IR observations and thermal modeling. In addition, it is often
assumed that the absolute nuclear magnitude can be determined from observations, either because 
the contribution of coma is thought to be negligible (e.g., observations close to the orbital 
aphelion), or because the nucleus appears to be resolved. On the other hand, the known sample of 
ECs with $D>10$ km and $q<2.5$ au may still be incomplete. Indeed, both 162P and 315P were only 
discovered in 2004.

As for HTCs, there are 1P/Halley ($D = 11$ km), C/1991 L3 Levy ($D=11.6$ km) and C/2001 OG108 LONEOS
($D=13.6$ km). The diameter of C/1991 L3 Levy was not measured in the thermal IR and is not reliable,
while the (effective) diameters of 1P/Halley and C/2001 OG108 LONEOS are well established. 109P/Swift-Tuttle with 
$D = 26$ km has semimajor axis $a=26.1$ au, and is outside the range considered here ($10<a<20$ au). 
We conclude that there are $\simeq$2-3 known HTCs with $D>10$ km, $q<2$ au, $a<20$ au. Again, it 
is not clear how complete this sample is, but it should probably be more incomplete than ECs. 
The large HTCs may therefore be as common as large ECs, if not even more common. 

Using Eq. (\ref{number}), we find from our simulations without P9 that $N_{\rm EC}=1$-2 for $D>10$~km, $2<T_{\rm J}<3$, 
$P<20$ yr, $q<2.5$ au, and $300<N_{\rm p}(2.5)<800$ (i.e., the range of $N_{\rm p}(2.5)$ required to fit 
the inclination distribution; Figure \ref{pops}a shows how $N_{\rm EC}$ depends on 
$N_{\rm p}(2.5)$).\footnote{The number of ECs obtained in our models with P9 is slightly lower, 
$N_{\rm EC}=0.7$-0.9.}
This is somewhat lower than the number of large ECs discussed above, thus indicating that our model 
may be anemic, by a factor of $\sim$2-4, when compared to observations. This factor would be larger 
if the observational incompleteness of large ECs is significant. With P9 and $300<N_{\rm p}(2.5)<800$,
we obtain $N_{\rm EC}=0.7$-1 for $D>10$ km, $2<T_{\rm J}<3$, $P<20$ yr and $q<2.5$ au. The number 
of ECs in a model with P9 is thus somewhat lower than in a model without P9. This is probably related 
to a larger excitation of the orbital inclinations of SDOs when P9 is present (see discussion in Section 
4.3).  

As for the HTCs, we find that the Oort cloud should produce $\simeq$2.3 HTCs in a steady state with 
$D>10$ km, $10<a<20$ au, $T_{\rm J}<2$ and $q<2$ au. This is right in the middle of the range inferred
from existing observations above, if the incompleteness of the existing sample could be ignored.  
This estimate was obtained for $N_{\rm p}(2.5)=3000$ and model C1G1S. The number of large HTCs obtained in 
our other Oort-cloud models is similar (2.7 in C1G1 and 1.7 in C1G2S). If $N_{\rm p}(2.5)\lesssim1000$ 
is assumed instead, then $N_{\rm HTC}\lesssim1$ (Fig. \ref{pops}b), at least $\simeq$2 times below 
the value indicated by observations. 

The P9 cloud is less efficient in producing HTCs than the Oort cloud. In particular, we find that
$N_{\rm HTC}=0.9$ with $D>10$ km, $10<a<20$ au, $T_{\rm J}<2$ and $q<2$ au in the 
C1M15 model and $N_{\rm p}(2.5)=3000$. The population estimates obtained in other models with P9
are similar. This is a factor of $\simeq$2.5 smaller than the number of HTCs obtained from the Oort 
cloud. The comparison of different models therefore shows that most HTCs should be coming from the Oort 
cloud and the contribution of P9, if real, should be relatively minor.
In the C1ALL model, where both P9 and the Oort cloud contribute 
to the population of HTCs, $N_{\rm HTC} = 2.1$ for $N_{\rm p}(2.5)=3000$, which is very similar to the 
estimates obtained without P9. There are fewer HTCs coming from the Oort cloud in the C1ALL model,
because the Oort cloud population is smaller (Section 4.1), but the contribution from the P9 cloud 
nearly compensates for the difference.  
\subsection{Source reservoirs}   
We used our simulation results to characterize the source reservoirs of SPCs. For that, we selected
ECs with $2<T_{\rm J}<3$, $P<20$ yr, $q<2.5$ au and $N_{\rm p}(2.5)=500$ and HTCs with $T_{\rm J}<2$, 
$10<a<20$ au, $q<2$ au and $N_{\rm p}(2.5)=3000$. The source orbits of selected ECs and HTCs in the C1G1S 
simulation are shown in Figure \ref{source}. For ECs, the orbits are shown at $t=1.5$ Gyr after the start of the 
C1G1S simulation, or roughly 3 Gyr ago. The migration phase of Neptune has ended at this point. 
For HTCs, we prefer to plot their orbits at $t=3.5$~Gyr, or roughly 1 Gyr ago. This is because the 
orbital structure of the Oort cloud, which is the main source reservoir of HTCs, continues to evolve 
over Gyrs. 

Most ECs ($\simeq$75\%) were produced from the scattered disk with $50<a<200$ au (Fig.~\ref{histo1}). 
About 20\% of ECs started with $a<50$ au. Of these, most bodies had stable orbits that remained 
with $a<50$ au from $t=1.5$ Gyr to $t=3.5$ Gyr. The classical KB, including various resonant populations 
below 50 au (about 4\% of ECs evolved from the Plutino population in the 3:2 resonance with Neptune), 
is therefore a relatively important source of ECs. Interestingly, $\simeq$3\% of our model ECs started in the 
Oort cloud (see also Emel'yanenko et al. 2013). The orbital evolution of these comets is similar to 
returning LPCs or HTCs, except that they were able to reach orbits with very low orbital periods and 
low inclinations. The median semimajor axis of source EC orbits is $\simeq$60 au. The median inclination of source 
EC orbits 1 Gyr ago was $\simeq$20$^\circ$.
   
Figure \ref{source}c,d shows that a great majority ($\simeq$95\%) of HTCs in C1G1S come from the Oort 
cloud, and only $\simeq$5\% from the $a<100$ au region. The inclination distribution of source orbits is 
slightly anisotropic with the median inclination $\simeq$$70^\circ$ (Fig. \ref{histo2}c). This is 
similar to the median inclination of new HTCs in our simulations. The inner and outer parts of the Oort cloud, as defined 
in Section 4.1 (1,$000<a<20,$000 au inner, $a>20,$000 au outer), contribute in nearly equal 
proportions to the HTC population. We see some exchange of orbits between the inner and outer Oort clouds 
in our simulations. The partition of source orbits therefore depends on the time when the source orbits 
are extracted. For example, if the orbits are extracted at $t=0.5$ Gyr, or roughly 4 Gyr ago, then
we find that $\simeq$70\% of current-day HTCs started in the inner Oort cloud (see also Kaib \& Quinn 2009).  

We can now estimate the number of bodies in the source reservoirs. Because of the uncertain nature of P9,
we limit the following discussion to our models without P9. Summarizing the findings discussed 
in the previous sections, we found that the inner SD ($50<a<200$ au) and Oort cloud ($a>10,000$ au)
represent the fractions $\simeq$$2.5\times10^{-3}$ and $\simeq$0.065 of the original planetesimal disk. 
The numbers of current-day ECs and HTCs in steady state are the fractions $\sim2.5\times10^{-10}$ and 
$\sim4.2\times10^{-10}$ of the original disk, respectively. The quoted fractions apply to active ECs
($N_{\rm p}(2.5)=500$) on orbits with $P<20$ yr, $2<T_{\rm J}<3$ and $q<2.5$~au, and to active 
HTCs ($N_{\rm p}(2.5)=3000$) on orbits with $10<a<20$ au, $T_{\rm J}<2$ and $q<2$~au. If $N_{\rm p}(2.5)=
3000$ is assumed for large ECs instead (see discussion in Section 5), the fraction becomes 
$\sim6.7\times10^{-10}$. 

From these, we estimate that the ratio of active ECs with $q<2.5$ au to inner SDOs is 
$\sim 2.5\times10^{-10}/2.5\times10^{-3}=1.0\times10^{-7}$ for $N_{\rm p}(2.5)=500$ or 
$\sim 2.7\times10^{-7}$ for $N_{\rm p}(2.5)=3000$. The former value is more similar to 
the fraction $6.7\times10^{-8}$ obtained in BM13 for $T_{\rm act}=12$,000~yr. 
The ratio of active HTCs ($q<2$ au, $10<a<20$ au) to Oort cloud bodies is $\sim$$4.2\times10^{-10}/0.065=
6.5\times10^{-9}$ for $N_{\rm p}(2.5)=3000$. Since, as we discussed in Section 4.6, there are some 4 
known active ECs with $D>10$ km, there should be $\sim 4/1.0\times10^{-7} = 4.0\times10^{7}$ $D>10$ 
km bodies in the inner SD if $N_{\rm p}(2.5)=500$ or $\sim 1.5\times10^{7}$ $D>10$ km inner SDOs 
if $N_{\rm p}(2.5)=3000$ (most of these have detached orbits; Section 4.1). We prefer the later estimate 
for reasons that will be explained in Section 5. Also, from 2-3 HTCs with $D>10$~km (Section 4.6), 
we estimate that the Oort cloud should contain $\sim 2.5/6.5\times10^{-9} = 3.8\times10^8$ $D>10$ km comets. 

According our work, the ratio of the Oort cloud to scattered disk should be ${\rm OC/SD} \sim 20$ (Section 4.1). 
BM13 obtained ${\rm OC/SD} = 12 \pm 1$ 
from their simulations based on the original Nice model, and inferred ${\rm OC/SD} \sim 44$ from 
observations. The later estimate has a large uncertainty mainly due to the uncertain size and number 
of new LPCs. For example, BM13 pointed out that the flux of 
new LPCs may be lower than assumed before, because some LPCs thought previously to be new are actually 
returning LPCs (Kr\'olikowska \& Dybczy\'{n}ski 2010). They ended up giving preference to ${\rm OC/SD} \sim 
23$, roughly in the middle of the values quoted above. Our new estimate, ${\rm OC/SD} \sim 20$,
is spot on their preferred value.
  
In previous works, briefly discussed in Section 1, various estimates were given for the number of 
$D>2$ km or $D>2.3$ km bodies in the scattered disk and Oort cloud. To be able to compare with these works,
we use the distribution shown in Fig. \ref{sfd}, where the ratio of $D>10$-km to $D>2$-km bodies is 
$\simeq$29, and the ratio of $D>10$-km to $D>2.3$-km bodies is $\simeq$22. 
From this we find that there should be $\sim 4.4\times10^8$ $D>2$-km bodies in the inner scattered disk,
and $\sim 1.1 \times 10^{10}$ $D>2$-km bodies in the Oort cloud. The former estimate is a factor of $\sim$2
lower that the one reported in Rickman et al. (2017), who found, combining several factors from LD97 
and other works, that the capture rate of JFCs requires $\sim 10^9$ $D>2$-km bodies in the scattered 
disk. Duncan \& Levison (1997) reported $\sim 6 \times 10^8$ SDOs from their modeling of ECs (the 
size range to which this estimate applies was not specified), which
is only slightly larger than our estimate for $D>2$ km.

BM13 and Brasser \& Wang (2015) obtained somewhat higher estimates:  
$\sim$$2\times10^9$ and $\sim$$6\times10^9$ SDOs with $D>2.3$ km, respectively. These 
estimates are a factor of $\sim$6-18 higher than ours. In addition, BM13 
found that the observed flux of new LPCs implies that there are 
$\sim4 \times 10^{10}$ to $\sim$$10^{11}$ $D>2.3$-km comets in the Oort cloud. These 
estimates are a factor of $\sim$5-12 higher than ours. Thus, while we agree with BM13  
on the OC/SD ratio, for some reason, our best estimates are at least a factor of $\sim$5 lower.  

Some of the difference quoted above can be explained by the uncertain relationship between 
total absolute magnitude and nuclear size. As we already mentioned,
we do not find any correlation when we compare $H_{\rm T}$ from the JPL database with the nuclear 
diameter estimates from Fern\'andez et al. (2013). This could mean that $H_{\rm T}$ expresses comet
activity rather than the nuclear size, perhaps because only a small part of a comet's surface is 
typically active (e.g., Sosa \& Fern\'andez 2011). 

If that's the case, it may be incorrect to use $H_{\rm T}$ as a proxy for size.
Brasser \& Wang (2015), for example, assumed that $H_{\rm T}=10.8$ corresponds to $D=2.3$ km and 
estimated that there are $\simeq$300 JFCs with $D>2.3$ km and $q<2.5$ au.\footnote{BM13, instead, 
assumed that $D=2.3$ km corresponds to $H_{\rm T}=9.3$ for JFCs and $H_{\rm T}=6.5$ 
for LPCs.} Adopting these numbers and assuming that there are $\sim$4 JFCs with $D>10$~km and 
$q<2.5$ au (Section 4.6), the cumulative power-law slope at $2<D<10$ km would be $\sim-3$, which 
is much steeper than $\sim-2$ typically found from observations (e.g., see Lamy et al. 2004 for a 
review). We therefore believe that the number of small JFCs is significantly lower, possibly 
$\gtrsim$4 times lower, than the one estimated in Brasser \& Wang (2015). 

Here we calibrated the number of large $D>10$ km SDOs and Oort cloud bodies from Jupiter 
Trojans and large ECs/HTCs for which the nuclear size is relatively well known from observations 
(e.g., thermal IR). These two calibrations are consistent in that they lead to the same 
population estimates. We then used the size distribution of Jupiter Trojans to extrapolate our estimates for 
$D>10$ km to $D>2$ km and $D>2.3$ km. This method should provide more robust results than the 
previous works, because it circumvents the problems with the uncertain relationship between 
$H_{\rm T}$ and nuclear size.    
\section{Discussion}
In Section 4.6 we found that our nominal model predicts $\sim$2-4 times fewer large ECs than are 
required to match observations. In addition, in Sections 4.3 and 4.4, we found that $N_{\rm p}(2.5)\simeq500$ 
is required to match the inclination distribution of known ECs, while $N_{\rm p}(2.5)>1000$ is 
required to match the number of known large HTCs. These results were obtained in a model without P9.
With P9, at least for the parameters of P9 investigated here (e.g., $200 < q_9 < 300$ au) we were 
unable to match the inclination distribution of ECs. This problem could be potentially resolved, 
for example, if $q_9 > 300$ au, because in such a case P9 would presumably not excite the orbits 
of inner SDO that much, resulting in a narrower inclination distribution for new ECs. It remains to 
be shown, however, whether P9 with $q_9 > 300$ au could be useful to explain other data, such as 
the orbits of extreme KBOs and solar obliquity. A detailed investigation of this issue is beyond 
the scope of this work. 

The discrepancy between the $N_{\rm p}$ values for ECs and HTCs is puzzling. Since both ECs and 
HTCs presumably formed in the same region, in the original planetesimal disk at $<30$~au, there is 
no a priori reason to think that their internal structures, and thus their physical lifetimes, should 
be different. Also, $N_{\rm p}(2.5)$ appears to be an adequate parametrization of the physical 
lifetime: the results for other parametrizations, such as $T(2.5)$ or $T_{\rm e}(2.5)$, are practically 
the same, indicating the same problem.

We believe that the low values of $N_{\rm p}$ found here for ECs cannot be a consequence of some problem
with the orbital distribution of SDOs produced in our model. This is because the inclination excitation 
of SDOs is produced by scattering encounters with Neptune over 4.5 Gyr. The effect of these encounters 
should be insensitive to our setup of early Neptune's migration and other simulation details.

Another, perhaps more plausible solution to this problem would be if the $N_{\rm p}(2.5)$ value of HTCs is 
shorter than found here. Since $N_{\rm p}(2.5)>1000$ of HTCs is driven by the population statistics (and 
not by orbital fits), better results would be obtained if the population of the Oort cloud could be 
increased by a factor of several.  The Oort cloud population could potentially be increased, for example, 
if the Sun captured comets from other stars during the embedded cluster stage (Levison et al. 2010). The magnitude 
of this effect is, however, uncertain and a significant enhancement may require special circumstances. 

Alternatively, we may have failed to properly calibrate the number of objects in the original planetesimal 
disk and the actual population of disk planetesimals was larger. Because the original disk was calibrated from 
Jupiter Trojans, this may have happened if the capture probability of stable Jupiter Trojans was significantly 
smaller than we assumed in Section 4.6. It is doubtful, however, whether the calibration issue could 
account for the full discrepancy, because other constraints, such as the total mass of the original disk 
estimated in NM12, cannot be easily tweaked to produce a factor of several.

In the model developed here, the Oort cloud was populated from the planetesimal disk at $\simeq$22-30 au.
We did not account for the disk above 30 au, because constraints from Neptune migration (Gomes et al. 2004)
and the population of CCs show that the extension of the planetesimal disk above 30 au was not very massive
(relative to the disk below 30 au). The outer disk extension should thus not substantially contribute 
to cometary populations. We also did not account for the disk below 22 au, because the NM12 model did not account for it either.
In retrospect, the contribution of the disk below 22 au to the Oort cloud needs to be reevaluated.
Dones et al. (2004) showed that Jupiter-scattered planetesimals typically do not end up in the
Oort cloud, because encounters with Jupiter are too powerful. Instead, the Oort cloud may have been 
populated from the planetesimal disk in the Saturn-Neptune zone ($\sim$10-20 au). If, for example,
the surface density of planetesimals was $\Sigma \propto 1/r$, the contribution from the 
Saturn-Neptune zone could easily double the population of comets in the Oort cloud. It would also probably 
increase the number of objects in the scattered disk and bring our model to a better agreement 
with the number of~large~ECs. 

Brasser et al. (2007) showed that km-sized comets in the Jupiter/Saturn zone cannot be ejected 
to the Oort cloud during the protoplanetary disk phase. This is because km-size bodies immersed in 
a gas nebula are strongly affected by the aerodynamic drag and their orbits, instead of being ejected 
to large heliocentric distances, tend to circularize near (on inside or outside of) planetary orbits. 
The Jupiter/Saturn zone should have thus been emptied of small planetesimals before 
the gas nebula was dispersed. The same should apply to the Uranus/Neptune zone if Uranus and Neptune 
formed early (e.g., Izidoro et al. 2015). If Uranus and Neptune formed relatively late (e.g., just 
before the nebular gas was removed), on the other hand, which may be required such that these 
planets did not acquire massive gas envelopes, a residual population of small planetesimals 
could have survived in the Uranus/Neptune zone.  

The inner part of the original planetesimal disk in the Uranus/Neptune zone was not considered in previous 
studies, because of concerns with the delay of planetary instability, which was thought to be needed to 
explain the Late Heavy Bombardment (LHB) of the Moon and terrestrial planets (e.g., Gomes et al. 2005, Levison 
et al. 2011, Bottke et al. 2012, Marchi et al. 2012, Morbidelli et al. 2012). 
If asteroids were not responsible for the LHB, as argued in Nesvorn\'y et al. (2017) 
and Morbidelli et al. (2017), the delay may not be needed. This motivates us to consider the planetesimal 
disk in the Uranus/Neptune zone. If this inner part of the disk was dispersed by planets before the main phase 
of Neptune's migration, its contribution to Jupiter Trojans may have been minor, which would leave the 
calibration of the outer disk roughly the same.

Another important issue is the potential dependence of $N_{\rm p}(2.5)$ on comet size. It is reasonable to expect
that small comets should have shorter physical lifetimes than large comets. This would have interesting 
consequences. First, the low value of $N_{\rm p}(2.5)\simeq500$ estimated here for ECs was driven
by the fit to the inclination distribution of ECs, with most contributing comets probably being
$\sim1$ km in size. The $D>10$-km class ECs, instead, could have longer physical lifetimes, which 
would be more consistent with $N_{\rm p}(2.5)>1000$ estimated from the population of $D>10$-km HTCs.
If we assume, for example, that $N_{\rm p}(2.5)=3000$ for large ECs, Fig. \ref{pops}a would imply 
that there should be $\simeq$4.3 $D>10$ km ECs with $q<2.5$ au, in excellent agreement with observations. 

There are several testable consequences of this hypothesis. We used the catalog of 98 ECs from 
Fern\'andez et al. (2013) and split it into two roughly equal parts corresponding to small ($D<3$ km) 
and large ($D>3$ km) comets. Figure \ref{bigsmall} shows their inclination distributions. The expectation
was, if $N_{\rm p}$ is indeed greater for larger comets, that large ECs should have a broader inclination 
distribution than small ECs (because the scattering encounters with Jupiter are given more time to act
with greater $N_{\rm p}$). Figure \ref{bigsmall} confirms this expectation. As a word of caution, 
we point out that the statistics in Fig. \ref{bigsmall} are relatively poor and affected by how 
Fern\'andez et al. selected the sample for their Spitzer observations. On the other hand, we tested the 
dependence on the diameter cut between the small and big comets and found that the results are 
relatively insensitive to it. Figure \ref{bigsmall} may thus really indicate that large comets 
stay active for a longer time than small comets. 

BM13 argued, to explain the observed flux of new LPCs, that typical LPCs 
must be much smaller than typical ECs. Here we confirm this result. The flux of new LPCs is 
estimated from observations to be $\sim$4 comets per year with $q<5$ au and $H_{\rm T}<11$ 
(e.g., Francis 2005). To obtain a similar flux of new LPCs in 
our C1G1S simulation, we find that typical LPCs must be $D<1$ km. In addition, in order to 
fit the ratio of the number of returning to new LPCs (e.g., Wiegert \& Tremaine 1999), we find that 
$N_{\rm p}(2.5)\lesssim10$ for LPCs, nearly two orders of magnitude below the $N_{\rm p}$ values required 
for ECs. These results will be reported in a subsequent publication. Here we just note that some of 
this difference may be related to the fact that LPCs typically reach orbits with lower perihelion 
distances than ECs; they are thus typically exposed to stronger heating during perihelion passages,
and may be more active (relatively to their size) than ECs (e.g., Sosa \& Fern\'andez 2011). Together, 
the stronger heating and 
presumably smaller sizes of typical LPCs could explain why they can survive only a few perihelion 
passages (see also Levison et al. 2002). Large LPCs, instead, may be active much longer (hundreds 
to thousands of perihelion passages), with some surviving long enough to reach the short-period 
orbits of HTCs. 

The small size ($D<1$ km) of new LPCs advocated here could appear to be in a conflict with the 
results of Sosa \& Fern\'andez (2011), who used measurements of the non-gravitational forces to 
infer cometary sizes and fractions of active surface areas, $f_{\rm act}$, and found that $1.3<D<3.6$ km 
for nine LPCs. This work, however, has several caveats. First, as a word of caution, we note that Sosa 
\& Fern\'andez assumed that the effective outflow speed of gas from a typical comet's surface is 
0.27 km s$^{-1}$. In reality, the effective outflow speed is unknown and depends on several parameters, 
including the degree of symmetry of the outflowing material. It may be possible that the LPC sizes were 
overestimated, for example, because the whole LPC surface is typically active, as found in Sosa 
\& Fern\'andez, and the outflow is more symmetrical than for SPCs. Second, only one (C/2007 W1 (Boattini)) 
out of nine LPCs reported in Table 3 of Sosa \& Fern\'andez (2011) had a hyperbolic orbit before 
entering the planetary region.\footnote{C/2007 W1 (Boattini) was estimated to have the smallest 
diameter ($D\simeq1.3$ km) and is one of the two comets in the whole Sosa \& Fern\'andez sample with 
$f_{\rm act}>2$ (see that paper for the meaning of $f_{\rm act}>1$).} The other LPCs in the 
Sosa \& Fern\'andez sample, including the giant comet C/1995 O1 (Hale-Bopp), are returning LPCs with 
$a<10,000$ au. These comets survived their previous perihelion passages, and should be, consistently with 
our hypothesis of the dependence of the cometary survival on size, larger than typical {\it new} 
LPCs. A fraction of them should evolve onto HTC orbits in the future.    

Figure \ref{fig21} illustrates the suggested dependence of $N_{\rm p}(2.5)$ on comet size. For 
$D>1$~km, the profile is constrained by the fit to the inclination distribution of ECs/JFCs 
(for $D \sim 1$~km) and by the number of large ECs and large HTCs (for $D \sim 10$ km). The dependence should 
be roughly linear for $D>1$ km with $N_{\rm p}(2.5) \sim 500 \times (D/1\ {\rm km})$. This would be consistent
with a mass loss driven by surface processes (e.g., sublimation of surface ices, outbursts driven
by sub-surface pressure build-up). If a $D=1$ km comet disappears on average in $\sim$500 perihelion 
passages, the implied average erosion rate is 2 meters per perihelion passage, or $\sim3\times10^9$
kg per perihelion passage for $D=1$ km and bulk density $\rho=500$ kg m$^{-3}$. For an EC orbit, 
this is roughly equivalent to an average loss rate of $\sim$10 kg s$^{-1}$. For comparison, Reach 
et al. (2007) found the loss rate $\sim$0.1-30 kg s$^{-1}$ from a survey of dust trails of 30 JFCs, 
with a median of $\sim$4 kg s$^{-1}$. 

Our results are also consistent with the measured mass loss of 67P/Churyumov-Gerasimenko as determined 
by the Rosetta radio science team, $\sim1.8\times10^{10}$ kg over the course of the escort phase, which may 
be taken as a proxy for the mass loss per orbit (Paetzold et al. 2016). If this is assumed to represent 
an average activity of 67P, then 67P with effective $D=3.3$ km loses about 0.2\% of its current mass per 
orbit. It should therefore last $\sim$500 orbits at this rate. Assuming instead that the activity of 67P
is driven by surface processes, and will diminish as the nucleus becomes smaller, we find from the current 
erosion rate of $\sim$1 meter per orbit that 67P will last $\sim$1600 orbits. These estimates of the 
physical lifetime of 67P favorably compare with those given in Fig. 20 for $D=3.3$ km.

The dependence of $N_{\rm p}(2.5)$ on comet size for $D<1$ km is poorly constrained, but the physical
lifetime should drop more steeply than a simple extrapolation from $D>1$~km to $D<1$ km would suggest
(Fig. \ref{fig21}). This is because $N_{\rm p}(2.5)\lesssim10$ to match the ratio of returning-to-new
LPCs, which presumably have $D<1$ km. We speculate that the hypothesized transition to very short 
physical lifetimes for comets below 1 km may be related to the rotational spin-up of small cometary
nucleii and their subsequent disruption by centrifugal force (e.g, Jewitt et al. 2016). The strong dependence 
on size would arise in this context because the $e$-folding timescale of rotational spin-up is $\propto\;D^2$
(Jewitt 1997). Alternatively, large comets may experience periods of very low activity when the 
dust expelled from active areas re-accretes and creates a protective layer on the surface. This 
process may not be effective for small comets because of their smaller gravity, thus implying a 
much shorter physical lifetime. Whatever is the cause, a dramatically shorter physical lifetime
of small comets could explain the relative paucity of ECs with $D<1$ km (e.g., Meech et al. 2004, 
Snodgrass et al. 2011, Fern\'andez et al. 2013).  

%
\section{Conclusions} 
\begin{figure}
\epsscale{0.43}
\plotone{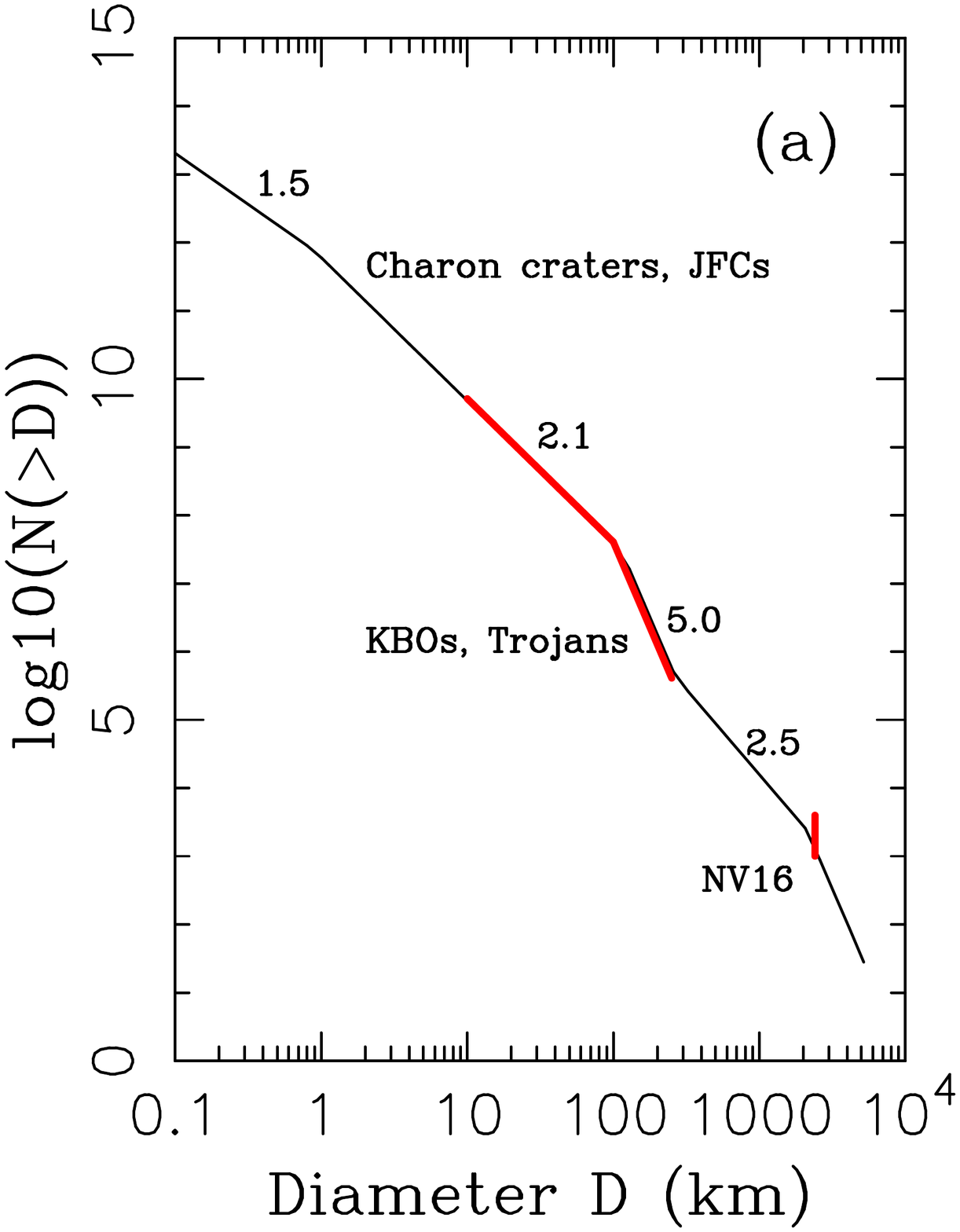}
\epsscale{0.43}
\plotone{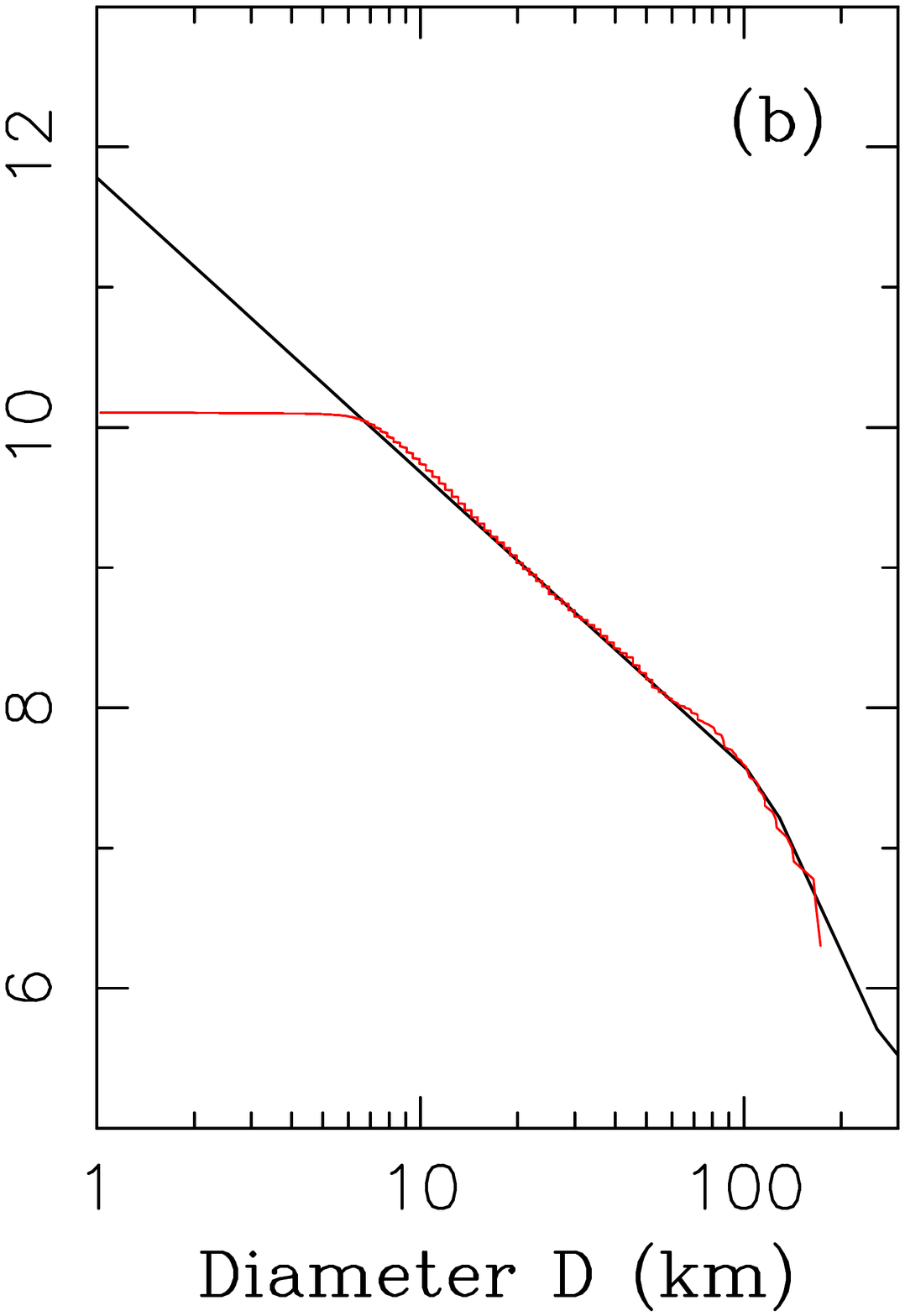}
\caption{The size distribution of the original planetesimal disk below 30 au (panel a). The red color 
denotes various constraints. The distribution for $10<D<300$ km was inferred from observations of Jupiter 
Trojans and KBOs. Panel (b) zooms in on the distribution of $1<D<250$ km planetesimals. The red line in
panel (b) shows the size distribution of known Jupiter Trojans (the sample is incomplete for $D<10$ km). The break 
between a shallow slope for small sizes and a steep slope for large sizes was fixed at $D=100$~km. The existence of 1000-4000 
Plutos in the original disk inferred in NV16 requires that the size distribution had a hump at $D>300$ km. 
The numbers above the reconstructed size distribution in panel (a) show the cumulative power index that was used for different 
segments. The total mass of the disk, here $M_{\rm disk}=20$ $M_\oplus$, is dominated by $\simeq$100-km-class bodies.}
\label{sfd}
\end{figure}

\begin{figure}
\epsscale{0.8}
\plotone{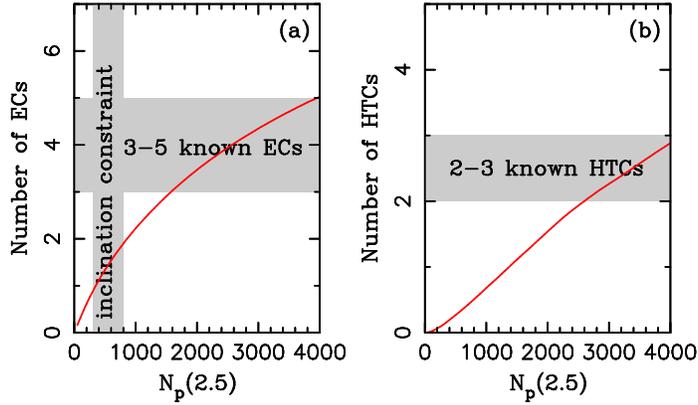}
\caption{The number of active SPCs expected in our model. In (a), we show the expected number of active ECs with $D>10$ km, 
$2<T_{\rm J}<3$, $P<20$ yr and $q<2.5$ au as a function of $N_{\rm p}(2.5)$. In (b), the expected number of active HTCs 
with $D>10$ km, $T_{\rm J}<2$, $10<a<20$ au and $q<2$~au is shown. These results were obtained for the C1G1S model.
The horizontal shaded areas show the number of known SPCs with $D>10$ km (and the same orbital cuts as in the model).
The vertical gray strip in panel (a) is where our model fits the observed inclination distribution of ECs 
($300<N_{\rm p}(2.5)<800$). Ideally, the red line in panel (a) should run through the intersection of the two constraints.}
\label{pops}
\end{figure}

\begin{figure}
\epsscale{0.42}
\plotone{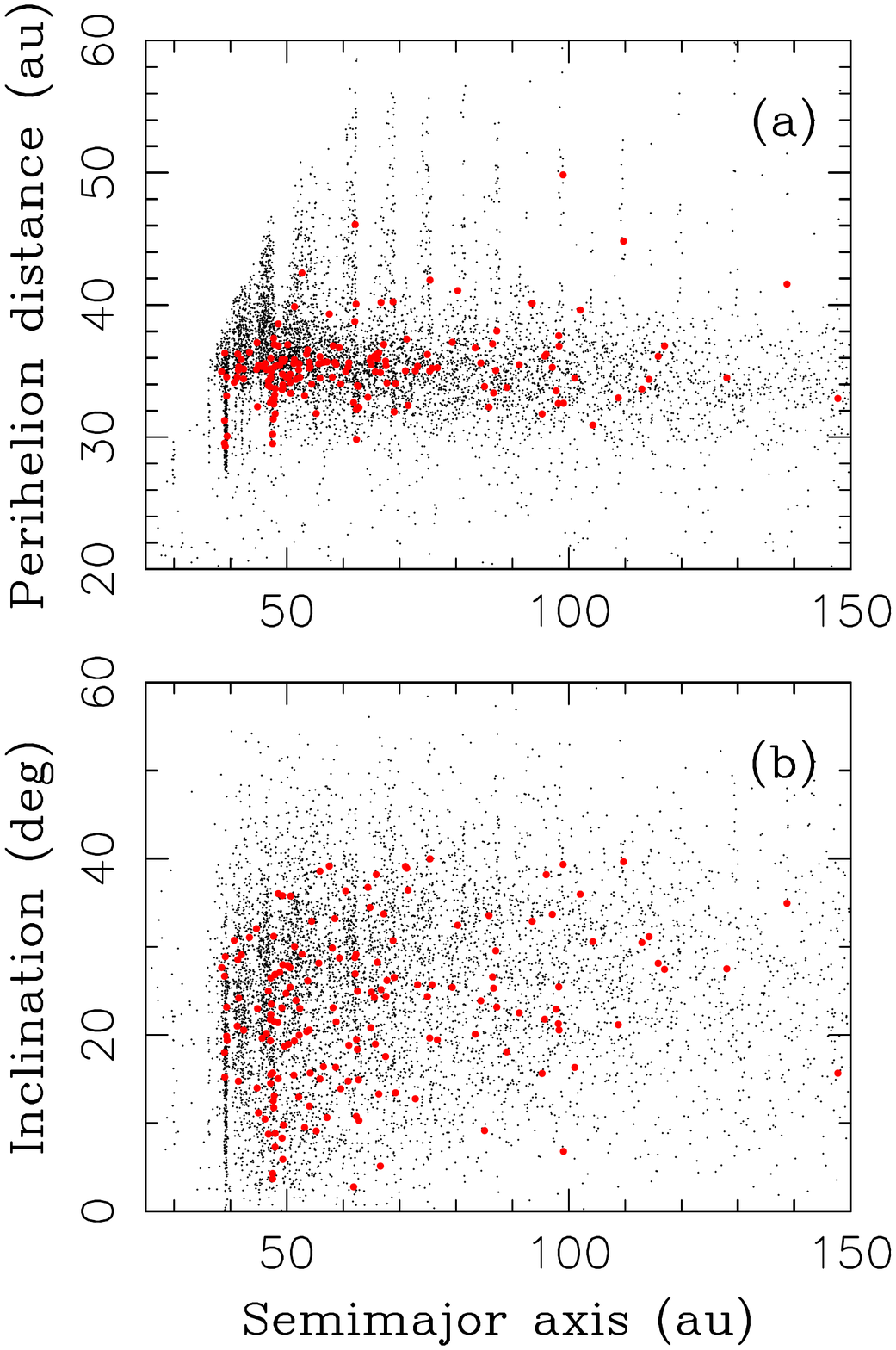}
\epsscale{0.42}
\plotone{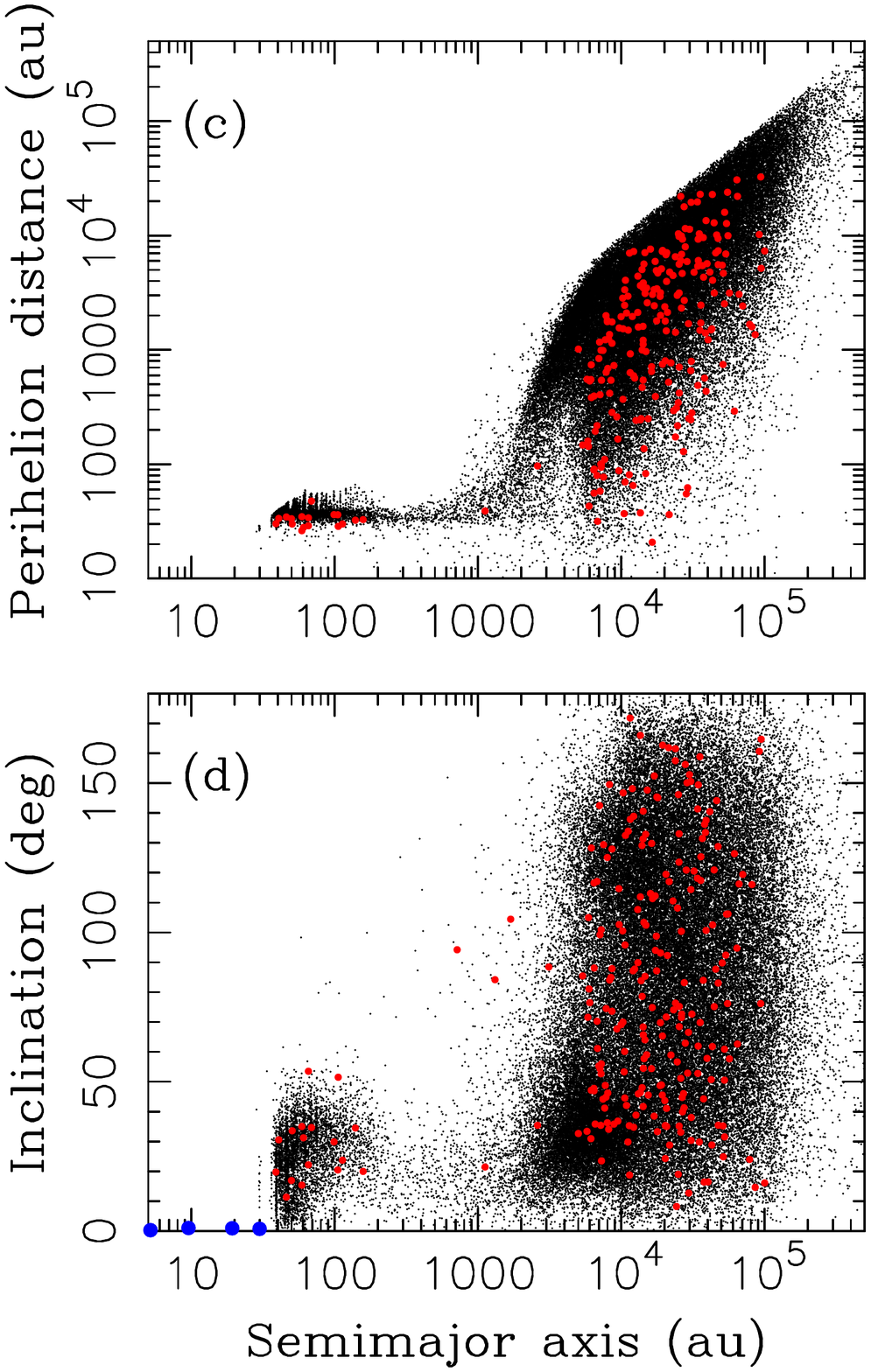}
\caption{The orbits of trans-Neptunian bodies that dynamically evolved to become SPCs in the C1G1S model. The 
source of ECs is shown on the left (panels a and b). The source of HTCs is shown on the right (panels c and d).
ECs were selected using $2<T_{\rm J}<3$, $P<20$~yr and $q<2.5$ au and $N_{\rm p}(2.5)=500$. We identified the 
source orbits of ECs at $t=1.5$ Gyr after the start of the C1G1S integration (i.e., about 3 Gyr ago), and plotted 
them here with red dots. HTCs were selected using $T_{\rm J}<2$, $10<a<20$ au and $q<2$ au and 
$N_{\rm p}(2.5)=3000$. The source orbits of HTCs are plotted at $t=3.5$ Gyr or about 1 Gyr ago.  Background 
orbits are denoted by black dots.}
\label{source}
\end{figure}

\begin{figure}
\epsscale{0.8}
\plotone{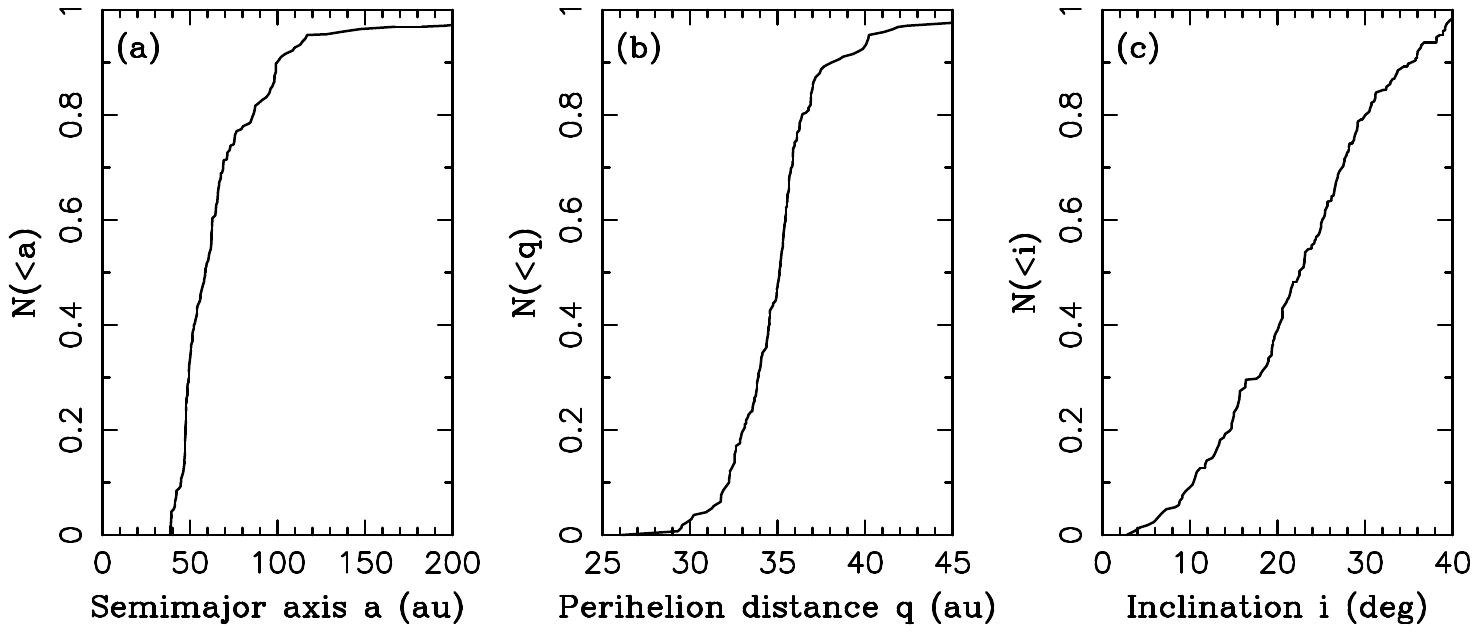}
\caption{The orbital distribution of trans-Neptunian bodies that dynamically evolved to become ECs in our C1G1S model. 
The panels show the cumulative distributions of the semimajor axes (panel a), perihelion distances (panel b), and 
inclinations (panel c). The orbital distributions are shown for $t=1.5$ Gyr or about 3 Gyr ago.}
\label{histo1}
\end{figure}

\begin{figure}
\epsscale{0.8}
\plotone{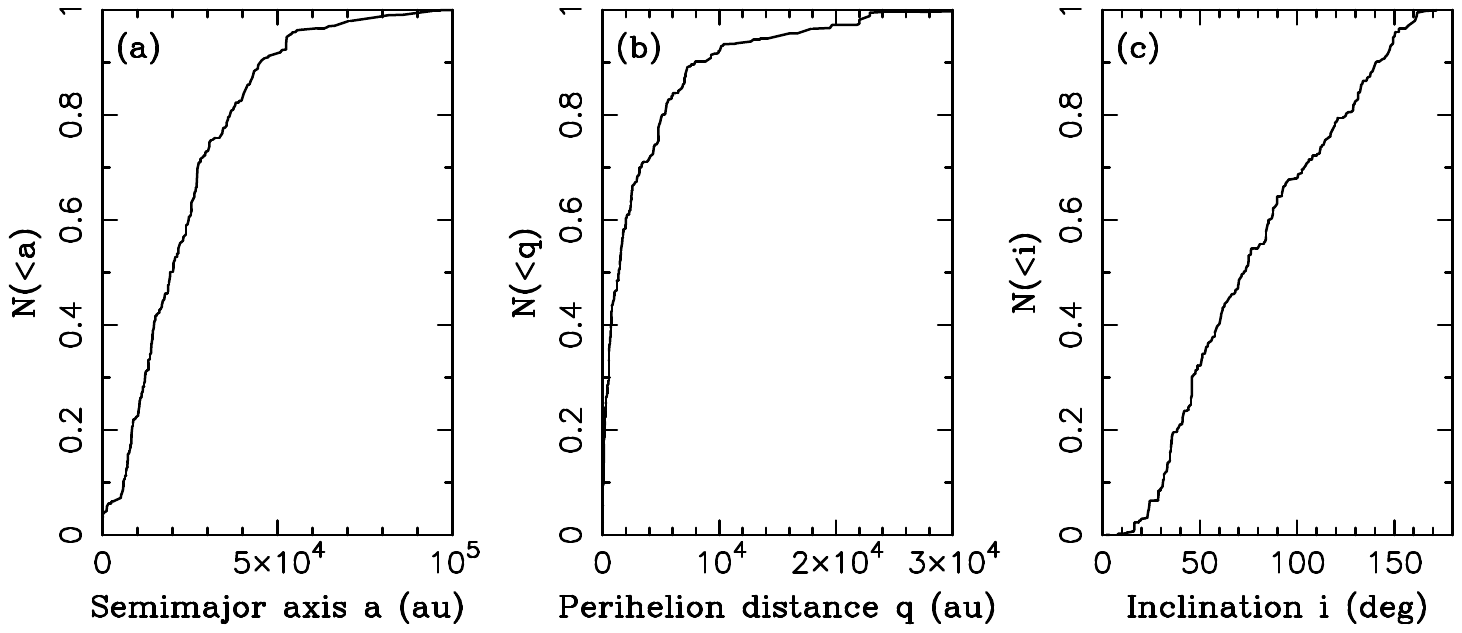}
\caption{The orbital distribution of trans-Neptunian bodies that dynamically evolved to become HTCs in our C1G1S model. 
The panels show the cumulative distributions of the semimajor axes (panel a), perihelion distances (panel b), and 
inclinations (panel c). The orbital distributions are shown for $t=3.5$ Gyr or about 1 Gyr ago.}
\label{histo2}
\end{figure}

\begin{figure}
\epsscale{0.7}
\plotone{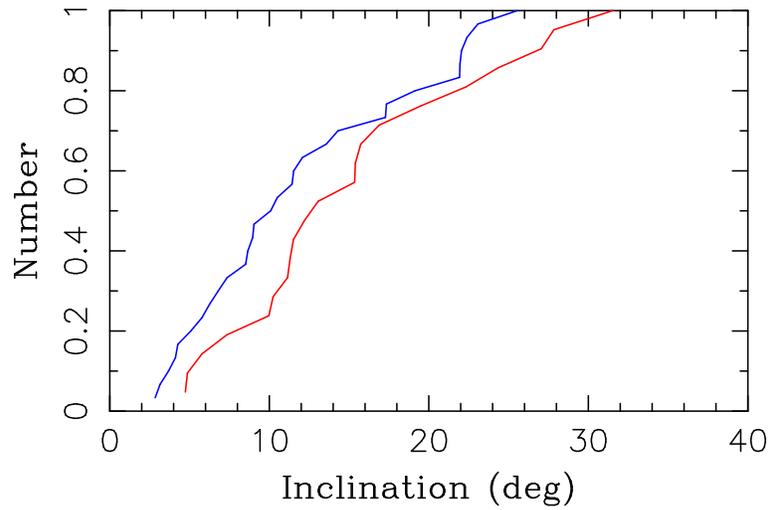}
\caption{The inclination distribution of ECs with $q<2.5$ au and sizes reported in Fern\'andez et al. (2013).
The red and blue lines show the distributions for large ($D>3$ km) and small ($D<3$ km) ECs. According to this 
plot, the inclination distribution of large ECs is broader than that of small ECs, as expected if the $N_{\rm p}$
value of large ECs is greater than that of small ECs.} 
\label{bigsmall}
\end{figure}

\begin{figure}
\epsscale{0.6}
\plotone{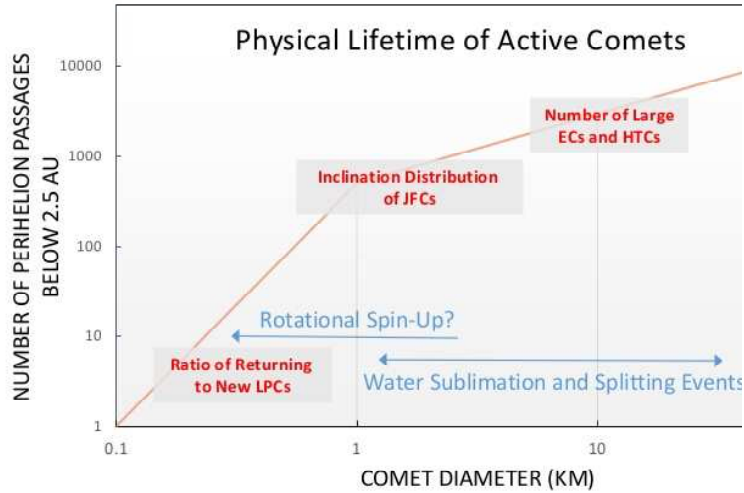}
\caption{A schematic plot showing the suggested dependence of the physical lifetime of comets as a function
of size. The physical lifetime is represented here by the number of perihelion passages below 2.5 au, 
$N_{\rm p}(2.5)$. The red text labels constraints from which the $N_{\rm p}(2.5)$ profile was obtained. 
As we discuss in the main text, $N_{\rm p}(2.5)>1000$ to fit the number of ECs and HTCs with $D>10$ km.
Also, $N_{\rm p}(2.5)=300$-800 to fit the inclination distribution of observed ECs, which are predominantly
$\sim$1 km to a few km in size. The dependence of $N_{\rm p}(2.5)$ on size below 1 km is uncertain. The 
observed ratio of returning-to-new LPCs implies that $N_{\rm p}(2.5)\lesssim10$ for the typical sizes of 
LPCs, which are assumed here to have $D<1$ km. The blue text lists plausible physical
mechanisms that may limit the physical lifetime of comets.} 
\label{fig21}
\end{figure}

The main conclusions of this work are:
\begin{enumerate}
\item The orbital distribution of ECs is well reproduced in our models without P9. With P9, the inclination 
distribution of model ECs is wider than the observed one. Models with $q_9>300$ au could resolve this
issue, but it is not clear whether they could also help to match other constraints (such as the orbits of 
extreme KBOs and the solar obliquity).
\item We find that known HTCs have a nearly isotropic inclination distribution, and appear in the model 
as an extension of the population of returning LPCs to shorter orbital periods. The contribution to 
HTCs from the P9 cloud, if real, would be relatively minor.  
\item The nominal model estimate of the number of large ECs falls short by a factor of $\sim$2-4 when compared 
to observations. This problem can be resolved if large comets have longer physical lifetimes (see below).
The number of large HTCs obtained in the model from the Oort cloud agrees well with observations. 
\item We demonstrate that the physical lifetime of active comets depends on their nuclear size and explain 
how this can help to produce the correct number of large ECs in the model. Combining the analysis of 
ECs, HTCs and LPCs, we estimate that comets a few hundred meters in size should only survive several 
perihelion passages, $\sim$1-km class comets should be active for hundreds of perihelion passages, and 
$\sim$10-km class comets should live for thousands of perihelion passages. [Previously, Di Sisto et al. 
(2009) and Rickman et al. (2017) considered the dependence of physical lifetime of comets on size in 
their models.] 
\item The inner scattered disk at $50<a<200$ au should contain $\sim 1.5\times10^{7}$ $D>10$~km bodies. 
The Oort cloud should contain $\sim 3.8\times10^8$ $D>10$ km comets. These estimates can be extrapolated 
to smaller or larger sizes using the size distribution of Jupiter Trojans (Fig. \ref{sfd}b).      
\end{enumerate}

\acknowledgements
The work of D.N. was supported by NASA's Emerging Worlds program. D.V. was supported by the Czech Grant Agency 
(grant GA13-01308S).  The CPU-expensive simulations were performed on NASA's Pleiades Supercomputer, and on
the computer cluster Tiger at the Institute of Astronomy of the Charles University, Prague. One $10^6$ particle 
simulation over 4.5 Gyr required about 600 hours on 25 Ivy-Bridge nodes (20 cores each). We greatly appreciate 
the support of the NASA Advanced Supercomputing Division. We thank Joel Parker for providing info about the mass
loss of 67P and anonymous reviewer for helpful suggestions to the manuscript.

\clearpage
\begin{table}
\centering
{
\begin{tabular}{lrrrrrrrr}
\hline \hline
                  & $\tau_1$ & $\tau_2$    & $M_9$       & $a_9$ & $e_9$ & $i_9$     & $\rho_0$              & SE \\  
                  & (Myr)    & (Myr)       & ($M_\oplus$) & (au) &       & ($^\circ$) & ($M_{\odot}/{\rm pc}^{3}$)   &     \\  
\hline
C1                & 30       & 100         & 0          & --   &  --   & --         & 0                    & no   \\
C1G1              & 30       & 100         & 0          & --   &  --   & --         & 0.1                  & no   \\
C1G1S             & 30       & 100         & 0          & --   &  --   & --         & 0.1                  & yes  \\
C1G2S             & 30       & 100         & 0          & --   &  --   & --         & 0.2                  & yes   \\
C1M15             & 30       & 100         & 15         & 700  &  0.6  & 30         & 0                    & no   \\
C1M20a            & 30       & 100         & 20         & 500  &  0.5  & 15         & 0                    & no   \\
C1M20b            & 30       & 100         & 20         & 700  &  0.6  & 30         & 0                    & no   \\
C1M20c            & 30       & 100         & 20         & 900  &  0.78 & 30         & 0                    & no   \\
C1I0              & 30       & 100         & 20         & 700  &  0.6  &  0         & 0                    & no   \\
C1ALL             & 30       & 100         & 15         & 700  &  0.6  & 30         & 0.1                  & yes  \\
C2                & 10       & 30          & 0          & --   &  --   & --         & 0                    & no   \\
C2M10             & 10       & 30          & 10         & 700  &  0.6  & 30         & 0                    & no   \\
C2M20             & 10       & 30          & 20         & 700  &  0.6  & 30         & 0                    & no   \\
C2M30             & 10       & 30          & 30         & 700  &  0.6  & 30         & 0                    & no   \\
\hline \hline
\end{tabular}
}
\caption{A summary of the numerical integrations performed in this work. Column SE indicates whether stellar 
encounters were included in each job.}
\end{table}

\clearpage

\clearpage

\clearpage

\clearpage

\clearpage

\clearpage

\clearpage

\clearpage

\clearpage

\clearpage

\clearpage

\clearpage

\clearpage


\clearpage

\clearpage

\clearpage

\clearpage

\clearpage

\clearpage

\clearpage


\begin{thebibliography}

\bibitem[Adams(2010)]{2010ARA&A..48...47A} Adams, F.~C.\ 2010, \araa, 48, 47 

\bibitem[Bailey et al.(2016)]{2016AJ....152..126B} Bailey, E., Batygin, K., \& Brown, M.~E.\ 2016, \aj, 152, 126 

\bibitem[Batygin \& Brown(2016)]{2016ApJ...833L...3B} Batygin, K., \& Brown, M.~E.\ 2016b, \apjl, 833, L3 

\bibitem[Batygin \& Brown(2016)]{2016AJ....151...22B} Batygin, K., \& Brown, M.~E.\ 2016a, \aj, 151, 22 

\bibitem[Batygin et al.(2012)]{2012ApJ...744L...3B} Batygin, K., Brown, M.~E., \& Betts, H.\ 2012, \apjl, 744, L3 

\bibitem[Bottke et al.(2012)]{2012Natur.485...78B} Bottke, W.~F., Vokrouhlick{\'y}, D., Minton, D., et al.\ 2012, \nat, 485, 78 

\bibitem[Bro{\v z} et al.(2013)]{2013A&A...551A.117B} Bro{\v z}, M., Morbidelli, A., Bottke, W.~F., et al.\ 2013, \aap, 551, A117 

\bibitem[Brasser \& Morbidelli(2013)]{2013Icar..225...40B} Brasser, R., \& Morbidelli, A.\ 2013 (BM13), Icarus, 225, 40 

\bibitem[Brasser \& Wang(2015)]{2015A&A...573A.102B} Brasser, R., \& Wang, J.-H.\ 2015, \aap, 573, A102 

\bibitem[Brasser et al.(2006)]{2006Icar..184...59B} Brasser, R., Duncan, M.~J., \& Levison, H.~F.\ 2006, Icarus, 184, 59 

\bibitem[Brasser et al.(2007)]{2007Icar..191..413B} Brasser, R., Duncan, M.~J., \& Levison, H.~F.\ 2007, Icarus, 191, 413 

\bibitem[Brasser et al.(2008)]{2008Icar..196..274B} Brasser, R., Duncan, M.~J., \& Levison, H.~F.\ 2008, Icarus, 196, 274 

\bibitem[Brasser et al.(2010)]{2010A&A...516A..72B} Brasser, R., Higuchi, A., \& Kaib, N.\ 2010, \aap, 516, A72 

\bibitem[Dawson \& Murray-Clay(2012)]{2012ApJ...750...43D} Dawson, R.~I., \& Murray-Clay, R.\ 2012, \apj, 750, 43 

\bibitem[Deienno et al.(2017)]{2017AJ....153..153D} Deienno, R., Morbidelli, A., Gomes, R.~S., \& Nesvorn{\'y}, D.\ 2017, \aj, 153, 153 

\bibitem[Di Sisto \& Brunini(2007)]{2007Icar..190..224D} Di Sisto, R.~P., \& Brunini, A.\ 2007, Icarus, 190, 224 

\bibitem[Di Sisto et al.(2009)]{2009Icar..203..140D} Di Sisto, R.~P., Fern{\'a}ndez, J.~A., \& Brunini, A.\ 2009, Icarus, 203, 140 

\bibitem[Dones et al.(2004)]{2004come.book..153D} Dones, L., Weissman, P.~R., Levison, H.~F., \& Duncan, M.~J.\ 2004, Comets II, 153 

\bibitem[Dones et al.(2015)]{2015SSRv..197..191D} Dones, L., Brasser, R., Kaib, N., \& Rickman, H.\ 2015, \ssr, 197, 191 

\bibitem[Duncan \& Levison(1997)]{1997Sci...276.1670D} Duncan, M.~J., \& Levison, H.~F.\ 1997, Science, 276, 1670 

\bibitem[Duncan et al.(1988)]{1988ApJ...328L..69D} Duncan, M., Quinn, T., \& Tremaine, S.\ 1988, \apjl, 328, L69 

\bibitem[Fern{\'a}ndez et al.(2013)]{2013Icar..226.1138F} Fern{\'a}ndez, Y.~R., Kelley, M.~S., Lamy, P.~L., et al.\ 2013, Icarus, 226, 1138 

\bibitem[Fernandez(1980)]{1980MNRAS.192..481F} Fern\'andez, J.~A.\ 1980, \mnras, 192, 481 

\bibitem[Fern{\'a}ndez et al.(2016)]{2016MNRAS.461.3075F} Fern{\'a}ndez, J.~A., Gallardo, T., \& Young, J.~D.\ 2016, \mnras, 461, 3075 

\bibitem[Francis(2005)]{2005ApJ...635.1348F} Francis, P.~J.\ 2005, \apj, 635, 1348 

\bibitem[Fraser et al.(2014)]{2014ApJ...782..100F} Fraser, W.~C., Brown, M.~E., Morbidelli, A., Parker, A., \& Batygin, K.\ 2014, \apj, 782, 100 

\bibitem[Gladman et al.(2008)]{2008ssbn.book...43G} Gladman, B., Marsden, B.~G., \& Vanlaerhoven, C.\ 2008, The Solar 
System Beyond Neptune, 43 

\bibitem[Gomes(2003)]{2003Icar..161..404G} Gomes, R.~S.\ 2003, Icarus, 161, 404 

\bibitem[Gomes et al.(2004)]{2004Icar..170..492G} Gomes, R.~S., Morbidelli, 
A., \& Levison, H.~F.\ 2004, Icarus, 170, 492 

\bibitem[Gomes et al.(2005)]{2005Natur.435..466G} Gomes, R., Levison, H.~F., Tsiganis, K., \& Morbidelli, A.\ 2005, \nat, 435, 466 

\bibitem[Gomes et al.(2015)]{2015Icar..258...37G} Gomes, R.~S., Soares, J.~S., \& Brasser, R.\ 2015, Icarus, 258, 37 

\bibitem[Gomes et al.(2017)]{2017AJ....153...27G} Gomes, R., Deienno, R., \& Morbidelli, A.\ 2017, \aj, 153, 27 

\bibitem[Hahn \& Malhotra(2005)]{2005AJ....130.2392H} Hahn, J.~M., \& Malhotra, R.\ 2005, \aj, 130, 2392 

\bibitem[Heisler \& Tremaine(1986)]{1986Icar...65...13H} Heisler, J., \& Tremaine, S.\ 1986, Icarus, 65, 13 

\bibitem[Heisler et al.(1987)]{1987Icar...70..269H} Heisler, J., Tremaine, S., \& Alcock, C.\ 1987, Icarus, 70, 269 

\bibitem[Higuchi et al.(2007)]{2007AJ....134.1693H} Higuchi, A., Kokubo, E., Kinoshita, H., \& Mukai, T.\ 2007, \aj, 134, 1693 

\bibitem[Izidoro et al.(2015)]{2015A&A...582A..99I} Izidoro, A., Morbidelli, A., Raymond, S.~N., Hersant, F., \& Pierens, A.\ 2015, \aap, 582, A99 

\bibitem[Jewitt(1997)]{1997EM&P...79...35J} Jewitt, D.\ 1997, Earth Moon and Planets, 79, 35 

\bibitem[Jewitt et al.(2015)]{2015aste.book..221J} Jewitt, D., Hsieh, H., \& Agarwal, J.\ 2015, Asteroids IV, Patrick Michel, 
Francesca E. DeMeo, and William F. Bottke (eds.), University of Arizona Press, Tucson, p. 221 

\bibitem[Jewitt et al.(2016)]{2016ApJ...829L...8J} Jewitt, D., Mutchler, M., Weaver, H., et al.\ 2016, \apjl, 829, L8 

\bibitem[Kaib \& Quinn(2008)]{2008Icar..197..221K} Kaib, N.~A., \& Quinn, T.\ 2008, Icarus, 197, 221 

\bibitem[Kaib \& Quinn(2009)]{2009Sci...325.1234K} Kaib, N.~A., \& Quinn, T.\ 2009, Science, 325, 1234 

\bibitem[Kaib \& Sheppard(2016)]{2016arXiv160701777K} Kaib, N.~A., \& Sheppard, S.~S.\ 2016, AJ, 152, 133 

\bibitem[Kaib et al.(2011)]{2011Icar..215..491K} Kaib, N.~A., Ro{\v s}kar, R., \& Quinn, T.\ 2011, Icarus, 215, 491 

\bibitem[Kenyon \& Bromley(2016)]{2016ApJ...825...33K} Kenyon, S.~J., \& Bromley, B.~C.\ 2016, \apj, 825, 33 

\bibitem[Lai(2016)]{2016AJ....152..215L} Lai, D.\ 2016, \aj, 152, 215 

\bibitem[Lamy et al.(2004)]{2004come.book..223L} Lamy, P.~L., Toth, I., Fernandez, Y.~R., \& Weaver, H.~A.\ 2004, Comets II, 223 

\bibitem[Levison \& Duncan(1994)]{1994Icar..108...18L} Levison, H.~F., \& Duncan, M.~J.\ 1994, Icarus, 108, 18 

\bibitem[Levison \& Duncan(1997)]{1997Icar..127...13L} Levison, H.~F., \& Duncan, M.~J.\ 1997 (LD97), Icarus, 127, 13 

\bibitem[Levison et al.(2001)]{2001AJ....121.2253L} Levison, H.~F., Dones, L., \& Duncan, M.~J.\ 2001, \aj, 121, 2253 

\bibitem[Levison et al.(2002)]{2002Sci...296.2212L} Levison, H.~F., Morbidelli, A., Dones, L., et al.\ 2002, Science, 296, 2212 

\bibitem[Levison et al.(2004)]{2004AJ....128.2553L} Levison, H.~F., Morbidelli, A., \& Dones, L.\ 2004, \aj, 128, 2553 

\bibitem[Levison et al.(2006)]{2006Icar..184..619L} Levison, H.~F., Duncan, M.~J., Dones, L., \& Gladman, B.~J.\ 2006, Icarus, 184, 619 

\bibitem[Levison et al.(2008)]{2008Icar..196..258L} Levison, H.~F., Morbidelli, A., Van Laerhoven, C., Gomes, R., \& Tsiganis, 
K.\ 2008, Icarus, 196, 258 

\bibitem[Levison et al.(2009)]{2009Natur.460..364L} Levison, H.~F., Bottke, W.~F., Gounelle, M., et al.\ 2009, \nat, 460, 364 

\bibitem[Levison et al.(2010)]{2010Sci...329..187L} Levison, H.~F., Duncan, M.~J., Brasser, R., \& Kaufmann, D.~E.\ 2010, Science, 329, 187 

\bibitem[Levison et al.(2011)]{2011AJ....142..152L} Levison, H.~F., Morbidelli, A., Tsiganis, K., Nesvorn{\'y}, D., \& Gomes, R.\ 2011, \aj, 142, 152 

\bibitem[Li \& Adams(2016)]{2016MNRAS.463..393L} Li, G., \& Adams, F.~C.\ 2016, \mnras, 463, 393 

\bibitem[Luu et al.(1997)]{1997Natur.387..573L} Luu, J., Marsden, B.~G., Jewitt, D., et al.\ 1997, \nat, 387, 573 

\bibitem[Malhotra(1993)]{1993Natur.365..819M} Malhotra, R.\ 1993, \nat, 365, 819 

\bibitem[Marchi et al.(2012)]{2012E&PSL.325...27M} Marchi, S., Bottke, W.~F., Kring, D.~A., \& Morbidelli, A.\ 2012, Earth and Planetary Science Letters, 325, 27 

\bibitem[Meech et al.(2004)]{2004Icar..170..463M} Meech, K.~J., Hainaut, O.~R., \& Marsden, B.~G.\ 2004, Icarus, 170, 463 

\bibitem[Morbidelli et al.(2005)]{2005Natur.435..462M} Morbidelli, A., Levison, H.~F., Tsiganis, K., \& Gomes, R.\ 2005, \nat, 435, 462 

\bibitem[Morbidelli et al.(2007)]{2007AJ....134.1790M} Morbidelli, A., Tsiganis, K., Crida, A., Levison, H.~F., \& Gomes, R.\ 2007, \aj, 134, 1790 

\bibitem[Morbidelli et al.(2009)]{2009Icar..202..310M} Morbidelli, A., Levison, H.~F., Bottke, W.~F., Dones, L., \& Nesvorn{\'y}, D.\ 2009, Icarus, 202, 310 

\bibitem[Morbidelli et al.(2012)]{2012E&PSL.355..144M} Morbidelli, A., Marchi, S., Bottke, W.~F., \& Kring, D.~A.\ 2012, Earth and Planetary Science Letters, 355, 144 

\bibitem[Naoz et al.(2013)]{2013MNRAS.431.2155N} Naoz, S., Farr, W.~M., Lithwick, Y., Rasio, F.~A., \& Teyssandier, J.\ 2013, 
\mnras, 431, 2155 

\bibitem[Nesvorn{\'y}(2011)]{2011ApJ...742L..22N} Nesvorn{\'y}, D.\ 2011, \apjl, 742, L22 

\bibitem[Nesvorn{\'y}(2015)]{2015AJ....150...73N} Nesvorn{\'y}, D.\ 2015a, \aj, 150, 73 

\bibitem[Nesvorn{\'y}(2015)]{2015AJ....150...68N} Nesvorn{\'y}, D.\ 2015b, \aj, 150, 68 

\bibitem[Nesvorn{\'y} 
\& Morbidelli(2012)]{2012AJ....144..117N} Nesvorn{\'y}, D., \& Morbidelli, A.\ 2012 (NM12), \aj, 144, 117 

\bibitem[Nesvorny \& Vokrouhlicky(2016)]{2016arXiv160206988N} Nesvorn\'y, D., \& Vokrouhlicky, D.\ 2016 (NV16), ApJ, 825, 94 

\bibitem[Nesvorn{\'y} et al.(2007)]{2007AJ....133.1962N} Nesvorn{\'y}, D., Vokrouhlick{\'y}, D., \& Morbidelli, A.\ 2007, \aj, 133, 1962 

\bibitem[Nesvorn{\'y} et al.(2013)]{2013ApJ...768...45N} Nesvorn{\'y}, D., Vokrouhlick{\'y}, D., \& Morbidelli, A.\ 2013, \apj, 768, 45 

\bibitem[Nesvorn{\'y} et al.(2014)]{2014ApJ...784...22N} Nesvorn{\'y}, D., Vokrouhlick{\'y}, D., \& Deienno, R.\ 2014, \apj, 784, 22 

\bibitem[Nesvorn{\'y} et al.(2016)]{2016ApJ...827L..35N} Nesvorn{\'y}, D., Vokrouhlick{\'y}, D., \& Roig, F.\ 2016, \apjl, 827, L35 

\bibitem[Nesvorn{\'y} et al.(2017)]{2017AJ....153..103N} Nesvorn{\'y}, D., Roig, F., \& Bottke, W.~F.\ 2017, \aj, 153, 103 

\bibitem[Nurmi et al.(2002)]{2002MNRAS.333..835N} Nurmi, P., Valtonen, M.~J., Zheng, J.~Q., \& Rickman, H.\ 2002, \mnras, 333, 835 

\bibitem[Paetzold et al.(2016)]{2016DPS....4811627P} Paetzold, M., Andert, T., Hahn, M., et al.\ 2016, AAS/Division for Planetary 
Sciences Meeting Abstracts, 48, 116.27 

\bibitem[Reach et al.(2007)]{2007Icar..191..298R} Reach, W.~T., Kelley, M.~S., \& Sykes, M.~V.\ 2007, Icarus, 191, 298 

\bibitem[Rickman et al.(2017)]{2017A&A...598A.110R} Rickman, H., Gabryszewski, R., Wajer, P., et al.\ 2017, \aap, 598, A110 

\bibitem[Shankman et al.(2017)]{2017AJ....153...63S} Shankman, C., Kavelaars, J.~J., Lawler, S.~M., Gladman, B.~J., \& Bannister, M.~T.\ 2017, \aj, 153, 63 

\bibitem[Snodgrass et al.(2011)]{2011MNRAS.414..458S} Snodgrass, C., Fitzsimmons, A., Lowry, S.~C., \& Weissman, P.\ 2011, \mnras, 414, 458 

\bibitem[Sosa \& Fern{\'a}ndez(2011)]{2011MNRAS.416..767S} Sosa, A., \& Fern{\'a}ndez, J.~A.\ 2011, \mnras, 416, 767 

\bibitem[Tisserand(1889)]{1889BuAsI...6..289T} Tisserand, F.\ 1889, Bulletin Astronomique, Serie I, 6, 289 

\bibitem[Trujillo \& Sheppard(2014)]{2014Natur.507..471T} Trujillo, C.~A., \& Sheppard, S.~S.\ 2014, \nat, 507, 471

\bibitem[Trujillo et al.(2001)]{2001AJ....122..457T} Trujillo, C.~A., Jewitt, D.~C., \& Luu, J.~X.\ 2001, \aj, 122, 457 

\bibitem[Tsiganis et al.(2005)]{2005Natur.435..459T} Tsiganis, K., Gomes, R., Morbidelli, A., \& Levison, H.~F.\ 2005, \nat, 435, 459 

\bibitem[Vokrouhlick{\'y} et al.(2016)]{2016AJ....152...39V} Vokrouhlick{\'y}, D., Bottke, W.~F., \& Nesvorn{\'y}, D.\ 2016, \aj, 152, 39 

\bibitem[Volk \& Malhotra(2008)]{2008ApJ...687..714V} Volk, K., \& Malhotra, R.\ 2008, \apj, 687, 714-725 

\bibitem[Wang \& Brasser(2014)]{2014A&A...563A.122W} Wang, J.-H., \& Brasser, R.\ 2014, \aap, 563, A122 

\bibitem[Wiegert \& Tremaine(1999)]{1999Icar..137...84W} Wiegert, P., \& Tremaine, S.\ 1999, Icarus, 137, 84 

\bibitem[Wong \& Brown(2015)]{2015AJ....150..174W} Wong, I., \& Brown, M.~E.\ 2015, \aj, 150, 174 

\end{thebibliography}
\end{document}